\documentclass[pre,twocolumn,floatfix, superscriptaddress,a4paper,nofootinbib]{revtex4-1}
\usepackage{amsmath,bm,graphicx}
\usepackage{amssymb}
\usepackage{amsfonts}
\usepackage{graphicx}
\usepackage{epstopdf}
\usepackage{amssymb}
\usepackage{mathrsfs}
\usepackage{amsmath}
\usepackage{color}
\usepackage{latexsym}
\usepackage{amsfonts}
\usepackage{hyperref}
\usepackage{subfigure}
\usepackage{wasysym}
\usepackage{dcolumn}
\usepackage{bm}
\usepackage{natbib}
\usepackage{cancel}
\usepackage{soul}
\usepackage{soul}
\newcommand{\nn}{\nonumber\\}
\newcommand{\eqr}[1]{Eq.~\eqref{#1}}
\usepackage{ulem}

\newcommand{\upd}{\mathrm{d}}
\newcommand{\cit}[1]{Ref.~\cite{#1}}

\begin{document}

%\title{Local electroneutrality breakdown for electrolytes embedded in nanopores}
\title{Driving an electrolyte through a corrugated nanopore}

\author{Paolo Malgaretti}
\email[Corresponding Author: ]{malgaretti@is.mpg.de }
\affiliation{Max-Planck-Institut f\"{u}r Intelligente Systeme, Heisenbergstr. 3, D-70569
Stuttgart, Germany}
\affiliation{IV. Institut f\"ur Theoretische Physik, Universit\"{a}t Stuttgart,
Pfaffenwaldring 57, D-70569 Stuttgart, Germany}
\author{Mathijs Janssen}
\affiliation{Max-Planck-Institut f\"{u}r Intelligente Systeme, Heisenbergstr. 3, D-70569
Stuttgart, Germany}
\affiliation{IV. Institut f\"ur Theoretische Physik, Universit\"{a}t Stuttgart,
Pfaffenwaldring 57, D-70569 Stuttgart, Germany}
\author{Ignacio Pagonabarraga}
\affiliation{Departament de Fisica de la Materia Condensada, Universitat de Barcelona, Carrer Mart\'{\i} i Franqu\'es, 08028-Barcelona, Spain}
\affiliation{CECAM, Centre Européen de Calcul Atomique et Moléculaire, École Polytechnique Fédérale de Lasuanne,
Batochime, Avenue Forel 2, 1015 Lausanne, Switzerland}
\affiliation{Universitat de Barcelona Institute of Complex Systems (UBICS), Universitat de Barcelona, 08028 Barcelona, Spain}
\author{J. Miguel Rubi}
\affiliation{Departament de Fisica de la Materia Condensada, Universitat de Barcelona, Carrer Mart\'{\i} i Franqu\'es, 08028-Barcelona, Spain}
\date{\today}

\begin{abstract}
We characterize the dynamics of a $z-z$ electrolyte embedded in a varying-section channel. 
In the linear response regime, by means of suitable approximations, we derive the  Onsager matrix associated to 
externally enforced gradients in electrostatic potential, chemical potential, and pressure, for both dielectric and conducting channel walls.
We show here that the linear transport coefficients are particularly sensitive to the geometry and the conductive properties of the channel walls  when the Debye length is comparable to the channel width.%, i.e. in the ``entropic electrokinetic regime'' [Phys. Rev. Lett. \textbf{113}, 128301 (2014)].
In this regime, we found that one pair of off-diagonal Onsager matrix elements increases with the corrugation of the channel transport, in contrast to all other elements which are either unaffected by or decrease with increasing corrugation.
Our results have a possible impact on the design of blue-energy devices as well as on the understanding of
biological ion channels through membranes
\end{abstract}

\maketitle

\section{Introduction}\label{sec:Introduction}
Many biological systems~\cite{Albers} and synthetic devices~\cite{Lyderic_Charlaix} rely on the dynamics of electrolytes confined within micro- and nano-pores. 
For example, ion channels~\cite{Calero,Roth2014}, membranes~\cite{Gracheva2017,Bacchin2018}, neuron signaling~\cite{Albers}, plant circulation~\cite{Strook2008}, and
lymphatic~\cite{Nipper2011} and interstitial~\cite{Wiig2012} systems rely on the transport of 
electrolytes across tortuous micro- and nano-pores. 
Recent technological advances have lead to the realization of nanotubes and nanopores of controllable shape~\cite{siria2013giant, secchi2016massive} 
that have been exploited to separate DNA, proteins~\cite{bonthuis2008conformation}, or colloids~\cite{Vinogradova2017}. Likewise, resistive-pulse sensing techniques have been developed to measure properties of tracers transported 
across charged nanopores~\cite{Saleh2003,Ito2004,Heins2005,Arjmandi2012}. 
Moreover, electrolyte-immersed electrodes have
been characterized~\cite{Trizac2016,Reindl2017_1,Reindl2017_2} and realized for novel energy-harvesting devices~\cite{Brogioli2009}. 
Recently it has been shown that novel dynamical regimes appear when the section of the confining vessel is not constant. 
Indeed, asymmetric pores have been used to pump~\cite{Yeh2015} and to rectify ionic currents~\cite{Siwy2005,hanggi,Gomez2015,Keyser2015,Wegrowe2016}. Moreover, recirculation has been reported for electrolytes confined between corrugated walls~\cite{Park2006,mani2009propagation,Malgaretti2014,Chinappi2018}, and the variation in the section of the channels can tune their permeability~\cite{Malgaretti2015,Malgaretti2016}. 
When an electrolyte is driven inside such conduits the local variations in the available space will couple to the local charge and ionic density distribution leading to 
modulations in the mesoscopic properties of the electrolyte such as the electrostatic decay length.

%\red{have others shown this? I thought this was one of the motivations for OUR work?}\textbf{precisely, this is a generic comment (that is true in general) that is like a preview of what we will discuss in the next paragraph...}
\begin{figure}[t]
 \includegraphics[scale=0.43]{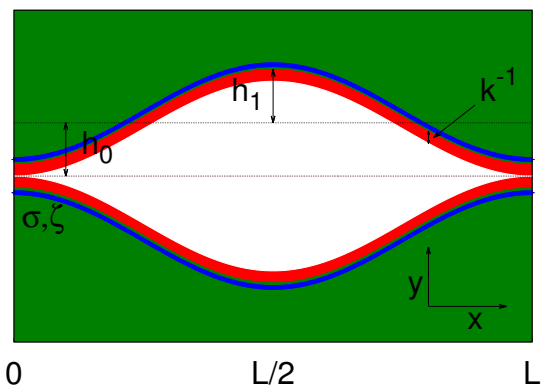}
 \caption{Schematic view of the system. The green regions are the channel walls, the red stripe represents the region, of size Debye length, where the electrostatic potential decays and the blue stripes represent the channel walls.}
 \label{fig:scheme}
\end{figure}
In this article, we show that analytical insight into such corrections can be obtained
for smoothly--varying channel sections. 
In this scenario, we exploit the lubrication approximation and we derive closed expressions for the geometrically--induced corrections to the local electrostatic potential, charge, and ionic density distributions and we identify the fluxes driven by weak external driving forces through applied electric fields, pressure, or salt concentration differences for both conducting and dielectric channel walls.
While for constant section channels the transpo coefficents are unaffected by the wall properties, for varying section channels , we show that the transport coefficeints are generally larger for dielectric walls. Moreover, upon increasing the corrugation of the channel we show that, as expected, the transport coefficients generally decrease. However, for some specific cases we find an increase of the transport coefficients upon increasing the corrugation of the channel.
% 
%we can identify induced first-order corrections to these quantities, 
%the relation between the applied forces and the fluxes thus driven. 

%Our analysis goes beyond the state of the art models that either rely on numerical simulations~\cite{} or on more 
The structure of the text is as follows. In section~\ref{sec:Model} we introduce our model
setup and the framework of electrokinetic equations that we use to describe solute and solvent fluxes.
In section~\ref{sec:Equilibrium} we determine the reference equilibrium scenario.
In section~\ref{sec:nonEquilibrium} we derive the linear transport coefficients 
of a channel driven out of equilibrium and show that the corresponding Onsager matrix is symmetric.
The Onsager matrix encodes for many physical scenarios of possible experimental interest. We discuss several of those scenarios in section~\ref{sec:Results}.
Finally, in section~\ref{sec:Conclusions} we present our conclusions. 

\section{Model}\label{sec:Model}
\subsection{Setup}\label{sec:Setup}
Throughout this article, we analyze a microfluidic channel filled with a $z-z$ electrolyte in a solvent of dielectric constant $\epsilon$. The solvent is incompressible and has a viscosity $\eta$. The channel of length $L$ along the $x$-direction is translationally invariant in the $z$-direction and has a varying pore width in the $y$-direction: 
the channel half section $h(x)$ depends solely on $x$. We write $\bar{h}\equiv(1/L)\int_{0}^{L}h(x)\,\upd x$ 
for the average pore section. %The general theoretical framework that we will develop does not make  further assumptions on the precise shape of the pore other than that the channel is symmetric around $y=0$ {\bf Where do we need symmetry about $y=0$??} and that it has smoothly varying walls [cf. section~\ref{sec:Lubrication_approximation}]. %\red{two immediate systems of interest hourglass-shaped pores and ``chamber pores'' with a widening of the pore width [see Fig.~\ref{fig:scheme}].
% To visualize our results, in the Results section we specify to the latter type of pore} [cf. \eqr{eq:coschannel}].

At $x=0$ and $x=L$, the channel is in contact with two chemostats at electrochemical potentials $\mu^\pm(x=0)$ and $\mu^\pm(x=L)$, respectively. 
Next to differences in these chemical potentials on either side of the channel, also a pressure difference $\Delta P$ or a potential difference $\Delta V$ can be applied.
We only consider isothermal systems at temperature $T$, 
which means that local heat generation \cite{janssen2017coulometry} is neglected. Moreover, our model does not account for  surface conduction in the Stern layer \cite{Werkhoven2018}.

\subsection{Electrokinetic equations}\label{sec:Electrokinetic_equations}
Under the above described conditions, the steady state of our system can be modeled by the classical electrokinetic equations: 
%which combine the Poisson, Nernst-Planck, and Stokes equation.
\begin{subequations}\label{eq:electrokinetic}
\begin{align}
 \epsilon\nabla^{2}\psi(x,y)&=-zeq(x,y)\,,\label{eq:poisson}\\
 \textbf{j}^{\pm}(x,y)&=\rho^{\pm}(x,y)\left[\textbf{v}(x,y)-D\beta \boldsymbol{\nabla} \mu^{\pm}(x,y)\right]\,,\label{eq:nernstplanck}\\
 \boldsymbol{\nabla}\cdot \textbf{j}^{\pm}(x,y)&=0\,,\label{eq:continuity}\\
 \eta\nabla^2 \textbf{v}(x,y)&=-\textbf{F}_{\text{tot}}(x,y)+\boldsymbol{\nabla}P_{\text{tot}}(x)\,,\label{eq:stokes}\\
 \boldsymbol{\nabla}\cdot \textbf{v}(x,y)&=0\,.\label{eq:incompressibility}
\end{align}
\end{subequations}
First, the electrostatic potential $\psi$ inside the channel is determined by the Poisson equation \eqref{eq:poisson}
with $e$ being the elementary charge and with $q(x,y)$ being the local charge number density (m$^{-3}$), which is nonzero whenever there is a difference between cationic  and anionic number densities ($\rho^{+}$ and $\rho^{-}$, respectively),
\begin{equation}
q(x,y)=\rho^{+}(x,y)-\rho^{-}(x,y)\,.
\label{eq:def-q}
\end{equation}
Second, we model the ionic currents $\textbf{j}^{\pm}$ by the Nernst-Planck equation \eqref{eq:nernstplanck}, 
which accounts for ionic transport by advection, diffusion, and electromigration. 
Equation~\eqref{eq:nernstplanck} describes the dynamics of point-like ions: 
this represents a fair approximation for dilute electrolytes in small electric fields. 
Third, when the system is driven by an external force such as a pressure drop or an electrostatic field, 
the electrolyte solution will flow ---at low Reynolds number--- according to the Stokes equation \eqref{eq:stokes}, 
%\begin{equation}
%\eta\nabla^2v_x(x,y)=-F_{\text{tot}}(x,y)+\partial_xP_{\text{tot}}(x)\,,
%\label{eq:stokes}
%\end{equation}
with $\partial_xP_{\text{tot}}(x)=\partial_xP(x)+\Delta P/L$, where $\partial_x P$ 
is the $x$-component of the geometrically-induced local pressure gradient that is determined by the boundary conditions and by fluid incompressibility and where
\begin{equation}
F_{\text{tot}}(x,y)=-zeq(x,y)\partial_x \psi(x,y)
\label{eq:def-tot-E0}
\end{equation}
is the $x$-component of the total electrostatic force density acting on the fluid. 
Finally, Eqs.~\eqref{eq:continuity} and \eqref{eq:incompressibility} represent 
the steady-state continuity equation and the incompressibility equation, respectively.

Equations~\eqref{eq:electrokinetic} are subject to the following boundary conditions
\begin{subequations}\label{eq:electrokineticbc}
\begin{align}
&\psi\to
\begin{cases}
\psi(x,\pm h(x))&=\zeta \quad\quad\text{cond}\,,\\
\mathbf{n}^\pm\cdot \nabla\psi(x,y)|_{y=\pm h(x)}&=\pm e\sigma \quad\text{diel}\,,
\end{cases}\\
&\mathbf{n}^\pm\cdot\mathbf{j}^{\pm}(x, \pm h(x))=0\,,\label{eq:noflux}\\%=j_{y}^{\pm}(x,-h(x))
&\textbf{v}(x, \pm h(x))=0\,,\label{eq:noslip}%=\textbf{v}^{\pm}(x,-h(x))
\end{align}
\end{subequations}
where $\mathbf{n}^\pm$ is the local normal at the channel walls.
Here, the boundary conditions on $\psi(x,y)$ depend on the conductive properties of the channel walls. 
For dielectric channel walls, the case for pores made from polymeric materials such as PDMS, we impose a constant surface charge $e\sigma$ whereas for conducting walls, such as carbon nanotubes, we impose a constant $\zeta$ potential. 
We denote the electrostatic potential in either case  $\psi^{\zeta}(x,y)$ or $\psi^{\sigma}(x,y)$, accordingly. 
Note that while for flat channels either choices are related via the capacitance, 
in the corrugated case, a constant $\zeta$ potential leads to a $x$-dependent surface charge $e\sigma(x)$ [cf.~\eqr{eq:sigma-cond}], 
while a constant surface charge gives rise to a varying surface potential $\psi^{\sigma}(x,h(x))$ [cf.~\eqr{eq:electric_potential-0-1}].
%The above superscript notation is also used for other variables: when $X$ is specified to either boundary condition we write $X^{\zeta}$ or $X^{\sigma}$, respectively.
The same superscript notation is also used for other variables whenever we specify quantities to either boundary conditions.
Equations \eqref{eq:noflux} and \eqref{eq:noslip} represent the no-flux and no-slip boundary conditions at the channel walls of the solute and solvent, respectively.

\subsection{Lubrication approximation}\label{sec:Lubrication_approximation}
In the following we restrict to pores whose section vary smoothly. This allows us to identify a separation between longitudinal and transverse length scales according to which changes of $\psi$ and $v_x$ along the $x$-direction are much smaller than those along the $y$-direction. This facilitates an essential simplification of Eqs.~(\ref{eq:poisson}) and (\ref{eq:stokes}) where $\partial_{x}^{2}$ terms therein become negligible as compared to $\partial_{y}^{2}$ terms.
Thanks to this ``lubrication-like'' approximation both \eqr{eq:poisson} and \eqr{eq:stokes} become analytically solvable.
In order to apply the lubrication approximation consistently to both the Stokes and Poisson equation, we need to identify a common small parameter. 
While the relevant longitudinal length scale of both the Stokes equation and the Poisson equation  is the channel length $L$, different transverse length scales appear in these equations: the average channel section $\bar{h}$, for the Stokes equation, and the screening length $\lambda$, for the Poisson equation. To proceed, we nondimensionalize the length scales via: $x=x_* L$ and $h(x)=h_*(x) \bar{h}$, while for the transverse direction we use either $y=y_* \bar{h}$ or $y=y_\star \lambda$. 
We then write the Stokes equation as
\begin{equation}
 \frac{\bar{h}^2}{L^2}\frac{\partial^2 v_x}{\partial x_*^2 }+\frac{\partial^2 v_x}{\partial y_*^2}=\frac{\bar{h}^2}{\eta}\left[-F_{\text{tot}}(x,y)+\partial_xP_{\text{tot}}(x)\right]\,,
 \label{eq:stokes2}
 \end{equation}
and the Poisson equation as
\begin{equation}
\frac{\lambda^2}{\bar{h}^2}\frac{\bar{h}^2}{L^2}\left[\frac{\partial^2 \psi}{\partial h_*^2 }\left(\frac{\partial h_*}{\partial x_*}\right)^{\!\!2}+\frac{\partial \psi}{\partial h_*}\frac{\partial^2 h_*}{\partial x_*^2}\right]+\frac{\partial^2 \psi}{\partial y_{\star}^{2}}=-\lambda^2\frac{zeq}{\epsilon}\,.
\label{eq:poisson2}
\end{equation}
A \textit{first order} lubrication approximation to \eqr{eq:stokes2} in the small parameter $\bar{h}/ L\ll1$
amounts to dropping the term of order $\mathcal{O}\left(\bar{h}^2/L^2\right)$ (first term on the left-hand side).
%, while the right hand side of order $\mathcal{O}\left(\bar{h}/L\right)$ is retained.
Similarly, in \eqr{eq:poisson2} we neglect the term of order $\mathcal{O}\left(\lambda^2/L^2\right)$ 
(first term on the left-hand side), 
requiring the smallness of $\lambda/L$; 
hence, this term is of $\mathcal{O}\left(\bar{h}/L\right)^2$ as compared to the second one, 
provided that $\lambda^2/\bar{h}^2 \lesssim \mathcal{O}(1)$. 
%\red{Relation to $\partial_{x}h\ll1$?}

We  have exploited the nondimensionalized Eqs.~\eqref{eq:stokes2} and \eqref{eq:poisson2} to identify the magnitude of the different terms when the pore section is smoothly varying.
However, since in the following we are going to make expansions in several small parameters, we continue our analysis with the dimensionful equations. 
This approach has the advantage that it allows us to keep track of all these small parameters on equal footing.

\section{Equilibrium}\label{sec:Equilibrium}
At equilibrium the electrochemical potential~\cite{RusselBook} is constant:
\begin{equation}
 \beta \mu_\text{eq}^{\pm} = \ln\left[\Lambda_{\pm}^3 \rho^{\pm}(x,y)\right]\pm \beta ze \psi(x,y)\,,
 \label{eq:def-mu}
\end{equation}
with $\beta=(k_BT)^{-1}$ being the inverse thermal energy and $\Lambda_{\pm}$ the cationic and anionic thermal De Broglie wavelengths, 
which we consider to be equal $\Lambda_{+}=\Lambda_{-}=\Lambda$. This implies that 
 \begin{align}
  \rho^{\pm}(x,y)&= \varrho^{\pm}(x)\exp{[\mp\beta ze \psi(x,y)]}\,,
 \label{eq:def-dens}
 \end{align}
with 
\begin{align}
  \varrho^{\pm}(x)=\frac{\exp{[\beta \mu^{\pm}_{\rm eq}]}}{\Lambda^{3}}\,.
 \label{eq:rho-pm-0}
\end{align}
In order to get analytical insight we assume that the electrostatic potential is weak $\beta ze\psi(x,y)\ll1$, 
i.e., we apply the Debye-H\"uckel approximation. For later convenience we retain contributions up to second order in $\beta ze\psi(x,y)$, hence the number densities of positive and negative ions read
% \begin{equation}
% \rho^{\pm}(x,y)  =   \varrho_{0}^{\pm}\left[1\mp\beta ze\psi(x,y)\right]\,.
% \label{eq:def-dens-lin}
% \end{equation}
% 
%  such that the ionic number densities read:
\begin{align}
\frac{\rho^{\pm}(x,y)}{ \varrho^{\pm}}  &=  1\mp\beta ze\psi(x,y)+\frac{1}{2}\left(\beta ze\psi(x,y)\right)^2%\nn&
+\mathcal{O}(\psi^3)\,.
\label{eq:def-dens-lin-0}
\end{align}
In order to simplify the notation we choose the zero of the electrostatic potential 
such that  we have $\mu^{+}_{\rm eq}=\mu^{-}_{\rm eq}\equiv\mu_{\rm eq}$ when the $z-z$ electrolyte is globally electroneutral in the reservoirs. Hence we have:
\begin{equation}
  \varrho^{+}= \varrho^{-}\equiv \varrho\,.
 \label{eq:psi-0}
\end{equation}
From hereon, we denote the expansion of a general variable $X(x,y)$ in the small parameter $\bar{h}/L$ as $X=X_0+X_1+\mathcal{O}(\bar{h}/L)^2$; hence, 
for instance $\psi(x,y)=\psi_{0}(x,y)+\mathcal{O}(\bar{h}/L)^2$ and $\rho^{\pm}(x,y)=\rho_{0}^{\pm}(x,y)+\mathcal{O}(\bar{h}/L)^2$.
Accordingly, at leading order in the lubrication expansion, we retain only the first terms of the above expansions and Eq.\eqref{eq:poisson} reads:
\begin{equation}
\partial_{y}^{2}\psi_0(x,y)= k_{0}^{2}\psi_0(x,y)+\mathcal{O}(\psi_0^3)\,,
\label{eq:poisson1-0}
\end{equation}
where $k_0=\sqrt{\beta (ze)^2 \gamma_0/\epsilon}$ is the inverse Debye length and $\gamma_0=2\varrho_0$ the salt number density. 
In order to keep notation as simple as possible, from hereon we omit the $\mathcal{O}(\psi_0^n)$ and we reintroduce it only when necessary.
Finally, the electrostatic potential for conducting channel walls reads:
\begin{equation}
 \psi^\zeta_0(x,y)=\zeta\frac{\cosh\left[k_0y\right]}{\cosh\left[k_0h(x)\right]}\,,
 \label{eq:electric_potential-cond}
\end{equation}
while for dielectric walls it reads:
\begin{equation}
\psi^\sigma_0(x,y)=\frac{e\sigma}{\epsilon k_0}\frac{\cosh\left[k_0y\right]}{\sinh\left[k_0h(x)\right]}\,.
\label{eq:electric_potential-0-1}
\end{equation}
While we enforced global electroneutrality [cf.~above \eqr{eq:psi-0}], local electroneutrality 
---the balance of the total ionic charge $ze\bar{q}(x)$ in a slab located at $x$ by a corresponding amount of opposite local surface charge $2e\sigma(x)$---  
can now be discussed.
%We denote the net ionic charge %$\mathfrak{Q}(x)$ 
% in a slab located at $x$ by , 
Here, $\bar{q}(x)$ is the cross-sectional total unit charge,  %(C m$^{-2}$)
\begin{align}
 &\bar{q}(x)=\int_{-h(x)}^{h(x)}\!\!q(x,y)\,\upd y\,.%=-\int_{-h(x)}^{h(x)}\!\frac{\epsilon k_0^2 \psi_{0}}{ze}\,dy\,.
  \label{eq:mathfrakQ}
\end{align}
%\red{there was a minus sign mistake in your (Paolo) version?}
For conducting walls, at lowest order in lubrication, this amounts with Eqs.~\eqref{eq:def-dens-lin-0} and \eqref{eq:electric_potential-cond} to
\begin{align}
 &\bar{q}_{0}^{\zeta}(x)=-2\epsilon k_0 \zeta \tanh[k_0 h(x)]\,.  
 \label{eq:q0cond} 
\end{align}
We remark that, at first order in lubrication, the surface charge at each conducting wall can be obtained by
\begin{equation}
e\sigma^\zeta(x)=\pm\epsilon \partial_y\psi_0^\zeta(x,y)|_{y=\pm h(x)}=\epsilon k_0 \zeta \tanh[k_0 h(x)]\,.
 \label{eq:sigma-cond}
\end{equation}
%\red{why this first equation? It now seems at though $\sigma$ is $x$ independent, while it is not?}
For dielectric walls we have:
\begin{align}
 &\bar{q}_{0}^{\sigma}(x)=-2e\sigma\,.
 \label{eq:q0diel} 
\end{align}
Eqs.~(\ref{eq:q0cond})-(\ref{eq:q0diel}) show that local charge neutrality is attained.
The presented theory is thus not able to reproduce the recently discovered electroneutrality breaking in narrow confinement \cite{luo2015electroneutrality}. To account for that, the authors of Refs.\cite{luo2015electroneutrality,colla2016charge} had to include additional interactions beyond the ones of our model.

\section{Transport}\label{sec:nonEquilibrium}
From hereon we characterize the electrolyte-filled 
corrugated nanochannel driven out of equilibrium by applied external forces $\Delta P/L$, $ze\Delta V/L$, and $\Delta \mu/L$. 
We assume these external forces to be small, which means that 
$\beta LL_z \bar{h} \Delta P\ll1$, $\beta ze\Delta V\ll1$, and $\beta \Delta \mu\ll1$, where $L_z$ is the 
thickness of the channel along the $z$ direction. 

\subsection{Stokes}
At leading order in lubrication, the solution $v_x(x,y)$ of the Stokes equation [Eq.~\eqref{eq:stokes}]
subject to no-slip boundary conditions  \eqr{eq:noslip} reads:
\begin{subequations}\label{eq:vel_field_x-y}
\begin{align}
v_x(x,y)&= u_{P}(x,y)+u_{\textrm{eo}}(x,y)\,,\\
u_{P}(x,y)&=\frac{\partial_xP_{\text{tot}}(x)}{2\eta}\left[y^{2}-h(x)^{2}\right]\,,\label{eq:vxpressure}\\
u_{\textrm{eo}}(x,y)&=\mathcal{U}(x,y)-\mathcal{U}(x,h(x))\,,\label{eq:ueo}\\
\mathcal{U}(x,y)&\equiv\frac{ze}{\eta}\int \upd y \int \upd y \,q(x,y) \partial_x\psi(x,y)\label{eq:curlyV}\,,	
\end{align}
\end{subequations}
where we partitioned the velocity $v_x(x,y)$ into a pressure-driven contribution $u_{P}$ and an electroosmotic contribution $u_{\textrm{eo}}$, that arrises when ions in an electric field drag along the fluid.
The local pressure gradient 
appearing in \eqr{eq:vxpressure}, $\partial_x P_\text{tot}(x)=\partial_xP(x)+\Delta P/L$,  
accounts for both the pressure drop $\Delta P$ from $x=L$ to $x=0$ as well as for the local pressure $P(x)$, that ensures fluid incompressibility [\eqr{eq:incompressibility}]. 
% In particular, $\mathcal{U}(x,y)$ is defined as 
% \begin{align}
% \mathcal{U}(x,y)&=\frac{ze}{\eta}\int dy \int dy \,q(x,y) \partial_x\psi(x,y)\,.
% \label{eq:curlyV}
% \end{align}
Inserting \eqr{eq:vel_field_x-y} into the volumetric fluid flow, 
\begin{equation}
 Q = \int_{-h(x)}^{h(x)} v_x(x,y)\,\upd y\,,%=\frac{\tilde Q}{L_z}\,.
 \label{eq:Q}
\end{equation}
 and performing the $y$-integral over $u_P$ leads to an expression for $\partial_x P_\text{tot}(x)$:
\begin{equation}
 \frac{2}{3\eta}\partial_x P_\text{tot}(x)=\frac{1}{h^3(x)}\left[\int_{-h(x)}^{h(x)}u_{\textrm{eo}}(x,y)\,\upd y-Q\right].
 \label{eq:partialxPtot}
\end{equation}
%is constant along the channel ($\partial_x Q=0$), and we moreover 
Integrating the last expression over $\int_{0}^{L}\upd x$, imposing fluid incompressibility
$\partial_x Q=0$, and using
\begin{equation}
\int_0^L \partial_x P_\text{tot}(x) \,\upd x =\Delta P\,,
\label{eq:BC-DP}
\end{equation}
which follows from the boundary conditions on the pressure, leads to 
\begin{subequations} \label{eq:Q-lin}
\begin{align}
 Q&\equiv Q_P+ Q_{\textrm{eo}}\,,\\
 Q_P&=-\frac{2}{3H_{3}}\frac{ \bar{h}^3 \Delta P}{\eta L}\,,\label{eq:QP}\\
 Q_{\textrm{eo}}&=\frac{\bar{h}^3}{H_{3}L}\int_0^L\frac{\upd x}{h^3(x)}\int_{-h(x)}^{h(x)}u_{\textrm{eo}}(x,y)\,\upd y\label{eq:QEO}\,,
\end{align}
\end{subequations}
where $Q_{P}$ is the pressure-driven volumetric fluid flow, $Q_{\textrm{eo}}$ the electroosmotic flow, and where
\begin{equation}\label{eq:H3}
H_{3}\equiv \frac{\bar{h}^{3}}{L}\int_{0}^{L}\!\!\frac{1}{h(x)^{3}}\,\upd x%\limits
\end{equation}
is a dimensionless geometrical measure for the corrugation of the channel. We find $H_{3}\ge 1$, with the equality holding when the channel is flat $[h(x)=\bar{h}]$. 
 %\red{This is somehow a harmonic mean of higher order. Or an $L_p$ norm with $p=-3$, Googling these things gives no results.}
Hence, for a flat channel, \eqr{eq:QP} simplifies to the standard result $Q_P=-2\bar{h}^3\Delta P/(3\eta L)$ of a Poisseuille flow between two flat plates.
Finally, in order to determine $u_{\rm eo}$ (and $Q_{\rm eo}$) we need to characterize the ionic transport. 
 
\subsection{Small-force expansions}
%We start by deriving a general solution to the Poisson equation when the ionic densities are weakly perturbed by the forcing.
For weak external forces, within the Debye-H\"uckel regime and at first order in lubrication, we expand the nonequilibrium 
electric potential, charge densities and electrochemical potential about their equilibrium values:
%\green{I think we should say: we expand phi and rho. Inserting this into the chemical potential we find an expansion of mu in terms of the expanded rho and phi. So at this point we do not mention mu yet}
\begin{subequations}
\begin{align}
 \psi(x,y)&=\psi_0(x,y)+\psi_{0,f}(x,y)+\mathcal{O}(f^2)+\mathcal{O}(\psi_0^3)\,,\label{eq:def-phi-exp-f}\\%+\mathcal{O}\left(\frac{\bar{h}}{L}\right)+\mathcal{O}(f)^2\\
 \rho^{\pm}(x,y) &=  \rho_{0}^{\pm}(x,y)+\rho^{\pm}_{0,f}(x,y)+\mathcal{O}(f^2)+\mathcal{O}(\psi_0^2).
\label{eq:def-rho-exp-f}
\end{align}
\end{subequations}
%While the chemical potential at equilibrium is exact, 
Here, both $\psi_0$ and $\rho^\pm_0$ carry corrections of $\mathcal{O}(\psi_0^3)$.
Hence, this expansion for small values of $f$ about the Debye-H\"uckel solution is meaningful provided that contributions of order $\mathcal{O}(f)$ are larger than those of order $\mathcal{O}(\psi_0^3)$.  For notation ease, in all $\mathcal{O}(f)$ 
terms that we write from hereon, we drop mentioning the lubrication approximation, 
in particular $\rho^{\pm}_{0,f}\to \rho^{\pm}_{f}$ and $\psi_{0,f}\to \psi_{f}$.
Inserting \eqr{eq:def-rho-exp-f} into \eqr{eq:def-mu} we find an expansion of the chemical potential,
 $\mu^\pm(x,y)=\mu^\pm_0+ \mu^\pm_f(x,y)+\mathcal{O}(f^2)$
with 
\begin{align}\label{eq:electrochemicalpotential}
 \beta \mu^\pm_f(x,y)&=\frac{\rho^\pm_f(x,y)}{\rho^\pm_0(x,y)}\pm \beta ze \psi_f(x,y)\,.
\end{align}
% \begin{align}\label{eq:Xi}
%  \Xi(x,y)&=\frac{1}{L}\int_{0}^{x} \chi(x',y) dx'\,,
% \end{align}
%with $\Xi(x)$ given in \eqr{eq:chi}. 
Assuming a small transverse Peclet number ($\bar{h} v_{y}/D\ll1$),
the steady state is achieved by systems that are in \textit{local equilibrium} $\partial_{y}\mu(x,y)=0$ in every section of the channel located $x$.
Accordingly, using \eqr{eq:def-dens-lin-0} leads, at linear order in $\psi_0$, 
to the density profiles 
\begin{align}
\rho^{\pm}_{f}(x,y)&=   \varrho_{0}\left[\beta \bar{\mu}^{\pm}_f(x)\mp\beta ze \psi_f(x,y)\right]\left[1\mp\beta ze\psi_0(x,y)\right]\,.\label{eq:def-rho-exp-f2}
%&
%& \quad+\mathcal{O}(f^2)+\mathcal{O}(f\psi_0^2)\,.\label{eq:def-rho-exp-f2}
\end{align}
With $\mu^\pm_f$ we define the intrinsic (electro)chemical potential as $ \bar{\mu}^{\pm}_f(x)\equiv \mu^\pm_f(x,y)$.
It is important to remark that the contribution contained in \eqr{eq:def-rho-exp-f2} are of lower order than those disregarded in  \eqr{eq:def-rho-exp-f} and that those contributions disregarded in \eqr{eq:def-rho-exp-f2} are of the same (or higher) order as those disregarded in \eqr{eq:def-rho-exp-f}.

% \cit{VanRoij2016}, and might be a lead into understanding the problem I sketch in Sec.\ref{sec:Peters}}

\subsection{Transport equations}
The steady-state continuity \mbox{equation \eqref{eq:continuity}}, together with the no-flux boundary condition 
\eqr{eq:noflux}, implies the $x$-independence of the following cross-sectional integrals 
\begin{align}
J^{\pm}&=\int_{-h(x)}^{h(x)}\! j_{x}^{\pm}(x,y) \,\upd y\,, %& = \frac{\tilde{J}_{\rho^{+}}}{L_z}= J_{c_{+}}
\label{eq:coupled_fokker_planck}
\end{align}
which represent the total ionic fluxes through a slab at $x$.
In Appendix~\ref{app:cfp} we find expressions for the solute $J_{c}=J^{+}+J^{-}$ 
and charge $J_{q}=J^{+}-J^{-}$ fluxes by inserting Eqs.~\eqref{eq:nernstplanck} and \eqref{eq:def-rho-exp-f2} into \eqr{eq:coupled_fokker_planck},
\begin{subequations}\label{eq:lmbd-eps-int}
\begin{align}
\frac{J_{c}}{D}&=\frac{ \gamma_{0}Q}{D}+\beta ze\overline{\psi_0}(x)\partial_x \xi_{f}(x)-2h(x)\partial_x \gamma_{f}(x)\nn
&\quad+\mathcal{O}(f^2)\,,\label{eq:lmbd-eps-int1}\\
\frac{J_{q}}{D}&=\frac{\mathcal{J}_{q} (x)}{D}+\beta ze\overline{\psi_0}(x)\partial_x \gamma_{f}(x)-2h(x)\partial_x \xi_{f}(x)\nn
&\quad+\mathcal{O}(f^2)\,,\label{eq:lmbd-eps-int2}
\end{align}
\end{subequations}
where $  \gamma_{f}(x),   \xi_{f}(x), \overline{\psi_0}(x)$, and $\mathcal{J}_{q} (x)$\footnote{$ze\mathcal{J}_{q}(x)$ coincides with the usual definition of the streaming current $I_{str}$ [see for instance Eq.~(1) of \cite{Dekker2005}] if the channel is flat, but differs from $I_{str}$ for a corrugated channel.} are defined as
\begin{subequations}
\begin{align}
  \gamma_{f}(x)&= \varrho_{0}\beta \left[\bar\mu^+_f(x)+\bar\mu^-_f(x)\right]\,,\label{eq:def-varphi}\\
  \xi_{f}(x)&= \varrho_{0}\beta\left[ \bar\mu^+_f(x)- \bar\mu^-_f(x)\right]\,,\label{eq:barmupm}\\
\overline{\psi_0}(x)&\equiv\int_{-h(x)}^{h(x)}\! \psi_{0}(x,y) \,\upd y\,,\\%\limits
 \mathcal{J}_{q} (x)&\equiv\int_{-h(x)}^{h(x)}q_0(x,y) v_{x}(x,y) \,\upd y\,.  \label{eq:Q-Q}
\end{align}
\end{subequations}
%We provide a physical interpretation of the function $\mathcal{J}_{q}$ in Appendix~\ref{sec:Jq}.

We now proceed as follows: from \eqr{eq:lmbd-eps-int} we will derive expressions for $ \gamma_{f}(x)$ and $ \xi_{f}(x)$
in terms of the fluxes $J_{c}$, $J_{q}$, and $Q$ [cf. Eqs.~\eqref{eq:cf} and \eqref{eq:qf}]. Since 
$\gamma_{f}(x)$ and $ \xi_{f}(x)$ are defined in terms of the intrinsic chemical potentials 
$\bar\mu^{\pm}_f(x)$---which must adhere to externally enforced boundary values $\bar\mu^{\pm}_f(0)$ and $\bar\mu^{\pm}_f(L)$, 
these expressions can in turn be inverted to yield the fluxes in terms of driving forces.  
%and $\mathcal{J}_{q} (x)$ [cf.~\eqr{eq:Q-Q}] in the first term of \eqr{eq-psi-phi2}(b).
%Neglecting from hereon $\mathcal{O}(f^2)$ terms, 
To do all that, we start by rewriting \eqr{eq:lmbd-eps-int2},
\begin{align}
\partial_x \xi_{f}(x)&=\frac{\beta ze\overline{\psi_0}(x)}{2h(x)}\partial_x \gamma_{f}(x)+\frac{\mathcal{J}_{q,f} (x)-J_{q,f}}{2Dh(x)}\,,\label{eq:partialpsi}
\end{align}
where $J_{q} =J_{q,f} +\mathcal{O}(f^2)$ and $\mathcal{J}_{q} =\mathcal{J}_{q,f} +\mathcal{O}(f^2)$. With a slight abuse of notation, we drop the subscript $f$ in $J_{q,f}$ and $\mathcal{J}_{q,f}$ from hereon, and we will do the same for $J_{c,f}$ and $Q_{f}$, which are defined analogously to $J_{q,f}$ and  $\mathcal{J}_{q,f}$.
Moreover, again for notation ease, instead of $J_{c}$ itself from hereon we will consider the ``excess'' solute flow not caused by advection,
$J_{c}'=J_{c}-\gamma_{0}Q$.
 %Section \ref{sec:Peters} discusses how the Onsager relations---which we will derive shortly---can be cast in terms of $J_{c}$ instead. 
Inserting \eqr{eq:partialpsi} into \eqr{eq:lmbd-eps-int1} we find 
\begin{align}\label{eq:partialvarphif}
\partial_x &\gamma_{f}(x)
=-\frac{1}{2Dh(x)}\left[J_{c}'+J_{q}\frac{\beta ze\overline{\psi_0}(x)}{2h(x)}\right]\,,
\end{align}
 which upon integrating yields
\begin{align}
   &\gamma_{f}(x)=\gamma_{f}(0)-\frac{J_{c}'}{2D}\int_0^x\!\frac{\upd x'}{h(x')}%\nn %
		 -\frac{J_{q}}{4D}\int_0^x\!\frac{\beta ze\overline{\psi_0}(x')}{h^{2}(x')}\,\upd x'\,.\label{eq:cf}
 \end{align}
Similarly, substituting \eqr{eq:partialvarphif} into \eqr{eq:partialpsi} and integrating, at leading order in $\psi_0$, yields
 \begin{align}
   \xi_{f}(x)&=\xi_{f}(0)-\frac{J_{q}}{2D}\int_0^x\!\frac{\upd x'}{h(x')}- \frac{J_{c}'}{4D}\int_0^x\!\frac{\beta ze\overline{\psi_0}(x')}{h^{2}(x')}\,\upd x'\nn
	      &\quad+\frac{1}{2D}\int_0^x\!\frac{\mathcal{J}_{q} (x')}{h(x')}\,\upd x' \,.\label{eq:qf}
 \end{align}
Evaluating the above two equations at $x=L$ gives
\begin{subequations}\label{eq:deltavarphiandpsi1}
 \begin{align}
  \frac{\Delta  \gamma}{L}&=-\frac{H_{1}}{2D\bar{h}}\left[J_{c}'+J_{q}\Phi \Upsilon_{1}\right]\label{eq:deltagammaf}\\
\frac{\Delta \xi}{L}&=-\frac{H_{1}}{2D\bar{h}}\left[J_{q}+J_{c}'\Phi \Upsilon_{1}\right]
+\frac{\Phi\Upsilon_{3}\gamma_{0}  }{Dk_{0}^2 \eta}\frac{\Delta P}{ L}\,,\label{eq:deltapsi1}
 \end{align}
\end{subequations}
%\end{widetext}
%where we multiplied both equations by $2D\bar{h}/(H_{1}L)$, where 
where we defined $\Delta  \gamma= \gamma_{f}(L)- \gamma_{f}(0)$, $\Delta  \xi= \xi_{f}(L)- \xi_{f}(0)$, 
%and $\Xi\equiv\Xi(L)$ [cf. \eqr{eq:Xi}] 
and where we used the following new functions:  
\begin{subequations}\label{eq:ips-def}
\begin{align}
H_{1}&\equiv \frac{\bar{h}}{L}\int_{0}^{L}\frac{1}{h(x)}\,\upd x\label{eq:H1}\,,\\
 \Phi &\equiv\beta ze\times\begin{cases}
                         \zeta& \quad\quad\text{cond}\,,\\
                         \displaystyle{\frac{e\sigma}{\epsilon k_{0}}}& \quad\quad\text{diel}\,,
   \end{cases}\\                        
\Upsilon_{1}&\equiv \frac{\bar{h}}{ H_{1}L}\int_0^L \!\!\upd x\,\frac{\beta ze\overline\psi_0(x)}{2 h^{2}(x)\Phi}\,,\label{eq:Upsilon1}\\
\Upsilon_{3}&\equiv \frac{\bar{h}^3}{H_{3}L}\int_{0}^{L}\!\!\upd x\,\frac{\beta ze[2h(x)\psi_0(x,h(x))-\overline\psi_0(x)]}{2 h^{4}(x)\Phi}\,.\label{eq:Upsilon3}
\end{align}
\end{subequations}
First, similar to $H_{3}$, $H_{1}$ is a measure for the corrugation of the channel. 
Second, $\Phi/(\beta ze)$ equals the surface potential $\psi^{\zeta}(x,h(x))$ for conducting walls, while for dielectric surfaces it differs from $\psi^{\sigma}(x,h(x))$ by a factor  $\coth[k_{0}h(x)]$ [cf.~\eqr{eq:electric_potential-0-1}]. Third, 
the $\Upsilon$ functions are dimensionless and depend solely on the parameter $k_{0}\bar{h}$ and the channel shape $h(x)$. We report their functional dependence on these parameters for both boundary conditions in \eqr{eq:Upsilonexplicit}.
Finally, the $\Upsilon_{3}$ 
term in \eqr{eq:deltapsi1} stems from the $\mathcal{J}_{q} (x)$ term in \eqr{eq:qf} [see Appendix~\ref{app:derivationtheta3}].

\subsection{Onsager matrix}
Reshuffling \eqr{eq:deltavarphiandpsi1} gives 
%\begin{widetext}}
\begin{subequations}
\begin{align}
J_{c}'&=-J_{q}\Phi\Upsilon_{1}-  \frac{2D\bar{h}}{H_{1}}\frac{\Delta  \gamma}{L}\label{eq:Jrho}\\
J_{q}&=-J_{c}'\Phi \Upsilon_{1}\,,
-\frac{2D\bar{h} }{H_{1}}\frac{\Delta  \xi}{L}+\frac{2\Phi \Upsilon_{3}}{H_{1}} \frac{\gamma_0 \bar{h}}{k_0^2\eta }\frac{\Delta P }{L} \,. \label{eq:Jq}
%&=-\frac{2D  }{H_{1}(L/\bar{h})}\left[\Delta  \xi_{f}+  \gamma_{0}\beta ze\Delta V\right]
% +\mathcal{O}\left(\psi_{0}\right)\,.
\end{align}
\end{subequations}

Using the formalism developed in this section, in Appendix \ref{app:Q-Q} we determine the missing piece of $Q$ [\eqr{eq:Q-lin}]:
\begin{align}
Q &=Q_P- J_{q}\frac{\Phi \Upsilon_{3}}{D\beta\eta k_0^2}\,,\label{eq:Qf}
 %=\red{Q_P-J_{q}\frac{\mu_{\rm eo}}{\mu_{\rm ion}} \frac{\Upsilon_{3} }{\gamma_0}}
\end{align}
for both electric boundary conditions [cf. \eqr{eq:Qeofzeta} and \eqr{eq:Qeofsigma}], provided that the channel satisfies $h(0)=h(L)$.
Inserting \eqr{eq:Jrho} into \eqr{eq:Jq},  \eqr{eq:Jq} into \eqr{eq:Jrho}, and \eqr{eq:Jq} into \eqr{eq:Qf}, at leading order in $\Phi$ leads to
\begin{subequations}\label{eq:fluxes-both}
 \begin{align}
 %&+\frac{1}{24}\frac{\epsilon}{\eta}\frac{1}{\beta ze \gamma_0}\frac{\zeta}{k_0}\Delta\xi_f\left[\frac{\Gamma(L)}{h^3(L)}-3\frac{\bar{h}}{H_{1}L}\int\limits_0^L\Gamma(x)\frac{\partial_x h(x)}{h^5(x)}dx\right]\nn
 J_q&=-\frac{2D\bar{h}\beta\gamma_{0} }{H_{1}}\frac{\Delta  \xi}{\beta\gamma_{0}L} +2\Phi \Upsilon_{1}\frac{D\bar{h}}{H_{1}}\frac{\Delta  \gamma}{L}+ \frac{2\Phi\Upsilon_{3}}{H_{1}}\frac{ \gamma_{0} \bar{h}}{k_{0}^2\eta}\frac{\Delta P }{ L}\,,\label{eq:JqH1} \\
 J_{c}'&=2\Phi\Upsilon_{1}\frac{ D\bar{h}\beta\gamma_{0} }{H_{1}}\frac{\Delta  \xi}{\beta\gamma_{0}L} -  \frac{2D\bar{h}}{H_{1}}\frac{\Delta  \gamma}{L}\,,\\
  Q&=\frac{2\Phi \Upsilon_{3}}{H_{1}}\frac{\gamma_{0}\bar{h} }{\eta k_0^2}\frac{\Delta  \xi}{\beta\gamma_{0}L}-\frac{2\bar{h}^3 }{3H_{3}\eta}\frac{ \Delta P}{ L} \,.
\end{align}
\end{subequations}
In \eqr{eq:fluxes-both} we identify three  effective force densities, 
namely $\Delta\xi/(\beta  \gamma_{0}L)$, $\Delta\gamma/(\beta  \gamma_{0}L)$ and $\Delta P/L$.
We use \eqr{eq:electrochemicalpotential} to rewrite $\Delta\xi=\beta\varrho_{0}[ \Delta \bar{\mu}_{f}^+ -\Delta \bar{\mu}_{f}^-]=\beta\gamma_0ze\Delta V$ 
and $\Delta\gamma=\beta\varrho_{0} [\Delta \bar{\mu}_{f}^+ +\Delta \bar{\mu}_{f}^-]=\beta\gamma_0\Delta \bar\mu$ in terms of the more 
familiar ionic chemical potential $\Delta \bar{\mu}$\footnote{From hereon we will omit subscripts $f$ when we denote chemical potential differences, because $\Delta\bar{\mu}$ is enforced upon the system, while the local perturbed chemical potential $\mu_{f}(x)$ is a reaction to that thermodynamic force.} and external potential drop $\Delta V$\footnote{With Eqs.~\eqref{eq:electrochemicalpotential} and \eqref{eq:def-rho-exp-f2} 
it is easy to show that $\Delta \gamma = \mu^+(x,y)+\mu^-(x,y)$ and $\Delta \xi = \mu^+(x,y)-\mu^-(x,y)$, i.e., that $\Delta \gamma$ and $\Delta \xi$ are the sum and difference of the full chemical potentials at order $\mathcal{O}(f)$.}.
%With \eqref{eq:barmupm} 
We can then relate the three fluxes and three forces in \eqr{eq:fluxes-both} via a $3\times3$ conductivity matrix $\mathcal{L}$,
the Onsager matrix of the out-of-equilibrium corrugated nanochannel:
\begin{align}\label{eq:onsagermatrix}
 \left( \begin{array}{c} \displaystyle{J_{q}} \\ J'_{c} \\ Q  \end{array} \right) 
 =& \begin{pmatrix} \mathcal{L}_{11} & \mathcal{L}_{12} &\mathcal{L}_{13}\\ 
			\mathcal{L}_{21} & \mathcal{L}_{22} & 0 \\ 
			\mathcal{L}_{31}& 0 & \mathcal{L}_{33}\end{pmatrix} 
 \left( \begin{array}{c} ze\displaystyle{\Delta V}\\\displaystyle{\Delta \bar\mu}\\ \displaystyle{\Delta P} \end{array} \right)\frac{1}{L}\,,%\nn
\end{align}
 where the coefficients read
 \begin{subequations}\label{eq:onsager_coeff}
 \begin{align}
\mathcal{L}_{11}&=\mathcal{L}_{22}=-2\gamma_0\frac{\bar h}{H_1}\frac{\mu_\text{ion}}{ze}\label{eq:L11=L22}\\
\mathcal{L}_{12}&=\mathcal{L}_{21}=- 2\gamma_0\frac{\bar h}{H_1}\frac{\mu_\text{ion}}{ze}\Phi\Upsilon_{1} \\
\mathcal{L}_{13}&=\mathcal{L}_{31}= - 2\gamma_0\frac{\bar h}{H_1}\frac{1}{\eta k_0^2}\Phi\Upsilon_{3}\\
&\hspace{0.5cm}\mathcal{L}_{33}=-\frac{2}{3H_{3}}\frac{\bar{h}^3}{\eta }\,,
\end{align}
\end{subequations}
where $\mu_{\rm ion}=D\beta ze$ is the ionic mobility.
Clearly, the matrix $\mathcal{L}$ in \eqr{eq:onsagermatrix} is symmetric; hence, Onsager's reciprocal relations are fulfilled.
Equation~ \eqref{eq:onsagermatrix} is the main results of this paper. We discuss its properties in the next section.

\section{Results}\label{sec:Results}

\subsection{General properties of the Onsager matrix}\label{Sec_Onsager}
We list a few general properties of the Onsager matrix:
\begin{enumerate}
\item{
Equation~\eqref{eq:onsagermatrix} relates three fluxes ($J_{q}, J'_{c}, Q$) to three thermodynamic forces 
($ze\Delta V/L, \Delta \bar{\mu}/L, \Delta P/L$) 
via four independent nonzero transport coefficients ($\mathcal{L}_{11},\mathcal{L}_{12},\mathcal{L}_{13},\mathcal{L}_{33}$). 
Note that the off-diagonal matrix elements vanish ($\mathcal{ L}_{12}=\mathcal{L}_{13}=0$)
when the channel walls are uncharged ($\Phi=0$). In that case, the charge flow $J_{q}$, the solute flow $J'_{c}$, and the fluid flow $Q$
respond solely to the electrostatic potential drop, chemical potential differences, and pressure differences, respectively. 
Conversely, for $\Phi\neq 0$ the off-diagonal terms of the Onsager matrix do not vanish 
($\mathcal{ L}_{i\neq j}\neq0$) and \eqr{eq:onsagermatrix} encodes a rich nonequilibrium behavior. }
\item{\label{appendix:L23}
In bulk electrolytes, a salt gradient does not drive a fluid flow. In the presence of a solid substrate, 
the interactions between the ions and the surface drive a phoretic flow $v\sim\nabla\mu \int \upd r\, r(\exp[\beta U(r)]-1)$ with $U$ the interaction potential between the ions and the walls.
Within the Deby-H\"uckel approximation the electrostatic potential is small. Hence, reversing the sign of the interaction leads to a reversal of the sign of the phoretic flow. 
This means that in the presence of a gradient $\nabla \mu$, the first nonzero contribution to the fluid flow is of 
$\mathcal{O}\left(f\psi^2\right)$, in agreement with Ref.~\cite{gross1968membrane}.}
\item{Equation~\eqref{eq:L11=L22} states that $\mathcal{L}_{11}=\mathcal{L}_{22}$. 
This implies that provided our approximations are valid, the knowledge of the diagonal coefficient, $\mathcal{L}_{11}$, 
associated to the electric current induced solely by a electrostatic potential drop, determines the diagonal coefficient 
$\mathcal{L}_{22}$ associated to the ionic current under the action of an ionic chemical potential imbalance $\Delta \bar{\mu}$
\footnote{Since a chemical potential drop alone cannot induce and solvent flow: $J'_c=J_c$.}.}
\item{The diagonal terms are controlled solely by $H_{1}$ and $H_3$. Since these functions do not depend on the boundary conditions (constant $\sigma$ or constant $\zeta$) on the electrostatic potential, nor do the diagonal terms.} %In contrast, the off-diagonal terms depends on $\Upsilon_{1,3}$ and therefore sensitive to the dielectric nature of the channel walls.}
\item{In contrast to the the diagonal elements, the off-diagonal terms are sensitive 
to the electrostatic boundary conditions.
%While the diagonal elements $\mathcal{ L}_{ii}$ depend on the channel shape through $H_{1}$ and $H_{3}$, they do not depend 
%on the boundary conditions on the electrostatic potential (constant $\sigma$ or constant $\zeta$).
%In contrast, the off-diagonal terms depend on $\Upsilon_{1,3}$ and .
This is evident after inserting Eqs.~\eqref{eq:electric_potential-cond} and \eqref{eq:electric_potential-0-1} 
into Eqs.~\eqref{eq:Upsilon1} and \eqref{eq:Upsilon3}:
\begin{subequations}\label{eq:Upsilonexplicit}
\begin{align}
\Upsilon^{\zeta}_{1}&=\frac{\bar{h} }{H_{1}L}\int_0^L\!\!\upd x\,\frac{\mathscr{G}[k_0h(x)]}{h(x)}\,, \\%\limits
\Upsilon^{\sigma}_{1}&=\frac{\bar{h} }{H_{1}L}\int_0^L\!\!\upd x\,\frac{1}{k_0h(x)^{2}}\,,\label{eq:upsilonsigma1}\\%\limits
\Upsilon^{\zeta}_{3}&=\frac{\bar{h}^3 }{H_{3}L}\int_0^L\!\!\upd x\,\frac{1-\mathscr{G}[k_0h(x)]}{h^3(x)}\,,\\ %\limits
\Upsilon^{\sigma}_{3}&=\frac{\bar{h}^3}{H_{3}L}\int_0^L\!\!\upd x\,\frac{\mathscr{L}[k_{0} h(x)]}{h^3(x)}\,,%\limits
\end{align} 
\end{subequations}
where
\begin{align}\label{eq:GandL}
\mathscr{G}(x)&=\frac{\tanh(x)}{x} \quad\quad,\quad\quad\mathscr{L}(x)=\coth(x)-\frac{1}{x}\,,
\end{align} 
with $\mathscr{L}(x)$ known as the Langevin function.
}
%dielectric nature of the channel walls.} 

%\item{The appearance of $H_{1}$ and $H_{3}$ the matrix elements $\mathcal{L}_{11}$, $\mathcal{L}_{13}$, $\mathcal{L}_{31}$, $\mathcal{L}_{33}$ confirms previous results of \cit{ajdari2001transverse,ghosal2002lubrication}. }
\end{enumerate}

\begin{figure}
 \includegraphics[scale=0.65]{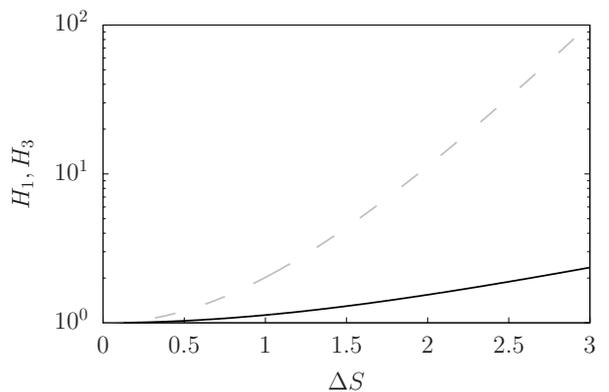}
 \caption{Plots of the functions $H_{1}$ [\eqr{eq:H1}] (black line) and $H_{3}$ [\eqr{eq:H3}] (gray dashed line)
 as a function of $\Delta S$ for the channel shape given in \eqr{eq:coschannel}.}
 \label{Fig_H1H3}
\end{figure}

In order to proceed with our analysis of the Onsager matrix and the functions $H_1$, $H_3$, $\Upsilon_1$, and $\Upsilon_3$ 
appearing therein, we need to restrict to a particular channel shape. Accodingly, we choose
\begin{equation}
  h(x)=\bar{h}+\Delta h\cos\left(2\pi\frac{x}{L}\right)\,.
  \label{eq:coschannel}
\end{equation}
A more general shape of the channel may include a ``phase'' in the argument of the cosine. 
However, Eqs.~(\ref{eq:onsager_coeff}) and (\ref{eq:ips-def}) show that the transport coefficients
depend solely on the integral of the channel shape and are thus phase independent\footnote{This differs from what has been reported for a tracer (see Ref.~\cite{Malgaretti2015}).
In the latter case a phase dependence arose because the concentration of tracers is not affecting the local electric field.}.
While the dimensionless combination $\Delta h/\bar{h}$ already gives a sense of the channel corrugation, 
in the following we prefer to use the related ``entropic barrier'' defined as
\begin{equation}
 \Delta S\equiv\ln \left[\frac{\bar{h}+\Delta h}{\bar{h}-\Delta h}\right]=2\tanh^{-1}(\Delta h/\bar{h})\,,
\end{equation}
which takes values between $\Delta S=0$ for flat channels and $\Delta S>0$ for corrugated channels.
%where $\alpha$ determines the ''phase`` of the channel with respect to the boundary conditions. For $\alpha=0$, the hour-glass shape is retreivede wheras for $\alpha=\pi$ the pore in Fig.~\ref{fig:scheme} is obtained.
%Since all the elements of the Onsager matrix are proportional $\Xi$ and to $H_1$,$H_3,$ and $\Upsilon_1$,$\Upsilon_3$, 
Inserting \eqr{eq:coschannel} into $H_{1}$ [\eqr{eq:H1}], $H_{3}$ [\eqr{eq:H3}], and $\Upsilon^{\sigma}_{1}$ [\eqr{eq:upsilonsigma1}] gives
\begin{subequations}\label{eq:H13explicit}
 \begin{align}
 H_{1}&=\cosh\frac{\Delta S}{2}\,, \\
 H_{3}&=\left[\frac{3}{2}\cosh^{2}\left(\frac{\Delta S}{2}\right)-\frac{1}{2}\right]\cosh^{3}\frac{\Delta S}{2}\,,\\ 
 \Upsilon^{\sigma}_{1}&=\frac{\cosh^{2}(\Delta S/2)}{k_{0}\bar{h}}\label{eq:upsilon1analytical}\,.
\end{align}
\end{subequations}
We plot $H_{1}$ and $H_{3}$ in Fig.~\ref{Fig_H1H3}.

\paragraph{Diagonal terms}
As noted earlier, the diagonal matrix elements $\mathcal{L}_{11}\sim1/H_{1}$ and $\mathcal{L}_{33}\sim1/H_{3}$ depend on the shape of the channel, 
 but not on the boundary conditions 
on the electrostatic potential at the channel walls\footnote{This is in agreement 
with Refs.~\cite{ajdari2001transverse,ghosal2002lubrication,bruus2008theoretical}}. 
Because $H_1$ and $H_3$ increase with $\Delta S$, $\mathcal{L}_{11}$ and $\mathcal{L}_{33}$ decrease therewith. 
This implies that the pressure-driven volumetric fluid flow $Q$ and the chemical potential-driven excess solute flow $J_c'$ 
 diminish with increasing $\Delta S$. 
 % when, respectively, induced by solely , 
%$\Delta P$ (with $\Delta \mu=\Delta V=0$), or , $\Delta \mu$ (with $\Delta P=\Delta V=0$), drop.

\begin{figure}
 \includegraphics[scale=0.65]{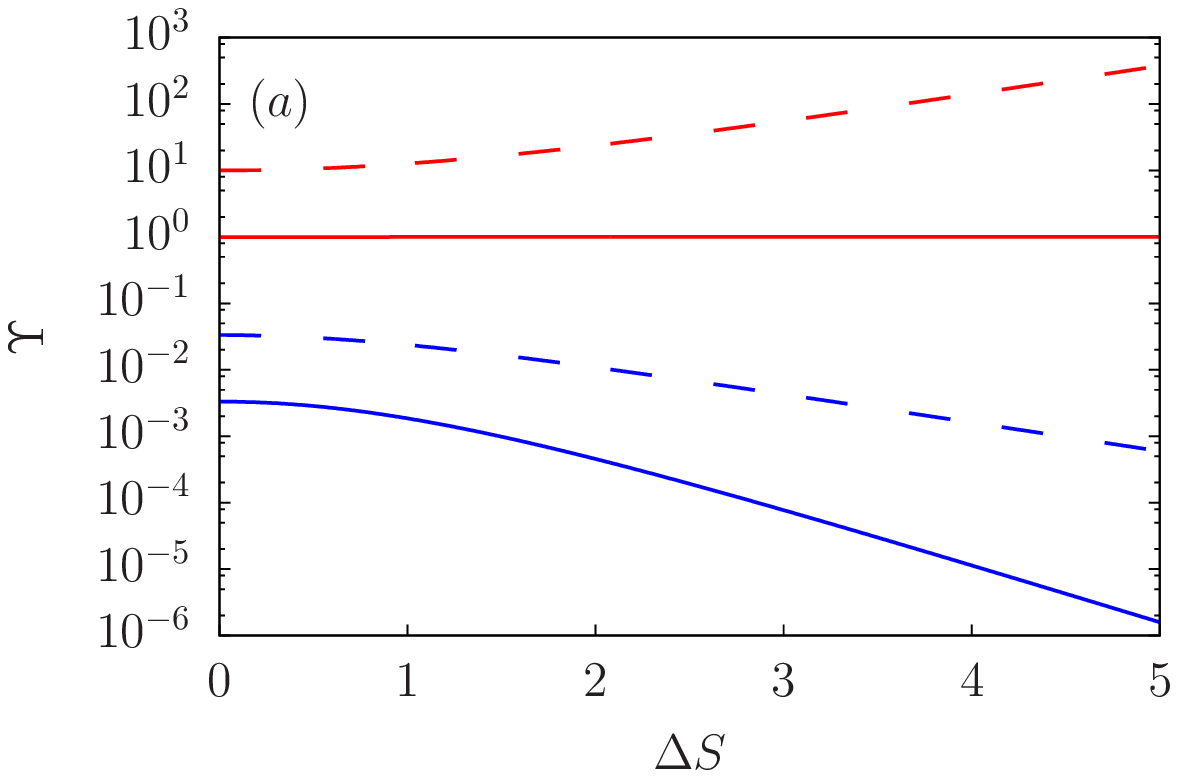}
 \includegraphics[scale=0.65]{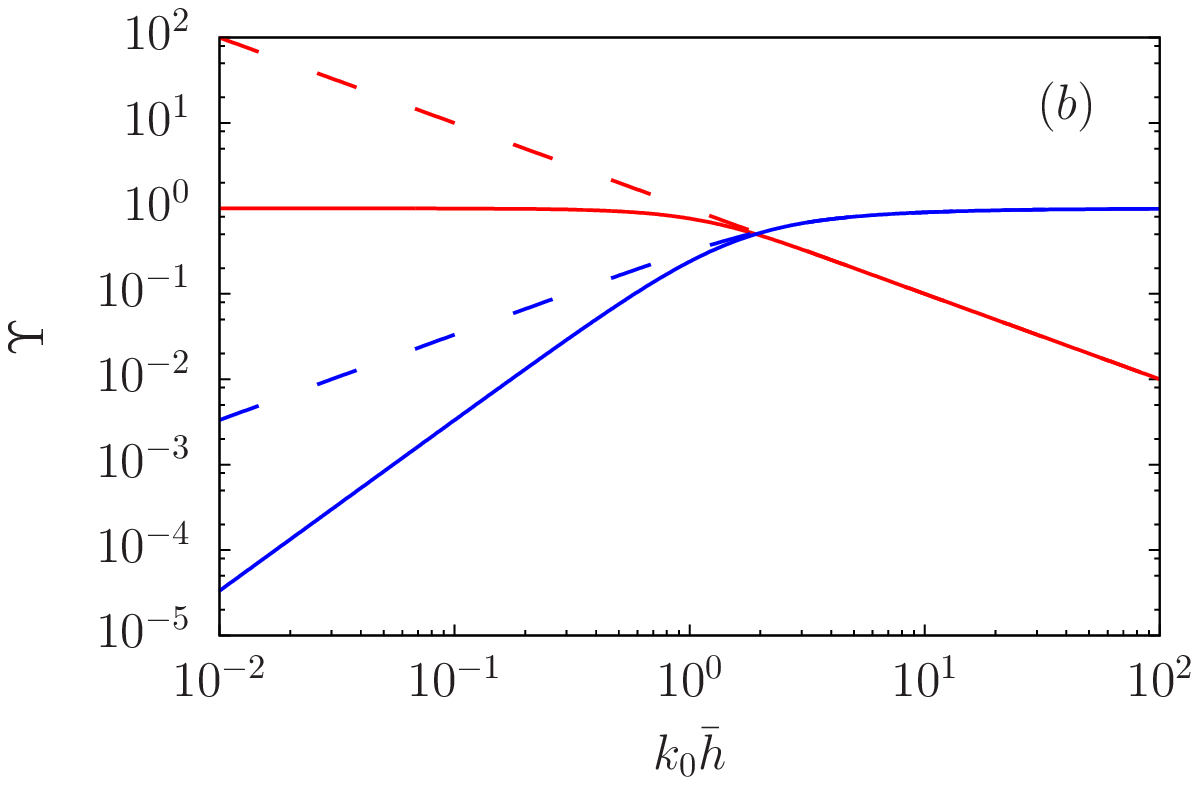}
 \caption{Plots of the functions $\Upsilon^{\zeta}_{1}$ (solid red), $\Upsilon^{\zeta}_{3}$ (solid blue), $\Upsilon^{\sigma}_{1}$ (dashed red), and $\Upsilon^{\sigma}_{3}$ (dashed blue) as a function of $\Delta S$
 (panel a) and $k_{0}\bar{h}$ (panel b) for the channel shape given in \eqr{eq:coschannel}. 
 %Black dotted lines indicate functions $\propto1/(k_{0}\bar{h})$, $\propto k_{0}\bar{h}$, and $\propto(k_{0}\bar{h})^2$, respectively.
 }
 \label{Fig_U}
\end{figure}

\paragraph{Off-diagonal terms}
The off-diagonal terms in \eqr{eq:onsagermatrix} are controlled by the functions $\Upsilon^{\sigma,\zeta}$ and by $H_1$. 
%As we discussed for the diagonal terms, $\Xi/H_1\simeq 1$ for $\epsilon_w\ll\epsilon$, whereas, $\Xi/H_1\simeq 1/H_1$ for $\epsilon_w\simeq\epsilon$ or porous materials. 
In the following we will focus on the dependence of $\mathcal{L}_{12}$ and $\mathcal{L}_{13}$ on 
$\Upsilon_{1}^{\sigma,\zeta}$ and $\Upsilon_{3}^{\sigma,\zeta}$, respectively. Figure~\ref{Fig_U}(a) 
shows for conducting channels walls that  $\Upsilon_{1}^{\zeta}$ 
(and thus $\mathcal{L}_{12}^\zeta$) is almost independent of $\Delta S$, whereas $\Upsilon_1^{\sigma}$ (and thus $\mathcal{L}_{12}^\sigma$) 
increases more drastically as $\Upsilon_1^{\sigma}\propto\exp{[\Delta S]}$ at large $\Delta S$ [cf.~\eqr{eq:upsilon1analytical}]. 
Hence, counterintuitively, 
increasing the corrugation of the channel enhances some off-diagonal transport coefficients.
Equation \eqr{eq:onsagermatrix} shows that this enhancement occurs for the electric current driven by a chemical potential drop $\Delta \mu$ (with $\Delta P=\Delta V=0$) and for the excess solute flow $J_{c}'$ driven by an external electrostatic field $\Delta V/L$ (with $\Delta P=\Delta \mu=0$). 

Figure~\ref{Fig_U}(b) shows that the sensitivity of $\Upsilon_{1}^{\zeta}$, and hence $\mathcal{L}_{12}$, 
on the boundary conditions disappears when the Debye length is much smaller than the channel section, $k_{0}\bar{h}\gg 1$,  
whereas it becomes significant when $k_{0}\bar{h} \lesssim 1$. In particular, Fig.~\ref{Fig_U}(b) shows that 
in the latter regime, $\mathcal{L}_{12}^\zeta$ keeps the linear dependence on $k_{0}\bar{h}$ whereas 
$\mathcal{L}_{12}^\sigma$  reaches a plateau.

The dependence of $\mathcal{L}_{13}$, on the top of its $H_1$-dependence, on both $\Delta S$ and $k_{0}\bar{h}$ is 
encoded in $\Upsilon^{\sigma,\zeta}_3$. Similar to $\mathcal{L}_{12}$, also $\mathcal{L}_{13}$ shows
an explicit dependence on the boundary conditions. In particular, Fig.~\ref{Fig_U}(a) shows that both $\mathcal{L}_{13}^{\sigma,\zeta}$ decrease with increasing $\Delta S$ and for large values 
of $\Delta S$ we have $\mathcal{L}_{13}^{\sigma}\propto \exp[-2\Delta S]$ and $\mathcal{L}_{13}^{\zeta}\propto \exp[-\Delta S]$. 
This means that the electric current induced by a pressure drop $\Delta P$ and the electroosmotic 
flow induced by $\Delta V$ decay exponentially with $\Delta S$.
The dependence of $\mathcal{L}^{\sigma,\zeta}_{13}$ on $k_{0}\bar{h}$ is shown in Fig.~\ref{Fig_U}(b). 
Interestingly, for large values of $k_{0}\bar{h}$ both  $\mathcal{L}^{\sigma}_{13}$ and  $\mathcal{L}^{\zeta}_{13}$ reach a plateau whereas for smaller values of $k_{0}\bar{h}$ they grow monotonically with $\mathcal{L}^{\sigma}_{13}\propto (k_{0}\bar{h})^2$ and  $\mathcal{L}^{\zeta}_{13}\propto k_{0}\bar{h}$.

To understand the influence of the channel walls (i.e., conducting or dielectric) on the $\Upsilon$ functions, 
 it is insightful to look at their relative differences through the combination
\begin{equation}
 %\Delta \Upsilon_{1,3}=\dfrac{\Upsilon^\sigma_{1,3}-\Upsilon^\zeta_{1,3}}{\dfrac{\Upsilon^\sigma_{1,3}+\Upsilon^\zeta_{1,3}}{2}}
 \Delta \Upsilon_{1,3}=2\,\dfrac{\Upsilon^\sigma_{1,3}-\Upsilon^\zeta_{1,3}}{\Upsilon^\sigma_{1,3}+\Upsilon^\zeta_{1,3}}\,.
 \label{eq:Diff-Ups}
\end{equation}
\begin{figure}
 \includegraphics[scale=0.65]{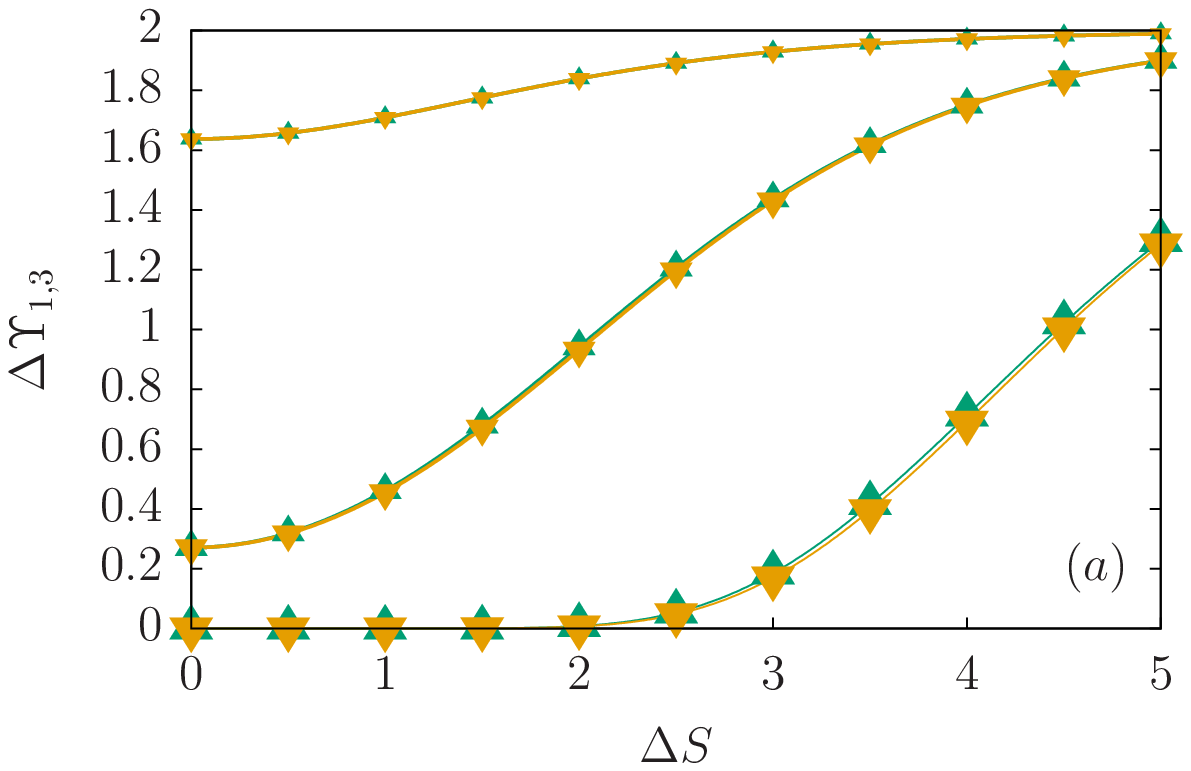}
 \includegraphics[scale=0.65]{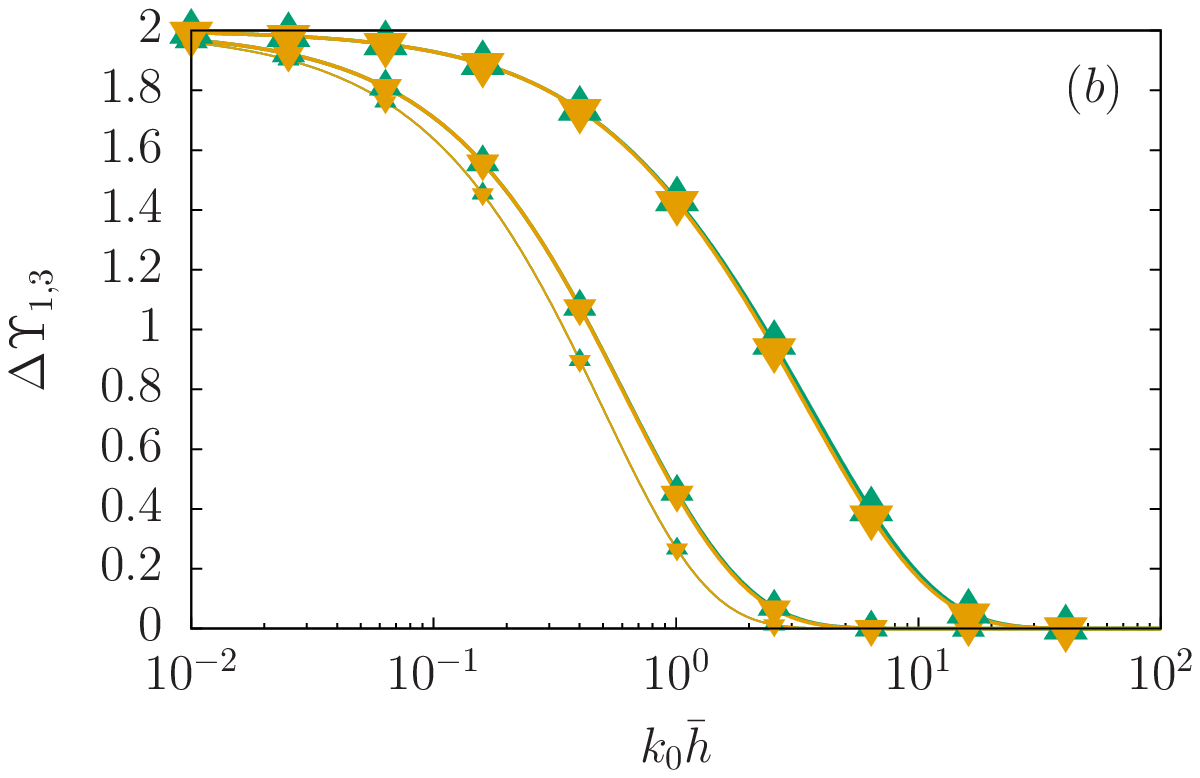}
 \caption{Plots of the functions $\Delta\Upsilon_{1}$ (downward triangles) and $\Delta\Upsilon_{3}$ (upward triangles), as a function of $\Delta S$
 (panel a) and $k_{0}\bar{h}$ (panel b) for the channel shape given in \eqr{eq:coschannel}. In (a) we use  $k_0 \bar{h}=0.1,1,10$: 
 larger points stand for larger values of $k_0 \bar{h}$. In (b) we use $\Delta S=0,1,3$: larger points stand for larger values of $\Delta S$.
 }
 \label{Fig_DiffU}
\end{figure}
Figure~\ref{Fig_DiffU} shows that the functions $\Upsilon_{1,3}$ 
have a surprisingly similar sensitivity to the boundary conditions: whenever $\Upsilon_1$ 
changes by switching from conducting to dielectric walls, so does $\Upsilon_3$, and, 
remarkably, by almost the same amount. This means that there is no regime in which some 
of the Onsager coefficients are more sensitive than others upon changing the electrostatic properties of the walls.
Finally, we notice from Fig.~\ref{Fig_DiffU} that $\Delta \Upsilon_{1,3}>0$, which means  [cf.~\eqr{eq:Diff-Ups}]
%Joining this data with  leads to the conclusion 
that the 
Onsager coefficients for dielectric walls are always larger than their counterparts for conducting walls.   

\subsection{Single external force}\label{Cases}
So far we have discussed the general properties of the Onsager matrix and their relation to some relevant cases.
In the following we discuss in detail several transport phenomena. In order to emphasize the role of the geometry and the onset of the entropic electrokinetic regime~\cite{Malgaretti2014,Malgaretti2015,Malgaretti2016,Chinappi2018} we normalize quantities by their corresponding values in a plane channel geometry with equal average section. 

\subsubsection{Electrostatic driven flows}%Imposed $\Delta V$}
\begin{figure}
    	\includegraphics[scale=0.65]{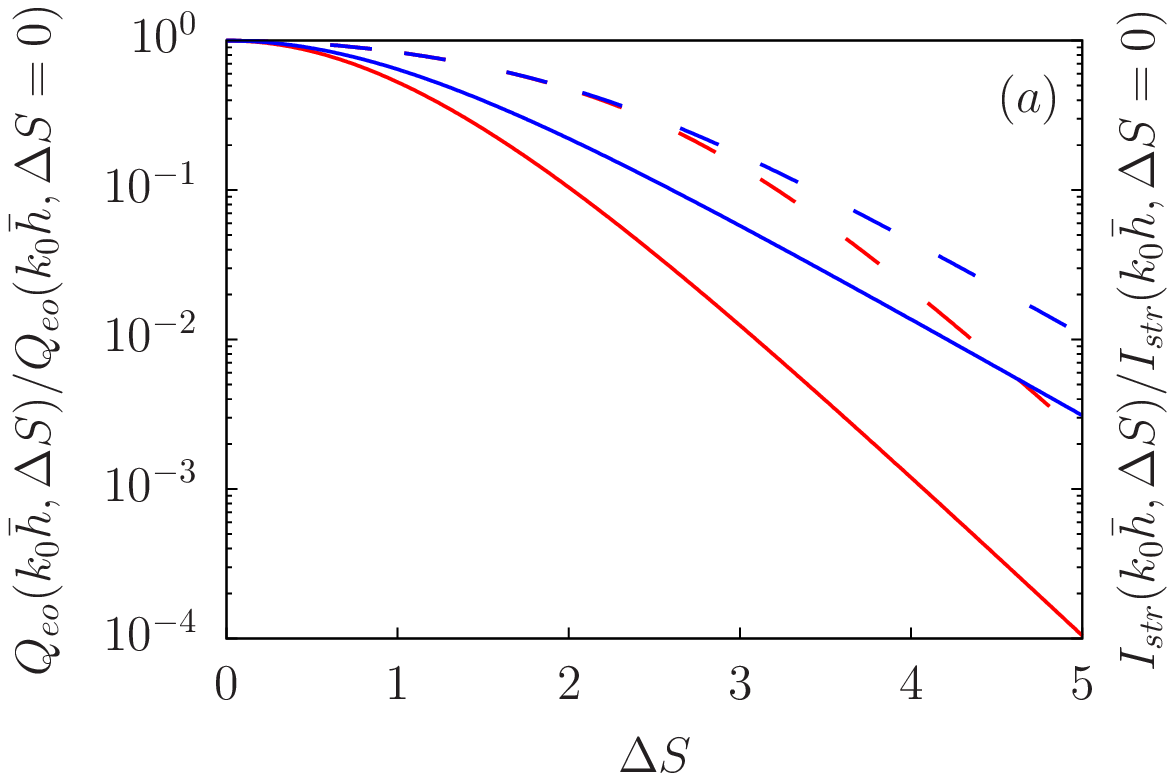}
    	\includegraphics[scale=0.65]{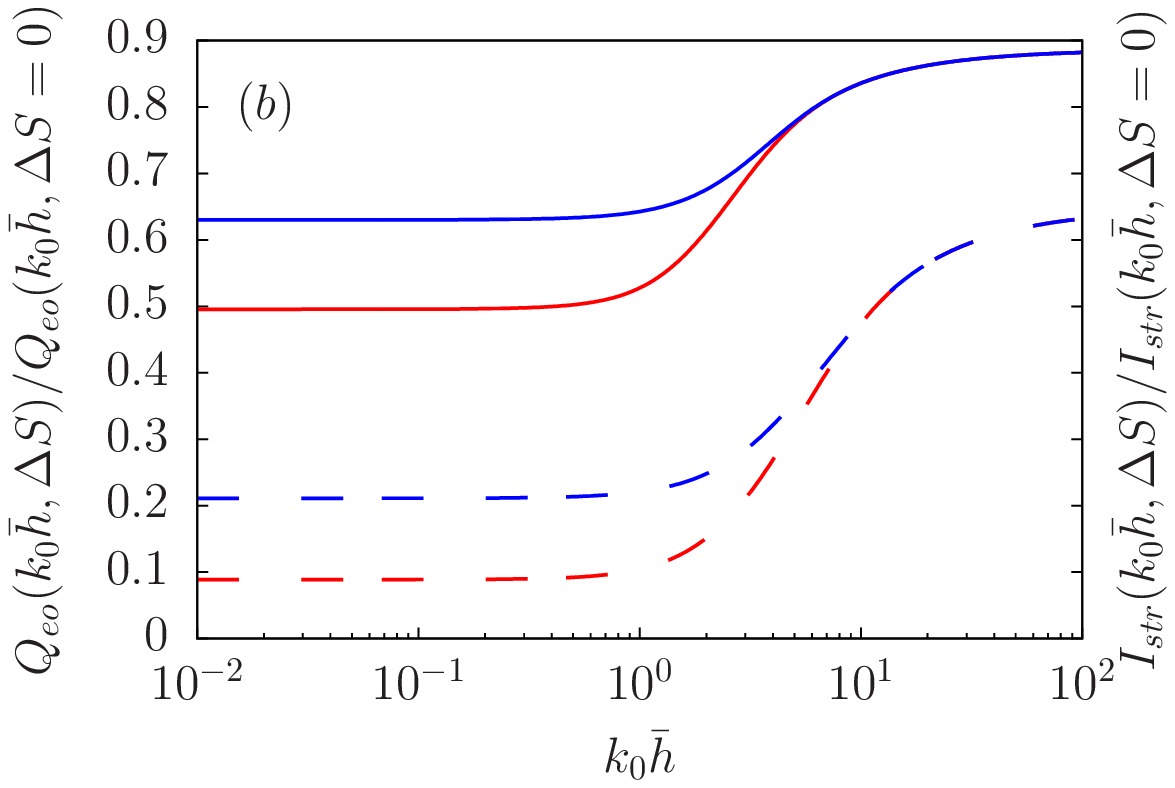}
   	\caption{Electroosmotic flow $Q_{\rm eo}$ [\eqr{eq:electroosmoticflowbc}] 
   	and streaming current $I_{\rm str}$ [\eqr{eq:streamingcurrentbc}] 
   	for conducting (red lines) and dielectric (blue lines) channel 
   	walls. 
   	Panel (a) shows $\Delta S$-dependence at $k_{0}\bar{h}=1$ (solid) and $k_{0}\bar{h}=10$ (dashed), respectively.
   	Panel (b) shows $\Delta V$-dependence at $\Delta S=1$ (solid) and $\Delta S=1$ (dashed), respectively. }
 \label{Fig_Qeo}
\end{figure}

In the case that $\Delta P=\Delta \bar{\mu}=0$ a potential difference $\Delta V\neq0$ drives an ionic current, 
salt flow, and electroosmotic fluid flow.  
In particular, the electroosmotic flow  (per unit length in the $z$-direction) reads:
\begin{align}\label{eq:electroosmoticflow}
Q_{\textrm{eo}}= \mathcal{L}_{13}\frac{ze\Delta V}{L}=\frac{2\Phi \epsilon \bar{h}}{\eta } \frac{\Delta V}{L}\times\frac{\Upsilon_{3}}{H_{1}}\,.
\end{align}
This amounts to
\begin{subequations}\label{eq:electroosmoticflowbc}
\begin{align}
Q_{\textrm{eo}}^{\zeta}&=\frac{2\zeta\epsilon \bar{h}}{\eta   } \frac{\Delta V}{L}\times\frac{\Upsilon^{\zeta}_{3}}{H_1}\,,\label{eq:electroosmoticflowzeta}\\
Q_{\textrm{eo}}^{\sigma}&=\frac{2e\sigma \bar{h}^2}{\eta  }\frac{\Delta V}{L}\times \frac{\Upsilon^{\sigma}_{3}}{k_0 \bar{h} H_{1}} \label{eq:electroosmoticflowsigma}\,.
\end{align} 
\end{subequations}
We note that \eqr{eq:electroosmoticflowsigma} coincides with Eq.~(40) of \cit{yoshida2016analysis} 
and that the combination $\zeta \epsilon/\eta$ is known as the ``electroosmotic mobility"~\cite{bruus2008theoretical}. 
For the channel as specified in \eqr{eq:coschannel}, 
we show \eqr{eq:electroosmoticflowbc} %these Onsager coefficients 
 as a function of  $\Delta S$ in Fig.~\ref{Fig_Qeo}(a) and as a function of $k_0 \bar{h}$ in Fig.~\ref{Fig_Qeo}(b). 
From this figure we see that both $Q_{\textrm{eo}}^{\zeta}$ and $Q_{\textrm{eo}}^{\sigma}$ 
vanish for highly corrugated channels (when $\Delta S\gg0$). 
Meanwhile, both $Q_{\textrm{eo}}^{\zeta}$ and $Q_{\textrm{eo}}^{\sigma}$
 diminish upon decreasing $k \bar{h}$. % i.e. with increasing double layer overlap. 
 Interestingly the transition between the two plateaus occurs for $k_0\bar{h}\in [1:10]$, the entropic electrokinetic regime~\cite{Malgaretti2014,Malgaretti2015,Malgaretti2016,Chinappi2018}.
For a straight channel, the integrals in $\Upsilon$ [\eqr{eq:Upsilonexplicit}] 
become trivial and \eqr{eq:electroosmoticflowbc} simplifies  to
\begin{subequations}\label{eq:electroosmoticflowflat}
\begin{align}
%=& -\frac{\bar{h}   \gamma_{0} \Upsilon_{3}}{\eta k_{0}^2} \frac{\Delta P}{L}\nn
%=& -\frac{  \gamma_{0} }{\eta } \frac{\beta ze}{k_0^2 } \frac{\Delta P}{L}\int_h^h\left[\psi_0(x,y)-\psi_0(x,h(x))\right]dy\nn
%=& -\frac{   \gamma_{0} }{\eta } \frac{\beta ze}{k_0^2 } \frac{\Delta P}{L}2\zeta\frac{\tanh k_0 \bar{h} -k_0 \bar{h}}{k_0} \nn
Q_{\textrm{eo}}^{\zeta}&=\frac{2\zeta  \epsilon \bar{h}}{\eta }\frac{\Delta V}{L}\times\left[1-\mathscr{G}(k_0 \bar{h}) \right]\,,\label{eq:electroosmoticflowflatzeta} \\
Q_{\textrm{eo}}^{\sigma}&=\frac{2e\sigma\bar{h}^2}{\eta}\frac{\Delta V}{L}\times\frac{\mathscr{L}(k_{0} \bar{h})}{k_{0} \bar{h}}\label{eq:electroosmoticflowflatsigma} \,.
%=& -\frac{ 2 \epsilon \zeta \bar{h}}{ze \eta }  \frac{\Delta P}{L}\left[1+\mathcal{O}\left(\frac{1}{k_0 \bar{h}}\right)\right]\,,
\end{align}
\end{subequations}
We note that \eqr{eq:electroosmoticflowflatzeta} is in agreement with Eq.~(50) of \cit{delgado2007measurement}.
Moreover, for a flat channel, $\psi^{\zeta}(x,y)=\psi^{\sigma}(x,y)$ provided that the surface potentials are the same: $\psi^{\zeta}(x,h)=\psi^{\sigma}(x,h)\Rightarrow \zeta=e\sigma \coth[k_0 h]/(\epsilon k_0)$. Inserting this into \eqr{eq:electroosmoticflowflatzeta} we confirm \eqr{eq:electroosmoticflowflatsigma}.

\subsubsection{Pressure driven flows}
In the case that $\Delta V=\Delta \bar{\mu}=0$, a pressure difference $\Delta P\neq0$ drives a 
\textit{streaming current} (per unit length in the $z$-direction):
\begin{align}\label{eq:streamingcurrent}
I_{\rm str}&= \mathcal{L}_{13}ze\frac{\Delta P}{L}=\frac{2\Phi\epsilon \bar{h}}{\eta} \frac{\Delta P}{L}\times\frac{\Upsilon_{3}}{ H_{1}}\,,
\end{align}
which amounts to
\begin{subequations}\label{eq:streamingcurrentbc}
\begin{align}
I^{\zeta}_{\rm str}&=\frac{2\zeta\epsilon \bar{h}}{\eta} \frac{\Delta P}{L}\times\frac{\Upsilon^{\zeta}_{3}}{H_{1}}\label{eq:streamingcurrentzeta}\\
I^{\sigma}_{\rm str}&=\frac{2e\sigma \bar{h}^2}{\eta} \frac{\Delta P}{L}\times\frac{\Upsilon^{\sigma}_{3}}{k_{0}\bar{h}H_{1}}\label{eq:streamingcurrentsigma}\,.
\end{align} 
\end{subequations}
Clearly, $I_{\rm str}$ is governed by the same matrix element $\mathcal{L}_{13}$ as 
the electroosmotic flow (its reciprocal phenomenon) discussed above. 
As a consequence, $Q^{\sigma}$ and $I_{\rm str}^{\sigma}$ 
share the same term $\Upsilon^{\sigma}_{3}/(k_{0}\bar{h}H_{1})$ and, hence, display 
the same $\Delta S$ and $k_0 \bar{h}$ dependence (see Fig.~\ref{Fig_Qeo}). 

From \eqr{eq:streamingcurrent} is easy to determine the streaming current 
between two parallel plates and to check that this agrees with Eq.~(37) of Ref.~\cite{delgado2007measurement}.

\subsubsection{Chemical potential steps $\Delta \bar{\mu}$}
Finally, we consider the case in which flows are driven solely by a chemical potential drop  $\Delta \bar{\mu}$.
Accordingly, the electric current reads:
\begin{equation}
 J_{q}=-2\gamma_0 \bar h\frac{\mu_\text{ion}}{ze}\Phi\Delta \bar{\mu}\times\frac{\Upsilon_{1}}{H_1}\,.
\end{equation}
which amounts to:
\begin{subequations}\label{eq:JqChemPot}
 \begin{align}
  J^\zeta_{q}&=-2\gamma_0 \bar{h}\beta \mu_\text{ion}\zeta\Delta \bar{\mu}\times\frac{\Upsilon^{\zeta}_{1}}{H_1}\,,\label{eq:JqChemPot-z}\\
  J^\sigma_{q}&=-2\gamma_0 \bar{h}^2\beta \mu_\text{ion} \frac{e\sigma}{\epsilon} \Delta \bar{\mu}\times\frac{\Upsilon^{\sigma}_{1}}{k_0 \bar h H_1}\,.\label{eq:JqChemPot-s}
 \end{align}
\end{subequations}
\begin{figure}
      \includegraphics[scale=0.65]{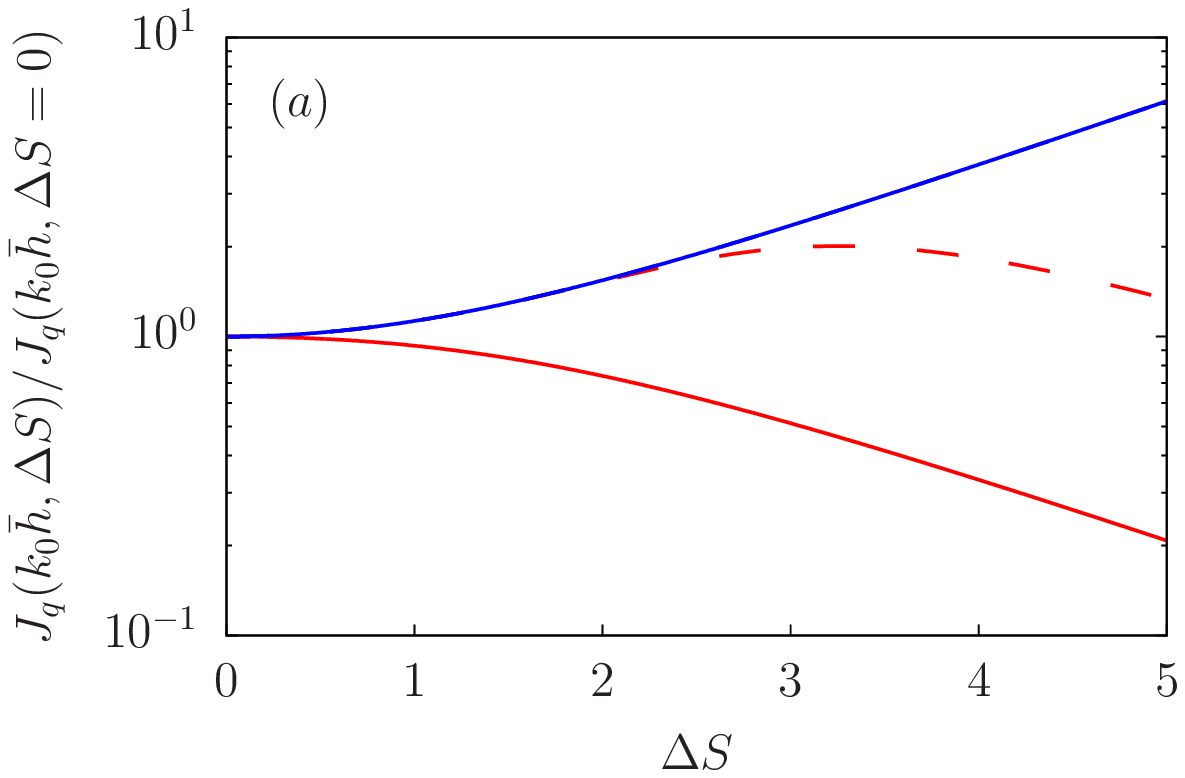}
      \includegraphics[scale=0.65]{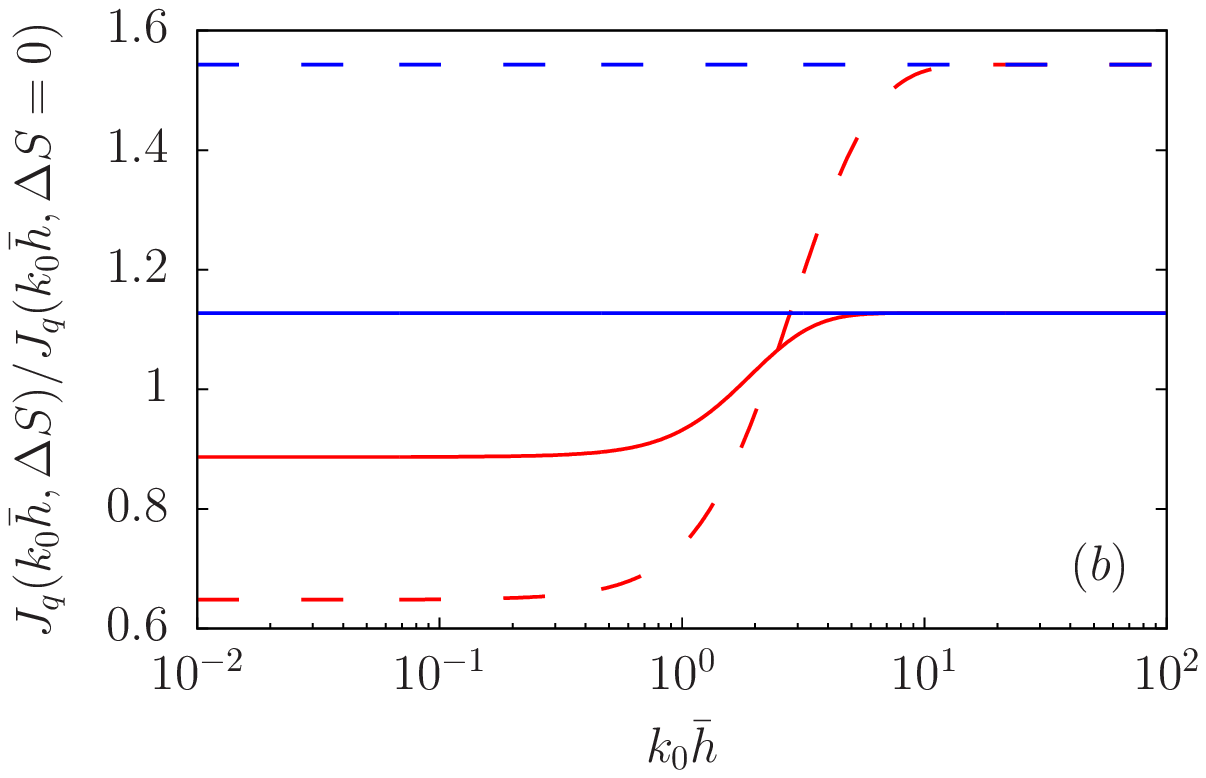}
   \caption{Electric current $J_q$ [Eqs.~\eqref{eq:JqChemPot}] driven by a chemical potential difference, as a function of $\Delta S$ (panel a) and $k_0\bar{h}$ (panel b) for conducting (red lines) and dielectric (blue lines) channel  walls. 
   %Panel a): $k_0 \bar h=1,10$ for respectively solid and dashed lines. 
   %Panel b): $\Delta S=1,2$ for respectively solid and dashed lines.
   Panel (a) shows $\Delta S$-dependence at $k_{0}\bar{h}=1$ (solid) and $k_{0}\bar{h}=10$ (dashed), respectively.
   Panel (b) shows $\Delta V$-dependence at $\Delta S=1$ (solid) and $\Delta S=1$ (dashed), respectively.}
 \label{Fig_JqChemPot}
\end{figure} 
%Interestingly, 
For dielectric channel walls, we find with \eqr{eq:upsilon1analytical} that $J^\sigma_q(\Delta S)/J^\sigma_q(\Delta S=0)=\cosh[\Delta S/2]$; hence,  
this ratio grows monotonically with increasing the corrugation of the channel $\Delta S$. 
Conversely, for conducting channel walls, Fig.~\ref{Fig_JqChemPot}(a) displays that $J^\sigma_q$ has a maximum for a finite value of $\Delta S$. The ionic charge currents as discussed in this section are the relevant physical phenomenon underlying reverse electrodialysis, whereby electrical energy is generated from a salt concentration difference~\cite{VanRoij2016}.

\subsection{Membrane}
In the following, we characterize several cases in which the channel is in
series with a membrane that selectively impedes the passage of (any combination of) solvent and ions. 
In this scenario we can control $Q$, and the fluxes of positive, $2J^+=J_c+J_q$  and negative, $2J_-=J_c-J_q$ ions. Due to its physcal interest, in this section we focus on the full solute flow $J_c=J'_c+\gamma_0 Q$ rather then on $J_c'$, .
%By solving the system\red{??}, at first order in $\Phi$, we obtain:
The general solution to \eqr{eq:onsagermatrix}, under the above constraints, reads:
\begin{subequations}\label{eq:membrane}
\begin{align}
\Delta\bar\mu&= \frac{J_c-\left(\mathcal{L}_{12}+\gamma_0\mathcal{L}_{13}\right)ze\Delta V-\gamma_0\mathcal{L}_{33}\Delta P}{\mathcal{L}_{22}}\,,\\
ze\Delta V&= \frac{\mathcal{L}_{22}J_q-\mathcal{L}_{12}J_c+\left(\mathcal{L}_{12}\gamma_0\mathcal{L}_{33}-\mathcal{L}_{22}\mathcal{L}_{13}\right)\Delta P}{\mathcal{L}^2_{22}}\,,\label{eq:str-pot}\\
Q&= \frac{\mathcal{L}_{13}}{\mathcal{L}_{22}}J_q+\mathcal{L}_{33}\Delta P\,.
\end{align}
\end{subequations}
\begin{figure}
      \includegraphics[scale=0.65]{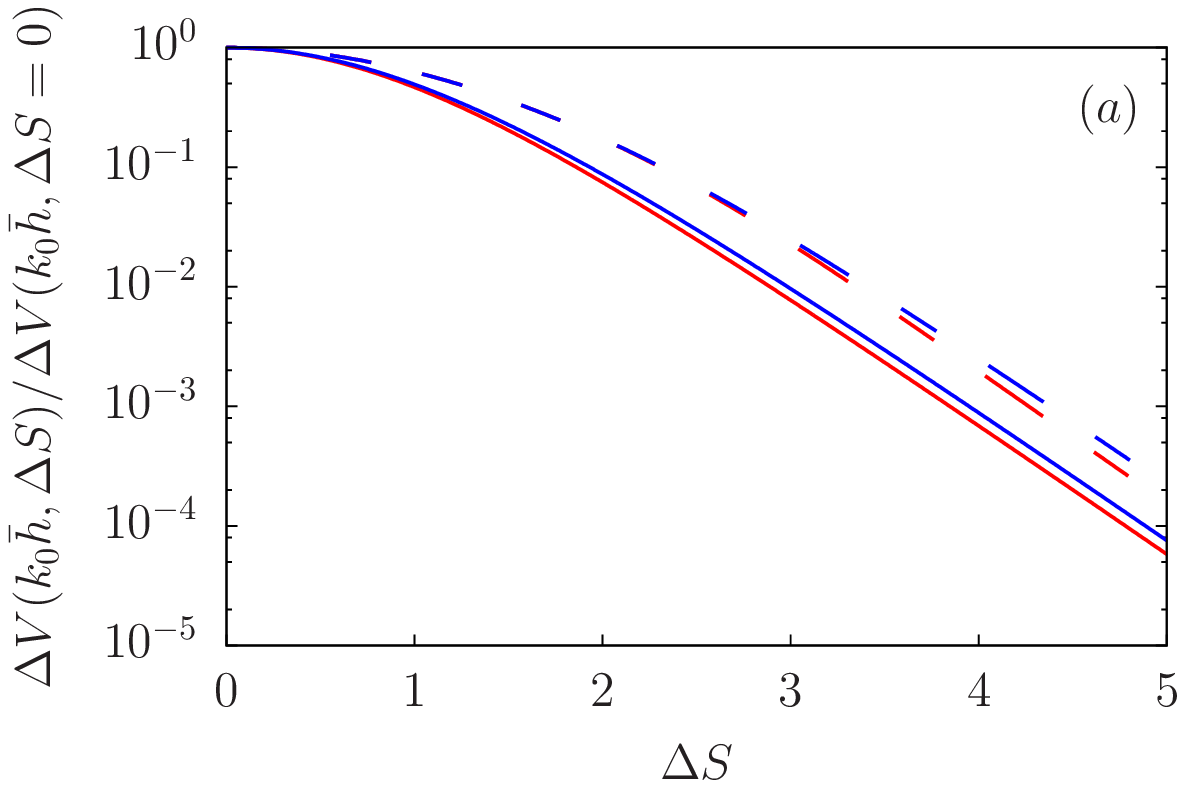}
      \includegraphics[scale=0.65]{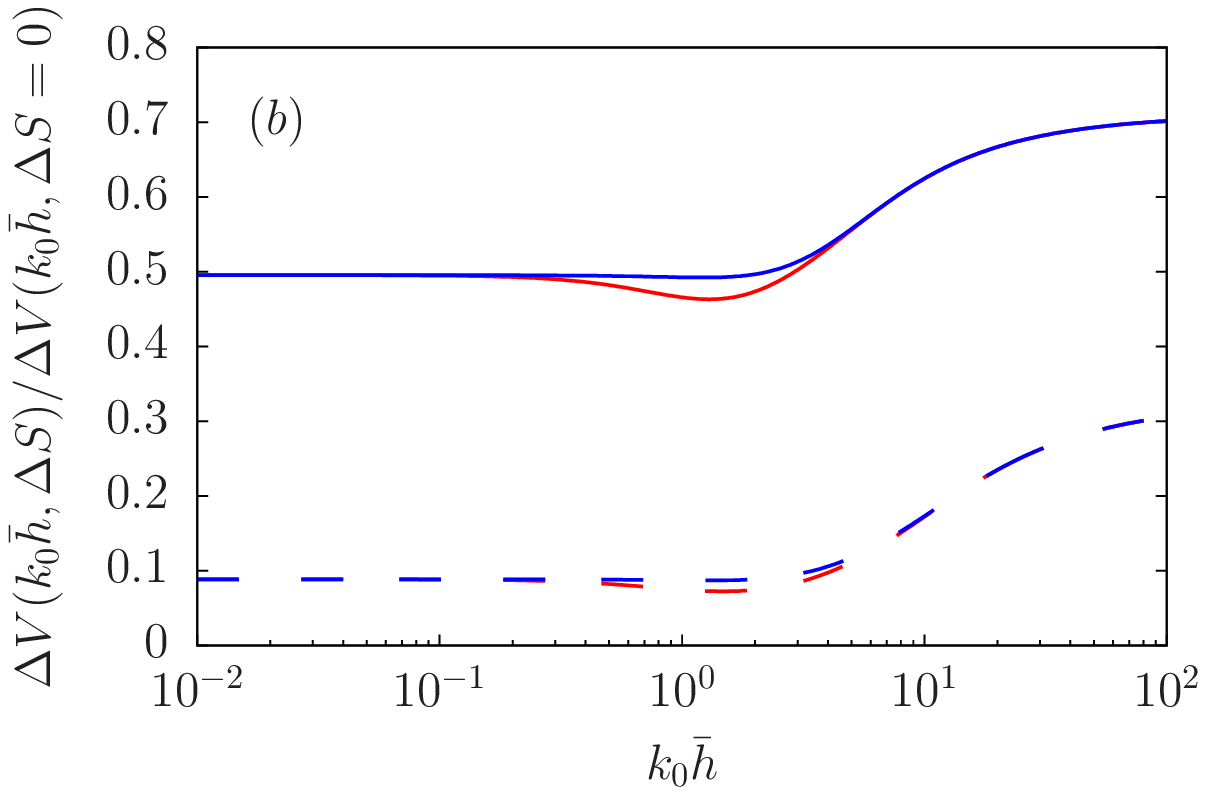}
   \caption{Electrostatic potential drop $\Delta V$ [Eq.\eqref{eq:DV-ImpMem}] across a solvent permeable membrane, as a function of $\Delta S$ (panel a) and $k_0\bar{h}$ (panel b) for conducting (red lines) and dielectric (blue lines) channel  walls. 
   %Panel a): $k_0 \bar h=1,10$ for respectively solid and dashed lines. 
   %Panel b): $\Delta S=1,2$ for respectively solid and dashed lines.
   Panel (a) shows $\Delta S$-dependence at $k_{0}\bar{h}=1$ (solid) and $k_{0}\bar{h}=10$ (dashed), respectively.
   Panel (b) shows $\Delta V$-dependence at $\Delta S=1$ (solid) and $\Delta S=1$ (dashed), respectively.}
 \label{Fig:ImpMem}
\end{figure} 

\subsubsection{Electrodes}
First, we consider a membrane that allows for a net electric current, $J_q\neq 0$ but not for mass fluxes, $Q=J_c=0$. This looks like having some electrodes that close the electric circuit at zero solvent and ionic flow.
When only $\Delta V$ is nonzero, Eqs.~\eqref{eq:membrane}, at linear order in $\Phi$, gives:
\begin{subequations}
 \begin{align}
  \Delta\bar{\mu}_V&=\frac{\mathcal{L}_{12}}{\mathcal{L}_{22}}ze \Delta V\,,\\
  \Delta P_V&=\frac{\mathcal{L}_{13}}{\mathcal{L}_{33}}ze \Delta V\,,\\
  J_q&=\mathcal{L}_{22}ze\Delta V\,,
 \end{align}
\end{subequations}
or, likewise,
\begin{subequations}
 \begin{align}
  \Delta\bar{\mu}_V&=ze \Delta V\Phi\times\Upsilon_1\,,\label{eq:elect1}\\
  \Delta P_V&=\frac{1}{3}\frac{\epsilon}{\beta ze}\frac{1}{\bar{h}^2} \Delta V\Phi\times\frac{H_3}{H_1}\Upsilon_3\,,\label{eq:DP-elctr}\\
  J_q&=-2\frac{\epsilon}{\beta ze}\bar h \frac{\mu_\text{ion}}{ze}\Delta V\times \frac{1}{H_1}\,.
 \end{align}
\end{subequations}
\begin{figure}
      \includegraphics[scale=0.65]{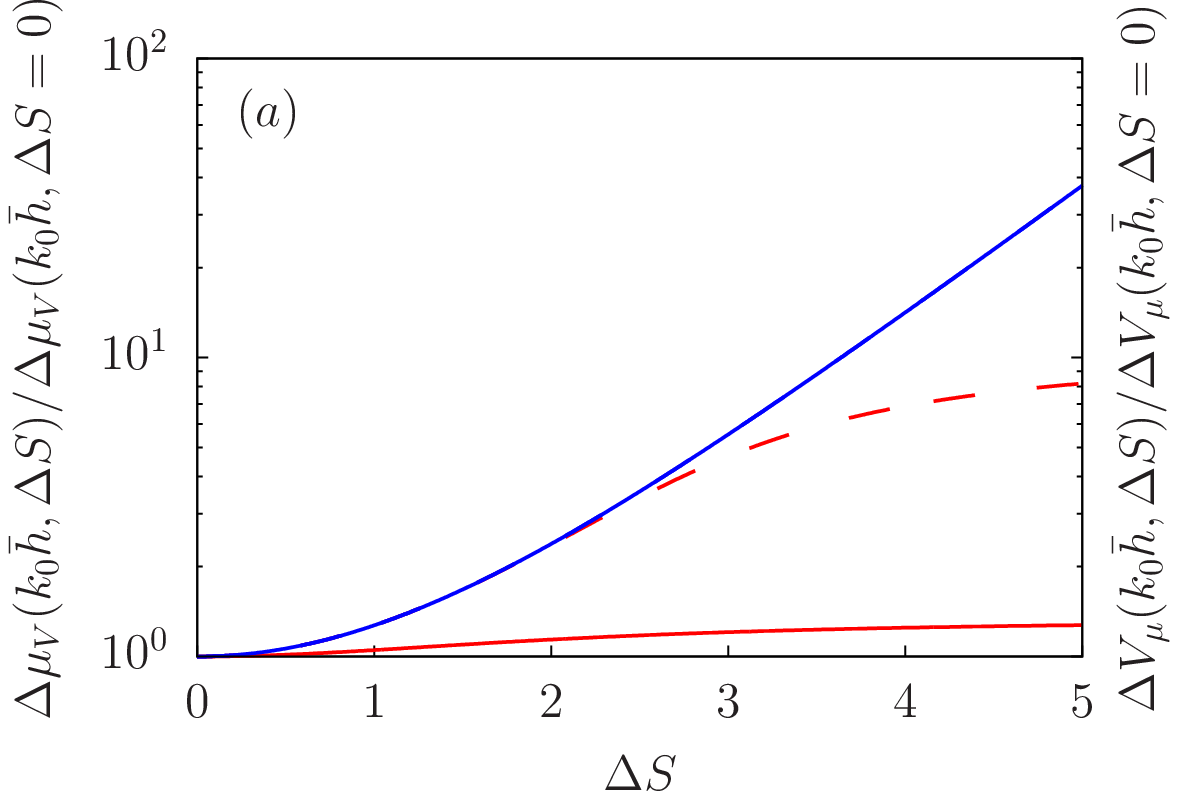}
      \includegraphics[scale=0.65]{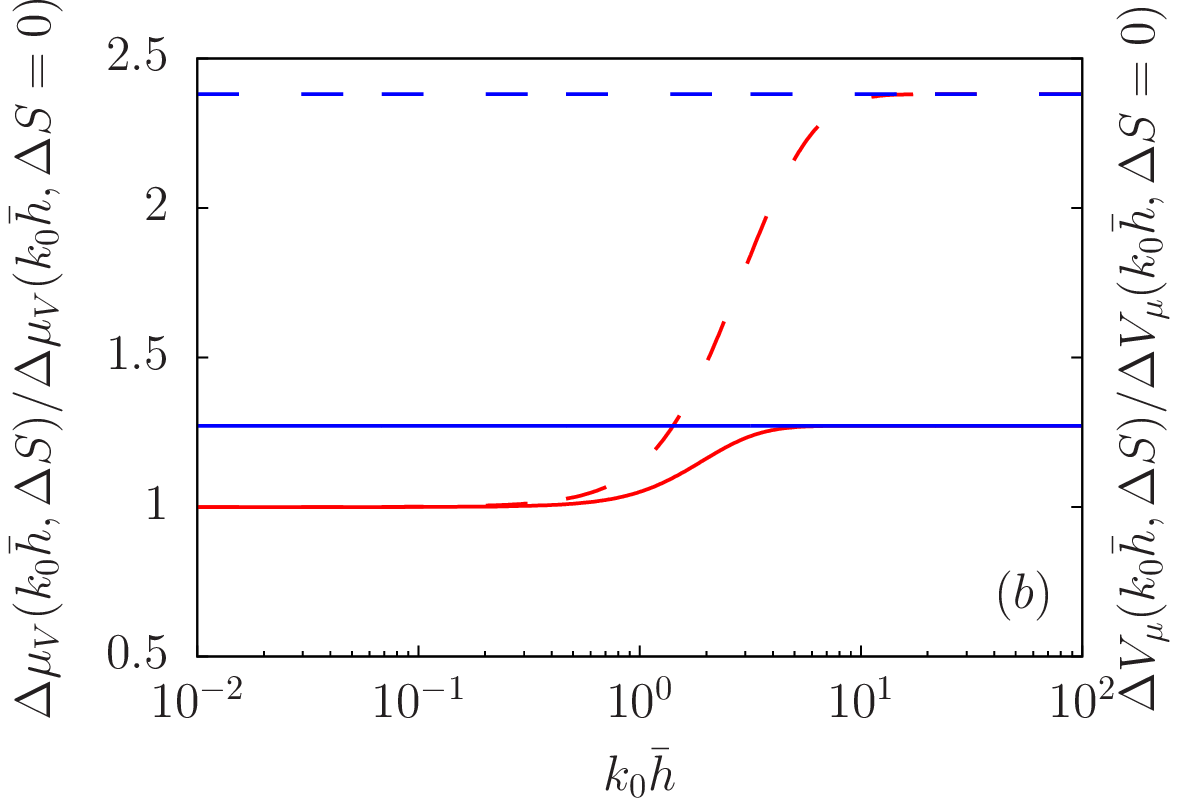}
   \caption{Chemical potential drop, $\Delta \mu$ [Eq.\eqref{eq:elect1}], as a function of $\Delta S$ (panel a) and $k_0\bar{h}$ (panel b) for conducting (red lines) and dielectric (blue lines) channel  walls. 
   %Panel a): $k_0 \bar h=1,10$ for respectively solid and dashed lines. 
   %Panel b): $\Delta S=1,2$ for respectively solid and dashed lines. 
   Panel (a) shows $\Delta S$-dependence at $k_{0}\bar{h}=1$ (solid) and $k_{0}\bar{h}=10$ (dashed), respectively.
   Panel (b) shows $\Delta V$-dependence at $\Delta S=1$ (solid) and $\Delta S=1$ (dashed), respectively.
   The same quantitative behavior holds for $\Delta V$ induced by an applied chemical potential, $\Delta \mu$, when the electric current is set to zero, $J q = 0$ (see text). }
 \label{Fig_U1}
\end{figure}
Figure~\ref{Fig_U1}(a) shows the dependence of $\Delta\bar{\mu}_V$ on $\Delta S$. 
In particular, for both conducting and dielectric channel walls $\Delta\bar{\mu}_V$ 
increases with the corrugation of the channel $\Delta S$. In contrast, 
the dependence of $\Delta\bar{\mu}_V$ on $k_0 \bar h$ is more sensitive
to the conductive properties of the channel walls. While for dielectric
walls $\Delta\bar{\mu}_V$ is independent on $k_0 \bar h$, for conducting
walls it reaches plateau for both $k_0 \bar h\ll 1$ and $k_0 \bar h\gg 1$ 
and it grows for $k_0 \bar h\in [1:10]$ [see Fig.~\ref{Fig_U1}(b)]. 
Interestingly, the very same behavior is observed for the electrostatic potential drop, $\Delta V_\mu$ 
induced by an applied chemical potential, $\Delta \bar \mu$ when the electric current is set to zero, $J_q=0$. In this case,
$\Delta V_\mu$ grows with $\Delta S$ for both kind of channel walls [see Fig.~\ref{Fig_U1}(a)]. 
Since both  $\Delta\bar{\mu}_V$ and $\Delta V_\mu$ are proportional to $\Upsilon_1$, their increas upon enlaring $\Delta S$ as already shown in Fig.~\ref{Fig_JqChemPot}.
Finally, we remark that $\Delta P_v$ [\eqr{eq:DP-elctr}] is also known as the \textit{electroosmotic back/counter pressure}~\cite{bruus2008theoretical}.
\begin{figure}
      \includegraphics[scale=0.65]{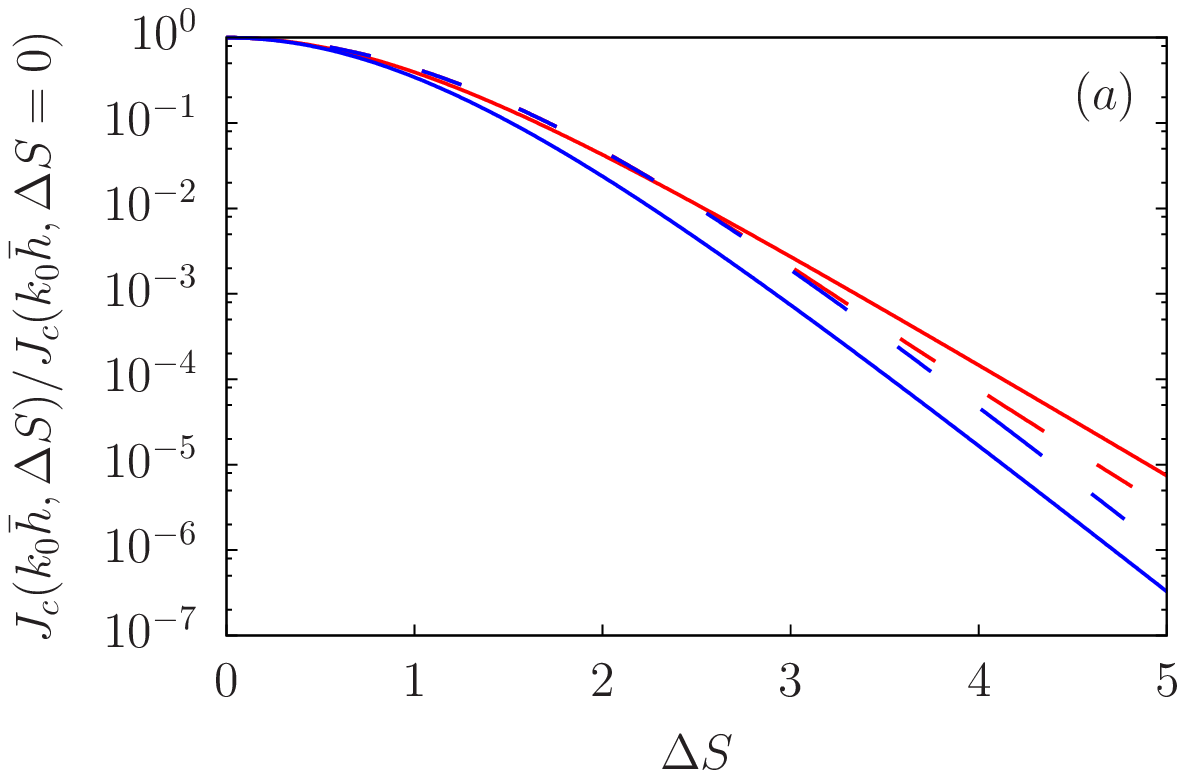}
      \includegraphics[scale=0.65]{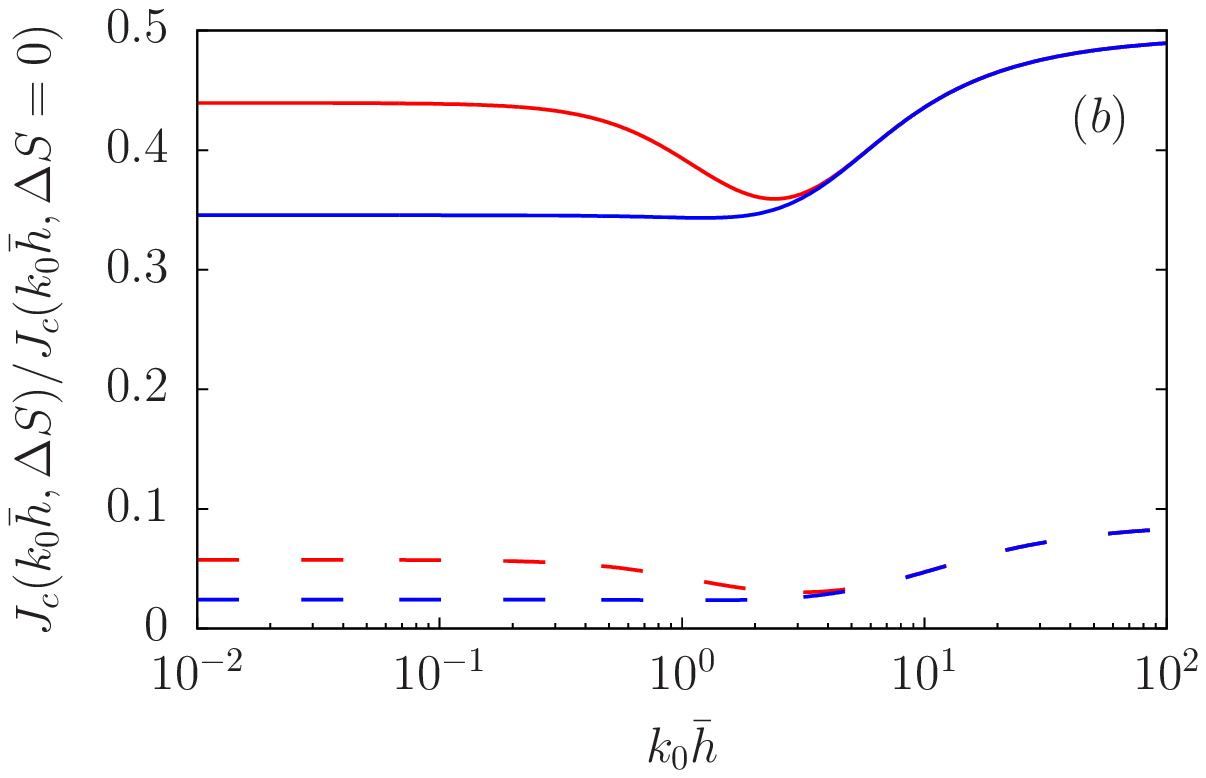}
   \caption{Solute current, $J_c$ [Eq.\eqref{eq:open-Jc}], as a function of $\Delta S$ (panel a) and $k_0\bar{h}$ (panel b) for conducting (red lines) and dielectric (blue lines) channel  walls. 
   %Panel a): $k_0 \bar h=1,10$ for respectively solid and dashed lines. 
   %Panel b): $\Delta S=1,2$ for respectively solid and dashed lines.
   Panel (a) shows $\Delta S$-dependence at $k_{0}\bar{h}=1$ (solid) and $k_{0}\bar{h}=10$ (dashed), respectively.
   	Panel (b) shows $\Delta V$-dependence at $\Delta S=1$ (solid) and $\Delta S=1$ (dashed), respectively.}
 \label{Fig:open-circuit}
\end{figure}

\subsubsection{Open electric circuit}
Second, we consider a membrane that allows for mass flows, $Q\neq 0$ and $J_c\neq 0$, but not for electric current, $J_q=0$. In this case, at linear order in $\Phi$, Eqs.~\eqref{eq:membrane} read:
\begin{subequations}\label{eq:openCircuit}
\begin{align}
\Delta\bar\mu&= \frac{J_c-\left(\mathcal{L}_{12}+\gamma_0\mathcal{L}_{13}\right)ze\Delta V-\gamma_0\mathcal{L}_{33}\Delta P}{\mathcal{L}_{22}}\,,\\
ze\Delta V&= \frac{-\mathcal{L}_{12}J_c+\left(\mathcal{L}_{12}\gamma_0\mathcal{L}_{33}-\mathcal{L}_{22}\mathcal{L}_{13}\right)\Delta P}{\mathcal{L}^2_{22}}\,,\\
Q&= \mathcal{L}_{33}\Delta P\,.
\end{align}
\end{subequations}
In particular, for $\Delta V=0$ this amounts to:
\begin{subequations}\label{eq:openCircuit2}
\begin{align}
\Delta\bar\mu&= -\frac{ze \bar{h}^2}{\mu_{\rm ion}\eta}\Delta P\times\frac{1}{k_0^2\bar{h}^2}\frac{\Upsilon_3}{\Upsilon_1}\,,\\
J_c&= -2\frac{\epsilon \bar{h}}{\beta (ze)^2 \eta}\Delta P\times\left[\frac{1}{3}\frac{k_0^2\bar{h}^2}{H_3}-\frac{1}{H_1}\frac{\Upsilon_3}{\Upsilon_1}\right]\,,\label{eq:open-Jc}\\
Q&= -\frac{2}{3}\frac{\bar{h}^3}{\eta }\Delta P\times \frac{1}{H_{3}}\,.
\end{align}
\end{subequations}
As shown in Fig.~\ref{Fig:open-circuit}(a), the solute current $J_c$ decreases monotonically for 
both dielectric and conducting channel walls upon increasing the channel corrugation $\Delta S$. 
More surprising is the dependence of $J_c$ on $k_0\bar h$. Indeed, Fig.~\ref{Fig:open-circuit}(b) 
shows that $J_c$ reaches a plateau for both $k_0\bar h \ll1 $ and $k_0\bar h \gg1 $ and 
for $k_0\bar h \simeq 1$ it displays a nonmonotonous dependence on $k_0\bar h$.

% \subsubsection{remainings}
% 
% Given a certain applied voltage $\Delta V$, the \textit{electroosmotic back/counter-pressure} $\Delta P_{\rm eo}$ is the pressure required to stop fluid flow: $Q=\mathcal{L}_{13}ze\Xi \Delta V/L+\mathcal{L}_{33}\Delta P_{\rm eo}/L=0$:
% \begin{equation}
% \Delta P_{\rm eo}=\frac{3\epsilon \zeta \Delta V}{\bar{h}^2}\times\frac{H_{3}\Upsilon_{3}\Xi}{H_{1}} \,.
% \end{equation}
% \red{Compare with Sec. 8.4  of Ref.~\cite{bruus2008theoretical}}
% 
% \textit{Electro-osmotic drag coefficient} 
%  $R=J^{\Delta V}_q/Q_{\textrm{eo}}$: $J_q=\mathcal{L}_{11}\Delta V/L$, $Q=\mathcal{L}_{13}\Delta V/L$, hence $R=\mathcal{L}_{11}/\mathcal{L}_{13}$: the inverse of  \eqr{eq:streamingpotential}.  compare to \cite{catalano2018ac}.
% (=electro-osmotic coupling?) Araki calls  $\mathcal{L}_{13}$ the electroosmotic coefficient.

\subsubsection{Solvent permeable membrane}
Third, we consider a membrane that selectively permits the flow of solute, but not of ions; 
hence, $J_c=J_q=0$ and $J'_c=\gamma_{0}Q$. Accordingly we obtain:
\begin{subequations}\label{eq:imp_mem}
\begin{align}
\Delta\bar\mu&= -\frac{\gamma_0\mathcal{L}_{33}}{\mathcal{L}_{22}}\Delta P\,,\\
ze\Delta V&= \frac{\mathcal{L}_{12}\gamma_0\mathcal{L}_{33}-\mathcal{L}_{22}\mathcal{L}_{13}}{\mathcal{L}^2_{22}}\Delta P\,,\\
Q&= \mathcal{L}_{33}\Delta P\,,
\end{align}
\end{subequations}
from \eqr{eq:membrane}. This amounts to:
\begin{subequations}\label{eq:imp_mem-2}
\begin{align}
\Delta\bar\mu&= -\frac{1}{3}\frac{\bar{h}^2}{\eta}\frac{ze}{\mu_{\rm ion}}\Delta P\times\frac{H_1}{H_3}\,,\label{eq:Dmu-ImpMem}\\
ze\Delta V&= \frac{\bar{h}^2}{\eta}\frac{ze}{\mu_{\rm ion}}\Phi\Delta P\times\left[\frac{1}{3}\frac{H_1}{H_3}\Upsilon_1-\frac{1}{k^2_0\bar{h}^2}\Upsilon_3\right]\,,\label{eq:DV-ImpMem}\\
Q&= -\frac{2}{3}\frac{\bar{h}^2}{\eta}\Delta P\times\frac{1}{H_3}\,.
\end{align}
\end{subequations}
Interestingly, in order to sustain a nonvanishing fluid flow, $Q\neq 0$, 
the system will excite all three external forces, i.e.,
we have $\Delta V\neq0$, $\Delta P\neq0$, and $\Delta \bar\mu\neq0$. 
In particular, when $\Delta P$ is the only externally applied force,
then from \eqr{eq:DV-ImpMem} we can read off the induced \textit{streaming potential}. 
Figure~\ref{Fig:ImpMem}(a) shows that  $\Delta V$ decays monotonically with $\Delta S$ 
for both conducting and dielectric channel walls whereas Fig.~\ref{Fig:ImpMem}(b) 
shows that $\Delta V$ reaches a plateau for both $k_0\bar{h}\ll 1$ and  $k_0\bar{h}\gg 1$ 
and that the sensitivity of $\Delta V$ on $k_0\bar{h}$ is maximum for $k_0\bar{h}\simeq 1$, 
i.e., in the entropic electrokinetic regime. 
Finally, by inverting \eqr{eq:Dmu-ImpMem}, Eqs.~\eqref{eq:imp_mem-2} show that a net fluid flow can be obtained by applying a chemical potential gradient $\Delta\bar\mu$. However, the net fluid flow is not a \textit{direct consequence} 
of $\Delta\bar\mu$ (we recall that $\mathcal{L}_{23}=\mathcal{L}_{32}=0$). Rather $\Delta\bar\mu$ induces an osmotic pressure drop that eventually drives the flow.

\subsubsection{Ion exchange membrane}
Finally, we consider the channel to be in series with a
membrane that selectively allows for flow of solvent and one ionic species, impeding the other species. 
Recalling that $J_c=(J^++J^-)/2$ and $J_q=(J^+-J^-)/2$, when only positive 
ions are flowing ($J^-=0$) we have, $J_c=J_q$, whereas when only negative ions are flowing ($J^+=0$) we have, $J_c=-J_q$.
Hence we have:
\begin{subequations}\label{eq:semi-membrane}
\begin{align}
J_c&= \frac{\mathcal{L}^2_{22}}{\pm\mathcal{L}_{22}-\mathcal{L}_{12}} ze\Delta V-\frac{\mathcal{L}_{12}\gamma_0\mathcal{L}_{33}-\mathcal{L}_{22}\mathcal{L}_{13}}{\pm\mathcal{L}_{22}-\mathcal{L}_{12}}\Delta P\,,\label{eq:J_c-mem}\\
\Delta\bar\mu&=-\frac{\mathcal{L}_{12}+\gamma_0\mathcal{L}_{13}}{\mathcal{L}_{22}}ze\Delta V-\frac{\gamma_0\mathcal{L}_{33}}{\mathcal{L}_{22}}\Delta P + J_c\,,\\
%\Delta\bar\mu&=\left[\frac{\mathcal{L}^2_{22}}{\pm\mathcal{L}_{22}-\mathcal{L}_{12}}-\frac{\mathcal{L}_{12}+\gamma_0\mathcal{L}_{13}}{\mathcal{L}_{22}}\right]ze\Delta V\nonumber\\
%&-\left[\frac{\gamma_0\mathcal{L}_{33}}{\mathcal{L}_{22}}+\frac{\mathcal{L}_{12}\gamma_0\mathcal{L}_{33}-\mathcal{L}_{22}\mathcal{L}_{13}}{\pm\mathcal{L}_{22}-\mathcal{L}_{12}}\right]\Delta P\\ 
Q&= \mathcal{L}_{13}\Delta V+\mathcal{L}_{33}\Delta P\,,
\end{align}
\end{subequations}
where the sign in front of $\mathcal{L}_{22}$ is positive when positive ions 
are flowing and negative otherwise. We remark that within linear response 
$\mathcal{L}_{22}  \geq \mathcal{L}_{12}$~\cite{DeGroot}.
In contrast to the solvent permeable membrane, for the ion exchange
membrane we can put one of the thermodynamic forces to zero. In particular, for 
electrostatic driven flows, $\Delta V\neq 0$ with $\Delta P=0$ Eqs.~\eqref{eq:semi-membrane} 
read\footnote{We remark that by using Stokes-Einstein, $D=k_BT/(6\pi \eta R)$ where $R\simeq 0.1~\text{nm}$
is the linear size of an ion, the prefactor in \eqr{eq:semi-perp2} reads 
$\epsilon/(\beta ze \mu_{\rm ion} \eta)\simeq 10^{-1}$.}: 
\begin{subequations}\label{eq:semi-perp}
 \begin{align}
  J_c&=  -2 \frac{\mu_{\rm ion}\epsilon}{\beta (ze)^2 \bar{h}} \Delta V\times \frac{k_0^2\bar{h}^2}{H_1}\frac{1}{\pm1 - \Phi\Upsilon_1}\label{eq:mem-Jc}\\
  \Delta\bar\mu&= ze\Delta V\Phi\times\left[\Upsilon_1+\frac{\epsilon}{\beta ze \mu_{\rm ion} \eta}\Upsilon_3\right]+J_c\label{eq:semi-perp2}\\
  Q&= - 2 \frac{\epsilon}{\beta ze} \Delta V\Phi \bar{h}\frac{1}{\eta}\times \frac{\Upsilon_{3}}{H_1}\,.
 \end{align}
\end{subequations}
Equation~\eqref{eq:mem-Jc} is remarkable for two reasons. First, as commented earlier, 
if it diverges when $\pm1 - \Phi\Upsilon_1\rightarrow 0$ it would 
require higher order corrections. Secondly, in \eqr{eq:mem-Jc} the 
dimensionless potential, $\Phi$, plays a major role, i.e., it modulates
the relative magnitude of the two terms in the denominator, and it is 
not simply a multiplicative constant as it is in all previous cases. 
In order to proceed with a numerical inspection of \eqr{eq:mem-Jc} we \textit{fix}
the  value of $\Phi$ such that $\pm1 - \Phi\Upsilon_1$ never vanishes. This allows us to fulfill the constraint $\mathcal{L}_{22}  \geq \mathcal{L}_{12}$. Fixing the value of $\Phi$ 
is crucial for the dielectric case since, in order to keep the magnitude of the potential fixed,  the surface charge density decreases with $k_0$. 
\begin{figure}
      \includegraphics[scale=0.65]{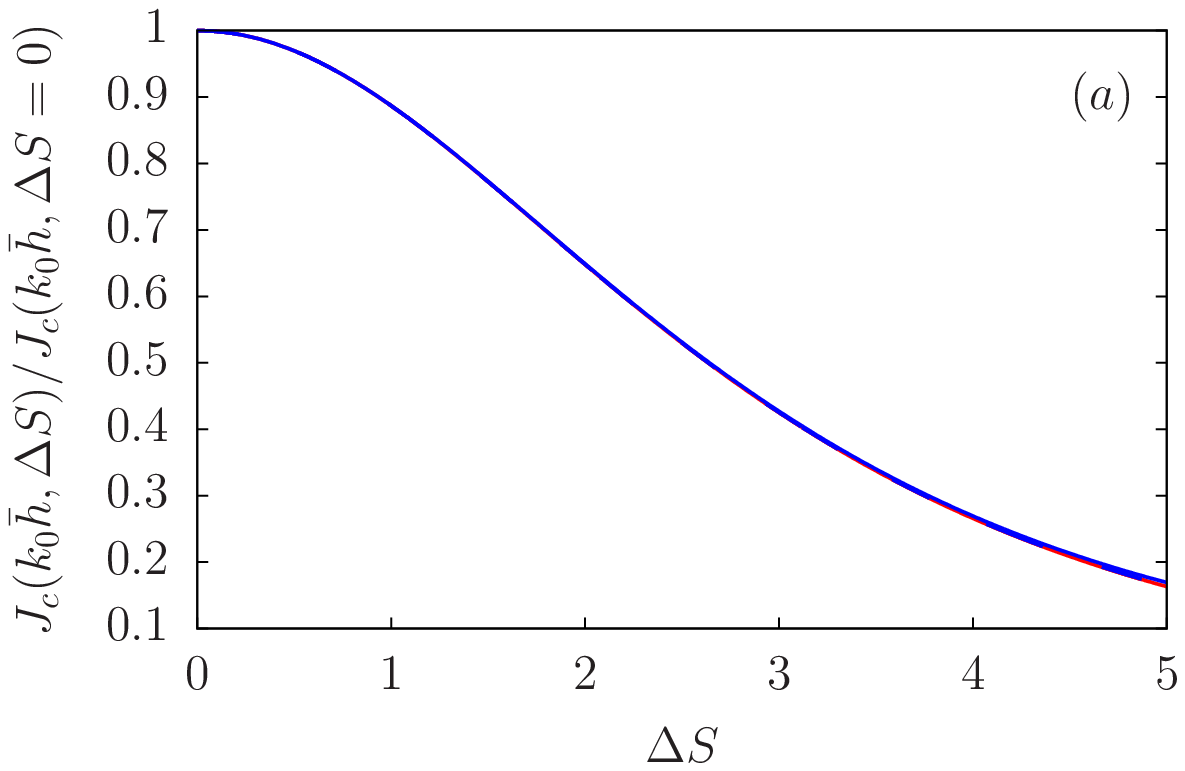}
      \includegraphics[scale=0.65]{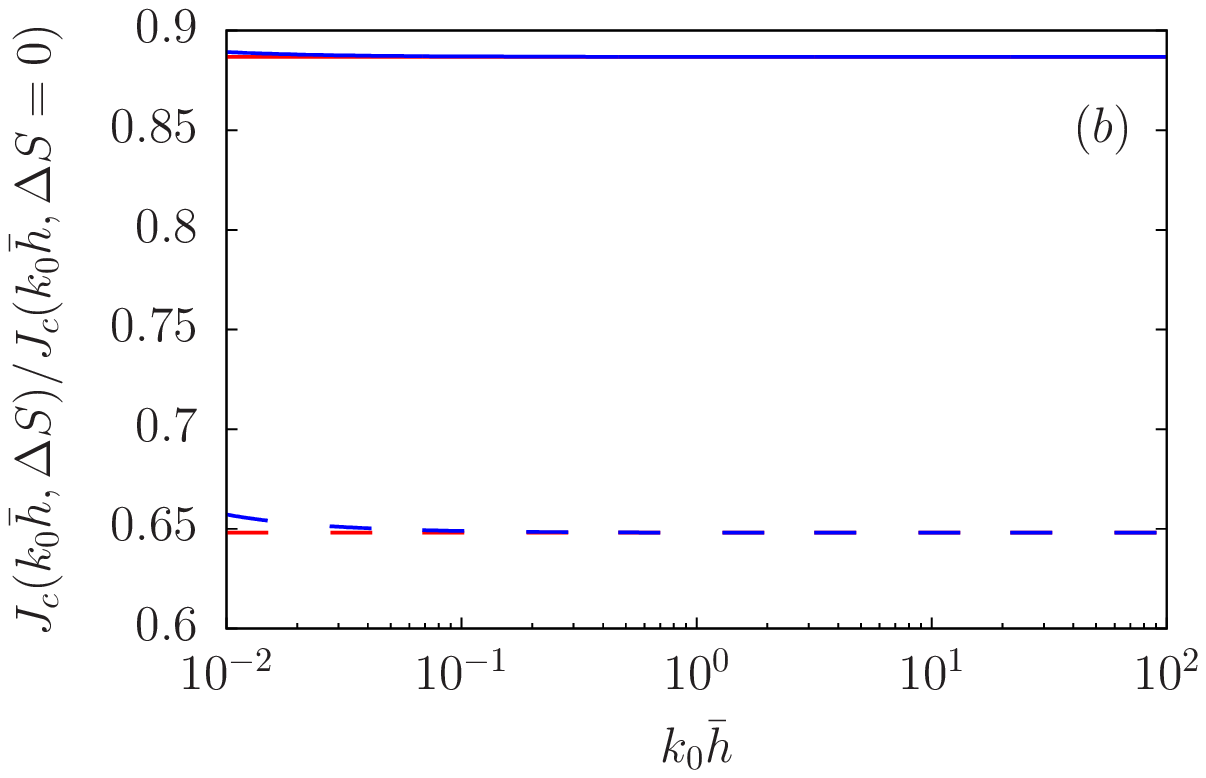}
   \caption{Solute flux $J_c$ [Eq.\eqref{eq:mem-Jc}] through a semipermeable membrane as a function of $\Delta S$ (panel a) 
   and $k_0\bar{h}$ (panel b) for conducting (red lines) and dielectric (blue lines) channel  walls. 
   %Panel a): $k_0 \bar h=1,10$ for respectively solid and dashed lines. 
   %Panel b): $\Delta S=1,2$ for respectively solid and dashed lines. $\Phi=10^{-4}$ in both panels.
   Panel (a) shows $\Delta S$-dependence at $k_{0}\bar{h}=1$ (solid) and $k_{0}\bar{h}=10$ (dashed), respectively.
   Panel (b) shows $\Delta V$-dependence at $\Delta S=1$ (solid) and $\Delta S=1$ (dashed), respectively.}
 \label{Fig:JqSemi}
\end{figure}

\section{Microscopic perspective}
%\red{We need this partfor Sec.\ref{sec:Jq}.}
So far we have discussed the macroscopic transport properties of the channel. 
However, our framework also allows us to discuss some microscopic details, such as the local electrostatic potential, $\psi_f(x,y)$, induced by the external forces modulated by the geometry of the channel.
At first order in lubrication, the leading order force-induced correction of the Poisson equation [\eqr{eq:poisson}] reads
\begin{align}
 %\epsilon\partial^2_y\psi_f(x,y)&=\beta (ze)^2\left[\psi_0(x,y) \gamma_{f}(x)+\psi_f(x,y)\gamma_{0}\right]\nn
  %    &-ze\xi_{f}(x)\,,
 \partial^2_y\psi_f(x,y)&=k_{0}^2\left[\psi_f(x,y)+\psi_0(x,y) \frac{\gamma_{f}(x)}{\gamma_{0}}\right]%\nn
      -\frac{ze\xi_{f}(x)}{\epsilon}\,,
      \label{eq:phi-f}
\end{align}
where the right hand side follows from the perturbed ionic charge density linear in the external force, $q_{f}=\rho^{+}_{f}-\rho^{-}_{f}$. 
Equation~(\ref{eq:phi-f}) is solved in Appendix~\ref{app:phifcon} for both dielectric and conducting boundary 
conditions. In the case of conducting walls, $\psi_{f}(x,y)$ reads
\begin{align}
\psi^{\zeta}_{f}(x,y)&=\frac{\psi^{\zeta}_{0}(x,y)}{2}\frac{ \gamma_{f}(x)}{ \gamma_{0}}\nn
&\times\left(k_{0} y \tanh[k_{0} y]-k_{0} h(x) \tanh[k_{0} h(x)]\right)\nn
&+\frac{1}{\beta ze}\frac{ \xi_{f}(x)}{ \gamma_{0}} \left(1-\frac{\cosh[k_{0} y]}{\cosh[k_{0} h(x)]}\right)\,, 
 \label{eq:phifcon}
\end{align}
while for dielectric walls we find
\begin{align}
 \psi^{\sigma}_f(x,y)&=\frac{\psi^{\sigma}_{0}(x,y)}{2}\frac{ \gamma_{f}(x)}{ \gamma_{0}}\nn
&\times\left(k_{0} y \tanh[k_{0} y]-k_{0} h(x)\coth [k_{0} h(x)] -1\right)\nn
&+\frac{1}{\beta ze}\frac{ \xi_{f}(x)}{ \gamma_{0}}\,, 
 \label{eq:phifdiel}
 \end{align}
with $\psi^{\zeta}_{0}(x,y)$ and $\psi^{\sigma}_{0}(x,y)$ given by \eqr{eq:electric_potential-cond} and \eqr{eq:electric_potential-0-1}, respectively.
We note that the contributions in Eqs.~\eqref{eq:phifcon} and \eqref{eq:phifdiel} are within the approximations in \eqr{eq:def-phi-exp-f}.

\subsection{Local charge electroneutrality}
At the end of Sec.~\ref{sec:Equilibrium} we showed that local charge neutrality is fulfilled at equilibrium in our system of interest.
For the out-of-equilibrium case discussed in this section one can show 
that local charge neutrality holds at lowest order in the applied forces as well: 
Using \eqr{eq:electric_potential-cond} and \eqr{eq:electric_potential-0-1} we show in Appendix~\ref{app:chargeneutrality}  that the total ionic charge 
 $q_{\rm tot,f}$ in a slab at $x$ precisely balances the local surface charge for both boundary conditions [$2e\sigma^{\zeta}_{f}(x)$, and $2e\sigma_{f}=0$, respectively]. 
Global charge neutral is then obviously satisfied at this order of approximation as well.

\subsection{Local Debye length}
Using Eqs.(\ref{eq:electrochemicalpotential}),(\ref{eq:def-varphi}) we can define a local Debye length
\begin{equation}
 k(x)=\sqrt{\frac{\beta (ze)^2}{\epsilon}(\gamma_0+\gamma_f(x))}\simeq k_0+k_1(x)\,,
\end{equation}
where 
\begin{align}
% k_0&=\sqrt{\frac{\beta (ze)^2}{\epsilon}\gamma_0}\\
 k_1(x)&=\frac{k_0}{2}\frac{\gamma_f(x)}{\gamma_0}\,.
\end{align}
Interestingly, assuming a local Debye length $k(x)=k_0 + k_1(x)$ and expanding 
Eqs.~\eqref{eq:electric_potential-cond} and \eqref{eq:electric_potential-0-1} 
for small values of $k_1(x)$, the terms proportional to $\gamma_f$ in Eqs.~\eqref{eq:phifcon} and \eqref{eq:phifdiel} 
are retrieved. Hence, our results show that the leading corrections 
to the local electrostatic potential  proportional to the local salt concentration 
$\gamma_f(x)$ can be interpreted as being caused by a local Debye length proportional to  $\gamma_f(x)$.

\section{Conclusions}\label{sec:Conclusions}
We have characterized the dynamics of a $z-z$ electrolyte embedded in a varying-section channel. 
%In order to gain analytical insight 
We focused our analysis on channels whose section
is varying smoothly enough so that we can apply a lubrication approximation to the (linearized)
Poisson-Boltzmann equation ---governing the electrostatic potential--- as well as for the Stokes equation ---governing the fluid flow. 
At equilibrium, we found that the Debye length stays constant up to first order in the lubrication expansion. 
For driven systems, we have focused on the linear response of the electrolyte to the external driving. 
Such a limit is relevant for weak external forces for which higher order contribution 
are negligibly small. Applying these approximations enabled us to derive analytical expressions for the corrections 
induced by the external driving to the local properties of the electrolyte. 
In such a regime, we have identified the set of thermodynamic forces and fluxes for which the Onsager matrix is symmetric.

Exploiting these results we have investigated several cases of experimental interest. 
In general, we found that increasing the channel corrugation $\Delta S$ 
leads to a decrease in the transport coefficients. However, our model shows
that there are a few counter-examples for which the opposite holds. 
Indeed, the electric current induced by an ionic chemical potential imbalance 
$\Delta \bar \mu$ grows with $\Delta S$ (Fig.~\ref{Fig_JqChemPot}). 
A similar effect can be obtained when multiple thermodynamic forces are applied. 
For example, both the ionic chemical potential drop, $\Delta \bar{\mu}_V$ 
induced by an applied voltage $\Delta V$ when $Q=J_c=0$ and the electric potential drop $\Delta V_\mu$ 
induced by a chemical potential drop $\Delta \bar \mu$ for $J_q=0$ grow with the channel corrugation $\Delta S$.

Finally, we have investigated the role of the conductive properties of the channel walls on the transport coefficients.
In contrast to the case of planar channel walls for which it is possible to map 
the solution for dielectric channel walls into that of conducting channel walls 
by properly rescaling the potential at the wall, when $\Delta S\neq 0$ this mapping 
does not hold anymore and different behavior appears.
Interestingly, our results show that the difference between the transport 
coefficients calculated for dielectric or conducting  channel walls 
can be significant (see Fig.~\ref{Fig_DiffU}). In particular, dielectric walls typically lead to larger transport coefficients than conducting walls. 
This difference is relevant only in the entropic electrokinetic regime, 
where the Debye length is comparable to the channel bottleneck, $k_0 h_\text{min}\simeq 1$, but not to the channel
widest section, $k_0 h_\text{max}\simeq 1$. This clearly requires $\Delta S=\ln[h_\text{max}/h_\text{min}]\neq 0$. Indeed, as shown in Fig.~\ref{Fig_DiffU} the difference between the
transport coefficient vanishes for $k_0 \bar h\rightarrow 0$, i.e., when
the Debye length is too small and also for $\Delta S\rightarrow 0$, i.e., for straight channels. 
In particular, the difference between the transport coefficients for conducting and 
dielectric walls becomes not only quantitative but also qualitative in the 
case of the electric current driven by an ionic chemical potential drop, $\Delta \bar{\mu}$ 
and with $Q=J_c=0$. For this case the current grows monotonically with $\Delta S$ for dielectric channel walls whereas it shows a maximum for conducting channel walls.

Our results show a rich dynamics of electrolyte embedded in varying-section channels. 
We believe that these results can open new routes for the realization of synthetic devices aiming at 
energy harvesting or water desalination and can be insightful for the understanding of 
biological processes such as ionic transport across pores and membranes.
%\green{Future work could: 1) consider a cylindrical geometry, to check and see if the small potential limit of \cit{VanRoij2016} coincides with our framework. This could solve our question in Sec.~\ref{sec:Peters}. 2) move beyond small surface potentials (towards \cite{burgreen1964electrokinetic}). See whether there's a solution to \ref{sec:nosaltflow} in that case. 3) nonlinear response.}

\section*{Acknowledgments}
This work has been supported by the COST Action MP1305 Flowing Matter. 
J.M.R. and I.P. acknowledge support from MINECO (Grant No. PGC2018-098373-B-100)) and DURSI (Grant No. 2017 SGR 884)

\appendix

\section{Derivation of \eqr{eq:lmbd-eps-int}\label{app:cfp}}
Inserting the chemical potential $\mu^{\pm}$ [\eqr{eq:electrochemicalpotential}] into the Nernst-Planck equation \eqref{eq:nernstplanck}], 
at linear order in the external 
force $f$ and up to quadratic in the equilibrium electrostatic potential $\psi_0$ (but disregarding $\mathcal{O}(f\psi_0^2)$) we find
\begin{align}
j_{x}^{\pm}(x,y)&=v_x(x,y)\rho_{0}^{\pm}(x,y)-D\beta \rho_{0}^{\pm}(x,y)\partial_x \mu^{\pm}_f(x,y)\nn
&\quad+\mathcal{O}(f^2)\,.
\end{align}
Using the local equilibrium approximation, $\partial_y\mu_f^\pm=0$,  we find
\begin{subequations}
 \begin{align}
 J^+&=\!\!\!\int\limits_{-h(x)}^{h(x)}\!\!\!\! \rho^+(x,y)v_x(x,y)\,\upd y-D\beta\partial_x \bar\mu^{+}_f(x)\!\!\!\!\int\limits_{-h(x)}^{h(x)}\!\!\!\! \rho^+(x,y)\,\upd y\,, \\
 J^-&=\!\!\!\int\limits_{-h(x)}^{h(x)}\!\!\!\! \rho^-(x,y)v_x(x,y)\,\upd y-D\beta\partial_x \bar\mu^{-}_f(x)\!\!\!\!\int\limits_{-h(x)}^{h(x)}\!\!\!\! \rho^-(x,y)\,\upd y \,.
 \end{align}
\end{subequations}
Recalling that $\rho^\pm_0(x,y)\simeq \varrho_0(1\mp\psi_0(x,y))$, $\gamma_0=2\varrho_0$ and defining $J^{\pm}(x)=\int_{-h(x)}^{h(x)} j_{x}^{\pm}(x,y)\, \upd y$ [\eqr{eq:coupled_fokker_planck}],  $J_{c}=J^{+}+J^{-}$ and $J_{q}=J^{+}-J^{-}$, we obtain:
\begin{subequations}\label{eq:lmbd-eps-int-app}
\begin{align}
\frac{J_{c}}{D}&=\beta ze\overline{\psi_0}(x)\partial_x \xi_{f}(x)-2h(x)\partial_x \gamma_{f}(x)+\frac{ \gamma_{0}Q}{D}\,, \label{eq:lmbd-epsa}\\
\frac{J_{q}}{D}&=\beta ze\overline{\psi_0}(x)\partial_x \gamma_{f}(x)-2h(x)\partial_x \xi_{f}(x)+\frac{\mathcal{J}_{q} (x)}{D}\,,%\label{eq:lmbd-eps-int2}
\end{align}
\end{subequations}
where we identified $Q$ [cf.~\eqr{eq:Q}] in the last term of \eqr{eq:lmbd-epsa}, and where we defined\footnote{We recall that $ze q_0(x,y)=-\epsilon k^2\psi_0(x,y)=-2\varrho\beta \psi_0(x,y)$.}
%Moreover, $\overline{\psi_0}(x)$ and $\mathcal{J}_{q} (x)$ are defined as
\begin{align}
%Q &=\int\limits_{-h(x)}^{h(x)}v_{x,f}(x,y) \gamma_{0}dy\\
\overline{\psi_0}(x)&=\int_{-h(x)}^{h(x)} \psi_{0}(x,y) \,\upd y\,,\\
%\overline{\chi\psi_0}(x) = & \int_{-h(x)}^{h(x)}\chi(x,y)\psi_0(x,y)dy\label{eq:over_chipsi}\\
%\overline{\chi}(x)= & \int_{-h(x)}^{h(x)}\chi(x,y)dy\label{eq:over_chi}\\
\mathcal{J}_{q} (x)&=\int_{-h(x)}^{h(x)}q_0(x,y) v_{x}(x,y) \,\upd y\,.  
%\label{eq:Q-Q}
\end{align}

\section{Derivation of \eqr{eq:deltapsi1}}\label{app:derivationtheta3}
Inserting $v_x(x,y)$ %$q_0(x,y)=-\gamma_{0}\beta ze\psi_0(x,y)$ 
[\eqr{eq:vel_field_x-y}] into $\mathcal{J}_{q} (x)$, to $\mathcal{O}(\psi_{0})$ only the pressure driven velocity $u_{P}$ remains, giving 
\begin{align}
 \mathcal{J}_{q} (x) &=\gamma_0 \beta ze \frac{\Delta P}{2\eta L}\frac{\bar{h}^3}{H_{3}h^{3}(x)}\nn
 &\quad\times\int_{-h(x)}^{h(x)}\left[h(x)^{2}-y^{2}\right] \psi_0(x,y)\,\upd y  
\label{eq:Q-Q-1}
\end{align}
Here we used that  $\partial_{x}P_{\textrm{tot},f}(x)=\Delta P \bar{h}^3/[H_{3}Lh^{3}(x)]+\mathcal{O}(\psi_{0})$, which follows from inserting \eqr{eq:Q-lin} into \eqr{eq:partialxPtot}.
We can rewrite the integral in \eqr{eq:Q-Q-1} by inserting the Poisson equation \eqref{eq:poisson1-0}. Two partial integrations then give
\begin{widetext}
\begin{align}\label{eq:app-Theta}
\int_{-h(x)}^{h(x)}k_{0}^2\left[h(x)^2-y^{2}\right] \psi_0(x,y)\,\upd y
&=h(x)^2\int_{-h(x)}^{h(x)}\partial_{y}^{2}\psi_0(x,y)\,\upd y-\int_{-h(x)}^{h(x)}y^{2}\partial^{2}_{y}\psi_0(x,y)\,\upd y\nn
&=\cancel{h(x)^2\int_{-h(x)}^{h(x)}\partial_{y}^{2}\psi_0(x,y) \,\upd y}-\cancel{y^{2}\partial_{y}\psi_0(x,y)\bigg|_{-h(x)}^{h(x)}}+2\int_{-h(x)}^{h(x)}y\partial_{y}\psi_0(x,y)\,\upd y\nn
%&=-2k_{0}y\psi_0(x,y)\bigg|_{-h(x)}^{h(x)}+2\int\limits_{-h(x)}^{h(x)}\psi_0(x,y)k_{0}dy\nn
&=2\int_{-h(x)}^{h(x)}[\psi_0(x,h(x))-\psi_0(x,y)]\,\upd y\,.
\end{align}
Evaluating \eqr{eq:qf} at $x=L$, a term containing $\int_{0}^{L}\upd x\,\mathcal{J}(x')/h(x')$ appears. With the above two equations and the definition of
$\Upsilon_{3}$ [cf. \eqr{eq:Upsilon3}] we find 
\begin{align}\label{eq:app_theta3}
\int_0^L\frac{\mathcal{J}_{q} (x)}{h(x)}\,\upd x
&=\gamma_0\frac{\Delta P}{k_{0}^2 \eta L}\frac{\bar{h}^3}{H_{3}}\beta ze\int_0^L \upd x\int_{-h(x)}^{h(x)}\frac{\psi_0(x,h(x))-\psi_0(x,y)}{h(x)^4}\,\upd y\nn
&=2\gamma_0\frac{\Delta P}{k_{0}^2 \eta }\Phi \Upsilon_{3}\,,  
\end{align}
\end{widetext}
which proves the appearance of $\Upsilon_{3}$ term in \eqr{eq:deltapsi1}. Note that the derivation 
in \eqr{eq:app-Theta} is valid for both boundary electric conditions; hence, so is \eqr{eq:app_theta3}.

\section{Derivation of \eqr{eq:Qf} \label{app:Q-Q}}
We recall the expression of the electroosmotic flow
\begin{align}\tag{\ref{eq:QEO}}
 Q_{\textrm{eo}}= \frac{\bar{h}^3}{H_{3}L}\int_0^L\frac{\upd x}{h^3(x)}\int_{-h(x)}^{h(x)}u_{\textrm{eo}}(x,y)\,\upd y\,,
 \label{eq:app-Q-lin}
\end{align}
where the electroosmotic fluid velocity [\eqr{eq:ueo}] reads:
\begin{align}
u_{\textrm{eo}}(x,y)&=\mathcal{U}(x,y)-\mathcal{U}(x,h(x))\,,
\label{eq:app-u}
\end{align}
with
\begin{align}\tag{\ref{eq:curlyV}}
\mathcal{U}(x,y)&=\frac{ze}{\eta}\int \upd y \int \upd y \,q(x,y)\partial_x\psi(x,y)\,,
\end{align}
the integrand of which, to first order in lubrication, $\psi_{0}$, and $f$, reads
\begin{align}\label{eq:curlyVintegrand}
q(x,y) \partial_x\psi(x,y)&= q_f(x,y)\partial_x\psi_0(x,y)\nn
&\quad\quad+q_0(x,y)\partial_x\psi_f(x,y)\,.
\end{align}
By inserting Eqs.(\ref{eq:phifcon}),(\ref{eq:phifdiel}) into \eqr{eq:curlyVintegrand}, we find $\mathcal{O}\left(\psi_0 f\right)$ 
expressions for the two terms in \eqr{eq:curlyVintegrand} for both boundary conditions, which we can write compactly as 
\begin{align}
 q_f(x,y)&\partial_x\psi_0(x,y)=\xi_{f}(x)\left[1 - \theta(x,y)\right]\partial_x\psi_0(x,y)\\
 q_0(x,y)&\partial_x\psi_f(x,y)=-\psi_0(x,y)\partial_x\left[\xi_{f}(x)\theta(x,y)\right]\,,
\end{align}
 with
\begin{align}\label{eq:def-theta}
 \theta(x,y)=
 \begin{cases}
\displaystyle 1-\frac{\cosh[k_{0} y]}{\cosh[k_{0} h(x)]}	\quad\quad&\textrm{cond,}\\
\displaystyle 1&\textrm{diel,}\,.
\end{cases}
\end{align}
After reshuffling $\mathcal{U}$ reads
\begin{align}
 \mathcal{U}_{f}(x,y)&=-\frac{ze}{\eta}\int\! \upd y\! \int\! \upd y\Bigl\{ \partial_x\left[\psi_0(x,y)\xi_f(x)(\theta(x,y)-1)\right]\nonumber\\
 &\quad\quad\quad\quad\quad\quad\quad\quad\quad+ \psi_0(x,y)\partial_x\xi_f(x)\Bigr\}\,.
 \label{eq:app-VV}
\end{align}
We can now determine $u_{\textrm{eo}}(x,y)$ for the two different boundary conditions explicitly (note that the first term of the above integrand is zero for dielectric walls and the last term can be explicitly integrated twice), 
\begin{widetext}
\begin{align}
 u^\zeta_{\textrm{eo}}(x,y)&=\frac{1}{\beta ze}\,\frac{\epsilon\zeta}{\eta}\,\left\{\frac{\partial_x\xi_f(x)}{\gamma_0}\left(1-\frac{\cosh[k_0 y]}{\cosh[k_0 h(x)]} \right)-\partial_x\left[\frac{\xi_f(x)}{\gamma_0}\frac{2k^2_0[y^2-h^2(x)]+\cosh[2k_0 y]-\cosh[2k_0 h(x)]}{8\cosh^2[k_0 h(x)]}\right]\right\}\,,\\
 %&+\frac{ze}{\eta}\frac{\Delta V}{L}\left(\widetilde{\chi\psi_0}(x,h(x))-\widetilde{\chi\psi_0}(x,y)\right)\\
u^\sigma_{\textrm{eo}}(x,y)&=\frac{e\sigma}{\eta k_0}\frac{1}{\beta ze}\frac{\partial_x\xi_f(x)}{\gamma_{0}}\frac{\cosh[k_0 h(x)]-\cosh[k_0 y]}{\sinh[k_0 h(x)]}\,.
\end{align}
Using $\mathscr{G}(x)=\tanh(x)/x$ and $\mathscr{L}(x)=\coth(x)-1/x$ as defined in \eqr{eq:GandL}, we obtain
\begin{subequations}\label{eq:intua}
\begin{align}
 \int_{-h(x)}^{h(x)}u^\zeta_{\textrm{eo}}(x,y) \,\upd y&=\frac{2h(x)}{\beta ze}\,\frac{\epsilon \zeta}{\eta }\,\frac{\partial_x\xi_f(x)}{\gamma_0}\left(1-\mathscr{G}[k_0h(x)]\right) - \frac{1}{\beta ze}\frac{\epsilon\zeta}{\eta k_0 }\frac{\partial_x\left[\xi_f(x)\Gamma(x)\right]}{\gamma_0}\,,\\
 %&+\frac{ze}{\eta}\frac{\Delta V}{L}\int_{-h(x)}^{h(x)}\left(\widetilde{\chi\psi_0}(x,h(x))-\widetilde{\chi\psi_0}(x,y)\right)dy\\
  \int_{-h(x)}^{h(x)}u^\sigma_{\textrm{eo}}(x,y) \,\upd y&=2h(x)\frac{e\sigma}{\eta k_0}\frac{\partial_x\xi_f(x)}{\beta ze \gamma_{0}}\mathscr{L}[k_{0} h(x)]\,,
\end{align}
\end{subequations}
with
\begin{align}
 \Gamma(x)&=\frac{1}{4}\tanh[k_0 h(x)]-\frac{k_0 h(x)}{2\cosh[k_0 h(x)]}+ \frac{3k_0 h(x)-4k_{0}^3h^3(x)}{12\cosh^2[k_0 h(x)]}
 %\widetilde{\chi\psi_0}(x,y)&=\int\! dy\! \int\! dy \psi_0(x,y) \chi(x,y)\,.
\end{align}
Using Eqs.~\eqref{eq:app-u} and \eqref{eq:intua} we find the following expressions for $Q_{\textrm{eo}}$ [cf. \eqr{eq:app-Q-lin}]
\begin{subequations}\label{QEOF2}
 \begin{align}
Q_{\textrm{eo}}^{\zeta}
&=2\,\frac{\epsilon\zeta}{\eta}\,\,\frac{\bar{h}^3}{H_{3}L}\int_0^L\frac{1}{\beta ze}\frac{\partial_x\xi_f(x)}{\gamma_0}\frac{1-\mathscr{G}[k_0h(x)]}{h^2(x)}\,\upd x-\frac{\epsilon\zeta}{\eta k_0}\frac{1}{\beta ze}\frac{\bar{h}^3}{H_{3}L}\int_0^L\frac{\partial_x\left[\xi_f(x)\Gamma(x)\right]}{ \gamma_0}\frac{\upd x}{h^3(x)}\,,\label{eq:app-dd}\\
Q_{\textrm{eo}}^{\sigma}&=2\frac{e\sigma}{\eta k_0}\frac{\bar{h}^3}{H_{3}L}\int_0^L \frac{1}{\beta ze}\frac{\partial_x\xi_f(x)}{\gamma_0}\frac{\mathscr{L}[k_{0} h(x)]}{h^2(x)}\,\upd x\,.\label{eq:Qfsigma}
\end{align}
\end{subequations}
%In which we recognize the Langevin function. 
We obtain an expression for $\partial_x\xi_f(x)$ by substituting \eqr{eq:partialvarphif} in \eqr{eq:partialpsi}. At order $\mathcal{O}((\psi_0)^{0})$, we obtain:
\begin{align}
 \partial_x \xi_{f}(x)=-\frac{J_{q}}{2Dh(x)}+\mathcal{O}(\psi_0)\,.\label{eq:app-xi-2}
%  \underbrace{\frac{1}{2Dh(x)}\left[(\gamma_0 Q-J_{c})\frac{\beta ze\overline{\psi_0}(x) }{2h(x)}  -\gamma_0\mathcal{J}_{q} (x)  \right] }_{\mathcal{O}(\psi_0)}
\end{align}
Inserting this expression into \eqr{QEOF2}, at linear order in $\psi_0$, we find:
\begin{subequations}\label{eq:app-Q}
\begin{align}
 Q_{\textrm{eo}}^\zeta&=-\frac{J_q}{D}\,\frac{\epsilon\zeta}{\eta}\,\,\frac{1}{\beta ze \gamma_0}\frac{\bar{h}^3}{H_{3}L}\int_0^L\frac{1-\mathscr{G}[k_0h(x)]}{h^3(x)}\,\upd x-\frac{\epsilon}{\eta}\frac{\zeta}{k_0}\frac{1}{\beta ze}\frac{\bar{h}^3}{H_{3}L}\int_0^L\frac{\partial_x\left[\xi_f(x)\Gamma(x)\right]}{ \gamma_0}\frac{\,\upd x}{h^3(x)}\,,\label{eq:app_Q-z}\\
 Q_{\textrm{eo}}^\sigma&=-2\frac{e\sigma}{\eta k_0}\frac{1}{\beta ze \gamma_0}\frac{\bar{h}^3}{H_{3}L}\int_0^L \frac{J_q}{2Dh(x)}\frac{\mathscr{L}[k_{0} h(x)]}{h^2(x)}\,\upd x\,.
 %&-\frac{1}{24}\frac{\epsilon}{\eta}\frac{\zeta}{k_0}\frac{\bar{h}^3}{H_{3}L}\frac{1}{\beta ze \gamma_0}\left[\frac{J_q}{D}+2\beta ze \Delta V\gamma_0\frac{\bar{h}}{L}\right]\left[G(h(L))-G(h(0))\right]
 % +\frac{1}{24}\frac{\epsilon}{\eta}\frac{\bar{h}^2}{H_{3}(L/\bar{h})}\frac{1}{\beta ze \gamma_0}\frac{\zeta}{k_0}\left[\frac{\xi(L)\Gamma(L)}{h^3(L)}-\frac{\xi(0)\Gamma(0)}{h^3(0)}+3\int_0^L\xi_f(x)\Gamma(x)\frac{\partial_x h(x)}{h^4(x)}dx\right]
\end{align}
\end{subequations}
\end{widetext}
Comparing the above equations to \eqr{eq:Upsilonexplicit} we see that we can write $Q_{\textrm{eo}}$ as
\begin{subequations}
\begin{align}
 Q_{\textrm{eo}}^\zeta&=-J_q\frac{\Phi \Upsilon_{3}^{\zeta}}{D\beta\eta k_0^2}-\frac{\epsilon}{\eta}\frac{\zeta}{k_0}\frac{1}{\beta ze}\frac{\bar{h}^3}{H_{3}L}\int_0^L\frac{\partial_x\left[\xi_f(x)\Gamma(x)\right]}{ \gamma_0 h^3(x)}\,\upd x\,,\label{eq:Qeofzeta}\\
 Q_{\textrm{eo}}^\sigma&=-J_q\frac{\Phi \Upsilon_{3}^{\sigma}}{D\beta\eta k_0^2}\,,\label{eq:Qeofsigma}
\end{align}
\end{subequations}
which proves \eqr{eq:Qf} up to the second term on the right hand side of \eqr{eq:Qeofzeta}. Since at $\mathcal{O}\left(\zeta^0\right)$ we have $\xi_f(x)=\xi_f[h(x)]$,
we can write
\begin{align}
 \int_0^L \frac{\partial_xF[h(x)]}{h^3(x)}\,\upd x=\int_0^L \frac{\delta F[h(x)]}{\delta h(x)} \frac{\partial_x h(x)}{h^3(x)}\,\upd x\,,
 \label{eq:app-dd-1}
\end{align}
with $F[h(x)]=\xi_f(x)\Gamma(x)$, 
i.e., $F[h(x)]$ depends on $x$ solely through $h(x)$.  Without loss of generality we can define a function $G[h(x)]$ such that
\begin{align}
\frac{1}{h^3(x)}\frac{\delta F[h(x)]}{\delta h(x)}&=\frac{\delta G[h(x)]}{\delta h(x)}\nn
 \Rightarrow\quad \int_0^L \frac{\partial_xF[h(x)]}{h^3(x)}\,\upd x&= \int_0^L \frac{\delta G[h(x)]}{\delta h(x)}\partial_x h(x) \,\upd x\nn
&= G[h(L)]-G[h(0)]\,,
\end{align}
i.e., for periodic channels, $h(L)=h(0)$, we have $G(L)=G(0)$ and the last term in \eqr{eq:Qeofzeta} vanishes. %\red{EO pump \cite{bruus2008theoretical, rinne2012nanoscale}?}

\begin{widetext}
\section{Derivation of \eqr{eq:phifcon} and \eqr{eq:phifdiel} \label{app:phifcon}}
The solution to \eqr{eq:phi-f} reads %\red{check factors of $z$ and $e$ in this section and the next: seems ok}
\begin{align}
 \psi_f(x,y)&=-A_0(x)\frac{ \gamma_{f}(x)}{4 \gamma_{0}}\left[\cosh(k_0 y)-2k_0y\sinh(k_0 y)\right]+A_f(x)\cosh(k_0 y) +\frac{1}{\beta ze}\frac{ \xi_{f}(x)}{ \gamma_{0}}\,,
\label{eq:sol-phi-f}
\end{align}
where $A_{0}$ is given for conducting and dielectric walls by $A^{\zeta}_{0}=\zeta/\cosh[k_{0}h(x)]$ and $A^{\zeta}_{0}=e\sigma/(\epsilon k_{0}\sinh[k_{0}h(x)])$, respectively.
The term $A_f(x)$ is obtained by imposing the suitable boundary conditions at the channel walls.
%In the following we derive the velocity profile at leading order in lubrication. Moreover, we assume that the external force is small enough so that we can disregard terms proportional to $\psi_f^2$\footnote{We recall that the equilibrium solution has been derived up to correction of the order of $\mathcal{O}(\bar{h}/L)^2$.}.

%\subsection{Conducting walls}
For conducting channel walls one has:
%\begin{align}
 $\psi_f(x,\pm h(x))=0$ %\\
% A_0(x)&=\frac{\zeta}{\cosh(k_0 h(x))}
%\end{align}
and hence
\begin{equation}
 A^{\zeta}_f(x)=\frac{1}{\cosh[k_0 h(x)]}\left[\frac{\zeta}{4}\frac{ \gamma_{f}(x)}{ \gamma_{0}}(1-2k_0h(x)\tanh[k_0 h(x)])-\frac{1}{\beta ze}\frac{ \xi_{f}(x)}{ \gamma_{0}}\right]\,.%\frac{\cosh(k_0 h(x))-2k_0h(x)\sinh(k_0 h(x))}{\cosh(k_0 h(x))}
 \label{eq:sol-phi-f-2-cond}
\end{equation}
%In the case of \textbf{conducting walls} ($A_{0}=\zeta /\cosh[\kappa h]$) we fix $A(x)$ by demanding $\xi^{(1)}(x,\pm h)=0$ to
% \begin{align}\label{eq:Aconst}
%A(x)&=\frac{1}{\cosh[\kappa h]}\left\{-\frac{\zeta}{4}\frac{ \gamma_{f}(x)}{2\rhos} 
%(2\kappa h \tanh[\kappa h]-1)-\frac{e \xi_{f}(x) }{\epsilon\kappa^2} \right\}. 
%\end{align}
Inserting $A^{\zeta}_f(x)$ into \eqr{eq:sol-phi-f} we find \eqr{eq:phifcon} of the main text.
 \begin{align}\label{eq:pot1cond}
\psi^{\zeta}_{f}(x,y)&=\frac{\zeta}{2}\frac{ \gamma_{f}(x)}{ \gamma_{0}} 
\left(k_{0} y \frac{\sinh[k_{0} y]}{\cosh[k_{0} h(x)]}-k_{0} h(x) \frac{\cosh[k_{0} y]\sinh[k_{0} h(x)]}{\cosh^{2}[k_{0} h(x)]}\right)
+\frac{1}{\beta ze}\frac{ \xi_{f}(x)}{ \gamma_{0}} \left(1-\frac{\cosh[k_{0} y]}{\cosh[k_{0} h(x)]}\right)\\
&=\frac{\psi^{\zeta}_{0}(x,y)}{2}\frac{ \gamma_{f}(x)}{ \gamma_{0}}
\left(k_{0} y \tanh[k_{0} y]-k_{0} h(x) \tanh[k_{0} h(x)]\right)
+\frac{1}{\beta ze}\frac{ \xi_{f}(x)}{ \gamma_{0}}\left(1-\frac{\cosh[k_{0} y]}{\cosh[k_{0} h(x)]}\right). 
\end{align}
%In contrast for dielectric channel walls one has:
%\begin{align}
%\left.\frac{\partial\psi_f(x,y)}{\partial y}\right|_{y=\pm h(x)}&=0
%A_0(x)&=\frac{\sigma}{\epsilon k_0}\frac{1}{|\sinh(k_0 h(x))|}
%\end{align}
%that leads to: 

%\subsection{Insulating walls}
In the case of dielectric walls we fix $A^{\sigma}_{f}(x)$ by demanding $\partial_{y}\psi_{f}(x,\pm h(x))=0$. First we find
 \begin{align}\label{eq:pot1}
\psi^{\sigma}_{f}(x,y)&=A^{\sigma}_{f}(x)\cosh[k_{0} y]-\frac{e\sigma}{4\epsilon k_{0}}\frac{ \gamma_{f}(x)}{ \gamma_{0}} 
\frac{\cosh[k_{0} y]-2k_{0} y \sinh[k_{0} y]}{\sinh[k_{0} h(x)]}+\frac{1}{\beta ze}\frac{ \xi_{f}(x)}{ \gamma_{0}} \nn
%\partial_{y}\psi^{\sigma}_{f}(x,y)&=A^{\sigma}_{f}(x)k_{0} \sinh[k_{0} y]+\frac{\sigma}{4\epsilon k_{0}}\frac{ \gamma_{f}(x)}{ \gamma_{0}} 
%\frac{k_{0} \sinh[k_{0} y]+2k_{0}^2 y \cosh[k_{0} y]}{\sinh[k_{0} h(x)]} \nn
\partial_{y}\psi^{\sigma}_{f}(x,h(x))&=A^{\sigma}_{f}(x)k_{0} \sinh[k_{0} h(x)]+\frac{e\sigma}{4\epsilon }
\frac{ \gamma_{f}(x)}{ \gamma_{0}} 
\left(1+2k_{0} h(x)\coth [k_{0} h(x)]\right). 
\end{align}
We now fix $A^{\sigma}_{f}(x)$ to
% \begin{align}\label{eq:Aconst}
%A(x)=-\frac{\sigma}{4\epsilon \kappa}
%\frac{ \gamma_{f}(x)}{ \gamma_{0}} 
%\left(1+\frac{2\kappa h}{\tanh\kappa h}\right)\frac{1}{\sinh[\kappa h]}
%\end{align}
\begin{equation}
 A^{\sigma}_f(x)=-\frac{e\sigma}{4\epsilon k_0}\frac{ \gamma_{f}(x)}{ \gamma_{0}}\frac{1+2k_0 h(x)\coth[k_0 h(x)]}{\sinh[k_0 h(x)]}\,.
\label{eq:sol-phi-f-2}
\end{equation}
Using this results we find \eqr{eq:phifdiel}:
\begin{align}\label{eq:pot1diel}
\psi^{\sigma}_{f}(x,y)&=-\frac{e\sigma}{4\epsilon k_{0} }
\frac{ \gamma_{f}(x)}{ \gamma_{0}} 
\left[\frac{\cosh[k_{0} y]-2k_{0} y \sinh[k_{0} y]}{\sinh[k_{0} h(x)]}+
\frac{\cosh[k_{0} y]}{\sinh[k_{0} h(x)]}+\frac{2k_{0} h(x)}{\tanh [k_{0} h(x)]}\frac{\cosh[k_{0} y]}{\sinh[k_{0} h(x)]}\right]
+\frac{1}{\beta ze}\frac{ \xi_{f}(x)}{ \gamma_{0}} \nn
&=\frac{e\sigma}{2\epsilon k_{0}}\frac{\cosh[k_{0} y]}{\sinh [k_{0} h(x)]}\frac{ \gamma_{f}(x)}{ \gamma_{0}} 
\left(k_{0} y \tanh[k_{0} y]-1-k_{0} h(x)\coth [k_{0} h(x)] \right)
+\frac{1}{\beta ze}\frac{ \xi_{f}(x)}{ \gamma_{0}}\nn
\psi^{\sigma}_{f}(x,y)&=\frac{\psi_{0}^{\sigma}(x,y)}{2}\frac{ \gamma_{f}(x)}{ \gamma_{0}}
\left(k_{0} y \tanh[k_{0} y]-1-k_{0} h(x)\coth [k_{0} h(x)] \right)
+\frac{1}{\beta ze}\frac{ \xi_{f}(x)}{ \gamma_{0}}\,. 
\end{align}

\section{Derivation of local charge neutrality at $\mathcal{O}(\psi f)$\label{app:chargeneutrality}}
The total ionic charge in a slab at $x$ is obtained by integrating the rhs of \eqr{eq:phi-f} along the transverse direction
\begin{align}
ze\bar{q}= ze\overline{q_f}(x)&=2zeh(x) \xi_{f}(x)-\beta (ze)^{2}\left[\overline{\psi_0}(x) \gamma_{f}(x)+\overline{\psi_f}(x) \gamma_{0}(x)\right]\,.
 \label{eq:overlineq-f}
\end{align}
Here,
$\overline{\psi^{\zeta}_{0}}=2\zeta\tanh[k_0 h(x)]/k_0 $ and %\quad,\quad
$\overline{\psi^{\sigma}_{0}}=2e\sigma/(\epsilon k_0^2)=2\sigma/(\beta z^2 e  \gamma_{0})$.
In the conducting case, 
\begin{align}
\overline{\psi^{\zeta}_{f}}(x,y)&=-\zeta\frac{ \gamma_{f}(x)}{2 \gamma_{0}}
\left( 2h(x) \tanh^{2}[k_{0} h(x)]-2h(x)+\frac{2\tanh[k_{0} h(x)]}{k_{0}}\right)
+\frac{1}{\beta ze}\frac{ \xi_{f}(x)}{ \gamma_{0}} \left(2h(x)-\frac{2}{k_{0}}\tanh[k_{0} h(x)]\right)\nn
&=-\zeta h(x)\frac{ \gamma_{f}(x)}{ \gamma_{0}} 
\left( \tanh^{2}[k_{0} h(x)]-1+\frac{\tanh[k_{0} h(x)]}{k_{0} h(x)}\right)
+\frac{2h(x)}{\beta ze}\frac{ \xi_{f}(x) }{ \gamma_{0}} \left(1-\frac{\tanh[k_{0} h(x)]}{k_{0} h(x)}\right). 
\end{align}
This gives
\begin{align}
  \overline{q^{\zeta}_f}(x,y)%&=2h(x) \xi_{f}(x)-\beta ze\left[ \gamma_{f}(x)\overline{\psi_0}(x,y)+ \gamma_{0}(x)\overline{\psi_f}(x,y)\right]\\
  &=\cancel{2h(x) \xi_{f}(x)}-\beta ze \gamma_{f}(x)\frac{2\zeta\tanh[k_0 h(x)]}{k_0}\nn	
  &\quad+\beta ze\zeta h(x) \gamma_{f}(x)\left(  \tanh^{2}[k_{0} h(x)]-1+\frac{\tanh[k_{0} h(x)]}{k_{0} h(x)}\right)
-2h(x) \xi_{f}(x)  \left(\cancel{1}-\frac{\tanh[k_{0} h(x)]}{k_{0} h(x)}\right)\\
ze\overline{q^{\zeta}_f}(x,y)&=\beta (ze)^2\zeta h(x) \gamma_{f}(x)\left(  \tanh^{2}[k_{0} h(x)]-1-\frac{\tanh[k_{0} h(x)]}{k_{0} h(x)}\right)
+2h(x) ze\xi_{f}(x) \frac{\tanh[k_{0} h(x)]}{k_{0} h(x)}\,.
\end{align}
Meanwhile the surface charge per surface is given by $e\sigma=-\epsilon\partial_{y}\psi_{f}(x,y=h(x))$.
We find for the two surfaces
\begin{align}\label{eq:surfacechargeconducting}
2e\sigma(x)%&=-2\epsilon\partial_{y}\psi_{f}(x,y=h(x))\nn
&=\epsilon\zeta\frac{ \gamma_{f}(x)}{ \gamma_{0}}
\left[k_{0}^2 h(x)\tanh^{2}[k_{0} h(x)]- k_{0}\tanh[k_{0} h(x)]-k_{0}^2 h(x)\right]
+2ze \xi_{f}(x)\frac{\tanh[k_{0} h(x)]}{k_{0}}\nn
2e\sigma^{\zeta}(x)&= \beta (ze)^2 \zeta h(x) \gamma_{f}(x)
\left[\tanh^{2}[k_{0} h(x)]- \frac{\tanh[k_{0} h(x)]}{k_{0} h(x)}-1\right]
+\frac{2ze \xi_{f}(x) }{k_{0}}\tanh[k_{0} h(x)]\,. 
\end{align}
Clearly, the local surface charge $2e\sigma^{\zeta}(x)$ balances the ionic charge $ze\overline{q^{\zeta}_f}(x,y)$ at each $x$.

%\subsection{Insulating walls}
In the dielectric case, there are no perturbations to the surface charge: $ -\epsilon\partial_{y}\psi_{f}(x,y)=0$. 
Which means that local charge neutrality is satisfied only if the total perturbed ionic density vanishes $ \overline{q^{\sigma}_f}(x,y)=0$. 
We find
\begin{align}
\psi^{\sigma}_{f}(x,y)&=-\frac{e\sigma}{2\epsilon k_{0}}\frac{ \gamma_{f}(x)}{ \gamma_{0}} 
\left( \frac{\cosh[k_{0} y]}{\sinh[k_{0} h(x)]}+k_{0} h(x)
\frac{\cosh[k_{0} y]\cosh[k_{0} h(x)]}{\sinh^{2}[k_{0} h(x)]}-k_{0} y \frac{\sinh[k_{0} y]}{\sinh[k_{0} h(x)]}\right)
+\frac{1}{\beta ze}\frac{ \xi_{f}(x)}{ \gamma_{0}} \nn
\Rightarrow\overline{\psi^{\sigma}_{f}}(x,y)&=-\frac{e\sigma}{2\epsilon k_{0}}\frac{ \gamma_{f}(x)}{ \gamma_{0}} 
\left(\frac{2}{k_{0}}+ \cancel{\frac{2h(x)}{\tanh[k_{0} h(x)]}}+\frac{2}{k_{0}}-\cancel{\frac{2h(x)}{\tanh[k_{0} h(x)]}}\right)
+\frac{2h(x)}{\beta ze}\frac{ \xi_{f}(x)}{ \gamma_{0}}\nn
\beta ze \gamma_{0}\overline{\psi^{\sigma}_{f}}(x,y)&=-\frac{2e\sigma}{ze}\frac{ \gamma_{f}(x)}{ \gamma_{0}}
+2h(x) \xi_{f}(x)\,.
\end{align}
With \eqr{eq:overlineq-f} we then indeed find that %This gives
\begin{align}
 \overline{q^{\sigma}_f}(x,y)&=2h(x) \xi_{f}(x)- \gamma_{f}(x)\frac{2\sigma}{z  \gamma_{0}}+\frac{2\sigma}{z}\frac{ \gamma_{f}(x)}{ \gamma_{0}}
-2h(x) \xi_{f}(x) =0\,.
\end{align}
\end{widetext}

\bibliography{bib_paper_entropy}

\begin{thebibliography}{47}
\expandafter\ifx\csname natexlab\endcsname\relax\def\natexlab#1{#1}\fi
\expandafter\ifx\csname bibnamefont\endcsname\relax
  \def\bibnamefont#1{#1}\fi
\expandafter\ifx\csname bibfnamefont\endcsname\relax
  \def\bibfnamefont#1{#1}\fi
\expandafter\ifx\csname citenamefont\endcsname\relax
  \def\citenamefont#1{#1}\fi
\expandafter\ifx\csname url\endcsname\relax
  \def\url#1{\texttt{#1}}\fi
\expandafter\ifx\csname urlprefix\endcsname\relax\def\urlprefix{URL }\fi
\providecommand{\bibinfo}[2]{#2}
\providecommand{\eprint}[2][]{\url{#2}}

\bibitem[{\citenamefont{Alberts et~al.}(2007)\citenamefont{Alberts, Johnson,
  Lewis, Raff, Roberts, and Walter}}]{Albers}
\bibinfo{author}{\bibfnamefont{B.}~\bibnamefont{Alberts}},
  \bibinfo{author}{\bibfnamefont{A.}~\bibnamefont{Johnson}},
  \bibinfo{author}{\bibfnamefont{J.}~\bibnamefont{Lewis}},
  \bibinfo{author}{\bibfnamefont{M.}~\bibnamefont{Raff}},
  \bibinfo{author}{\bibfnamefont{K.}~\bibnamefont{Roberts}}, \bibnamefont{and}
  \bibinfo{author}{\bibfnamefont{P.}~\bibnamefont{Walter}},
  \emph{\bibinfo{title}{Molecular Biology of the Cell}}
  (\bibinfo{publisher}{Garland Science}, \bibinfo{address}{Oxford},
  \bibinfo{year}{2007}).

\bibitem[{\citenamefont{Bocquet and Charlaix}(2010)}]{Lyderic_Charlaix}
\bibinfo{author}{\bibfnamefont{L.}~\bibnamefont{Bocquet}} \bibnamefont{and}
  \bibinfo{author}{\bibfnamefont{E.}~\bibnamefont{Charlaix}},
  \bibinfo{journal}{Chem. Soc. Rev.} \textbf{\bibinfo{volume}{39}},
  \bibinfo{pages}{1073} (\bibinfo{year}{2010}).

\bibitem[{\citenamefont{Calero et~al.}(2011)\citenamefont{Calero, Faraudo, and
  Aguilella-Arzo}}]{Calero}
\bibinfo{author}{\bibfnamefont{C.}~\bibnamefont{Calero}},
  \bibinfo{author}{\bibfnamefont{J.}~\bibnamefont{Faraudo}}, \bibnamefont{and}
  \bibinfo{author}{\bibfnamefont{M.}~\bibnamefont{Aguilella-Arzo}},
  \bibinfo{journal}{Phys. Rev. E} \textbf{\bibinfo{volume}{83}},
  \bibinfo{pages}{021908} (\bibinfo{year}{2011}).

\bibitem[{\citenamefont{Peyser et~al.}(2014)\citenamefont{Peyser, Gillespie,
  Roth, and Nonner}}]{Roth2014}
\bibinfo{author}{\bibfnamefont{A.}~\bibnamefont{Peyser}},
  \bibinfo{author}{\bibfnamefont{D.}~\bibnamefont{Gillespie}},
  \bibinfo{author}{\bibfnamefont{R.}~\bibnamefont{Roth}}, \bibnamefont{and}
  \bibinfo{author}{\bibfnamefont{W.}~\bibnamefont{Nonner}},
  \bibinfo{journal}{Biophysical Journal} \textbf{\bibinfo{volume}{107}},
  \bibinfo{pages}{1841} (\bibinfo{year}{2014}).

\bibitem[{\citenamefont{Melnikov et~al.}(2017)\citenamefont{Melnikov, Hulings,
  and Gracheva}}]{Gracheva2017}
\bibinfo{author}{\bibfnamefont{D.~V.} \bibnamefont{Melnikov}},
  \bibinfo{author}{\bibfnamefont{Z.~K.} \bibnamefont{Hulings}},
  \bibnamefont{and} \bibinfo{author}{\bibfnamefont{M.~E.}
  \bibnamefont{Gracheva}}, \bibinfo{journal}{Physical Review E}
  \textbf{\bibinfo{volume}{95}}, \bibinfo{pages}{063105}
  (\bibinfo{year}{2017}).

\bibitem[{\citenamefont{Bacchin}(2018)}]{Bacchin2018}
\bibinfo{author}{\bibfnamefont{P.}~\bibnamefont{Bacchin}},
  \bibinfo{journal}{Membranes} \textbf{\bibinfo{volume}{8}}
  (\bibinfo{year}{2018}).

\bibitem[{\citenamefont{Wheeler and Stroock}(2008)}]{Strook2008}
\bibinfo{author}{\bibfnamefont{T.}~\bibnamefont{Wheeler}} \bibnamefont{and}
  \bibinfo{author}{\bibfnamefont{A.}~\bibnamefont{Stroock}},
  \bibinfo{journal}{Nature} \textbf{\bibinfo{volume}{455}},
  \bibinfo{pages}{208} (\bibinfo{year}{2008}).

\bibitem[{\citenamefont{Nipper and Dixon}(2011)}]{Nipper2011}
\bibinfo{author}{\bibfnamefont{M.}~\bibnamefont{Nipper}} \bibnamefont{and}
  \bibinfo{author}{\bibfnamefont{J.}~\bibnamefont{Dixon}},
  \bibinfo{journal}{Cardiovasc. Eng. Technol.} \textbf{\bibinfo{volume}{2}},
  \bibinfo{pages}{296} (\bibinfo{year}{2011}).

\bibitem[{\citenamefont{Wiig and Swartz}(2012)}]{Wiig2012}
\bibinfo{author}{\bibfnamefont{H.}~\bibnamefont{Wiig}} \bibnamefont{and}
  \bibinfo{author}{\bibfnamefont{M.}~\bibnamefont{Swartz}},
  \bibinfo{journal}{Physiol. Rev.} p. \bibinfo{pages}{1005}
  (\bibinfo{year}{2012}).

\bibitem[{\citenamefont{Siria et~al.}(2013)\citenamefont{Siria, Poncharal,
  Biance, Fulcrand, Blase, Purcell, and Bocquet}}]{siria2013giant}
\bibinfo{author}{\bibfnamefont{A.}~\bibnamefont{Siria}},
  \bibinfo{author}{\bibfnamefont{P.}~\bibnamefont{Poncharal}},
  \bibinfo{author}{\bibfnamefont{A.-L.} \bibnamefont{Biance}},
  \bibinfo{author}{\bibfnamefont{R.}~\bibnamefont{Fulcrand}},
  \bibinfo{author}{\bibfnamefont{X.}~\bibnamefont{Blase}},
  \bibinfo{author}{\bibfnamefont{S.~T.} \bibnamefont{Purcell}},
  \bibnamefont{and} \bibinfo{author}{\bibfnamefont{L.}~\bibnamefont{Bocquet}},
  \bibinfo{journal}{Nature} \textbf{\bibinfo{volume}{494}},
  \bibinfo{pages}{455} (\bibinfo{year}{2013}).

\bibitem[{\citenamefont{Secchi et~al.}(2016)\citenamefont{Secchi, Marbach,
  Nigu{\`e}s, Stein, Siria, and Bocquet}}]{secchi2016massive}
\bibinfo{author}{\bibfnamefont{E.}~\bibnamefont{Secchi}},
  \bibinfo{author}{\bibfnamefont{S.}~\bibnamefont{Marbach}},
  \bibinfo{author}{\bibfnamefont{A.}~\bibnamefont{Nigu{\`e}s}},
  \bibinfo{author}{\bibfnamefont{D.}~\bibnamefont{Stein}},
  \bibinfo{author}{\bibfnamefont{A.}~\bibnamefont{Siria}}, \bibnamefont{and}
  \bibinfo{author}{\bibfnamefont{L.}~\bibnamefont{Bocquet}},
  \bibinfo{journal}{Nature} \textbf{\bibinfo{volume}{537}},
  \bibinfo{pages}{210} (\bibinfo{year}{2016}).

\bibitem[{\citenamefont{Bonthuis et~al.}(2008)\citenamefont{Bonthuis, Meyer,
  Stein, and Dekker}}]{bonthuis2008conformation}
\bibinfo{author}{\bibfnamefont{D.~J.} \bibnamefont{Bonthuis}},
  \bibinfo{author}{\bibfnamefont{C.}~\bibnamefont{Meyer}},
  \bibinfo{author}{\bibfnamefont{D.}~\bibnamefont{Stein}}, \bibnamefont{and}
  \bibinfo{author}{\bibfnamefont{C.}~\bibnamefont{Dekker}},
  \bibinfo{journal}{Physical review letters} \textbf{\bibinfo{volume}{101}},
  \bibinfo{pages}{108303} (\bibinfo{year}{2008}).

\bibitem[{\citenamefont{Dubov et~al.}(2017)\citenamefont{Dubov, Molotilin, and
  Vinogradova}}]{Vinogradova2017}
\bibinfo{author}{\bibfnamefont{A.~L.} \bibnamefont{Dubov}},
  \bibinfo{author}{\bibfnamefont{T.~Y.} \bibnamefont{Molotilin}},
  \bibnamefont{and} \bibinfo{author}{\bibfnamefont{O.~I.}
  \bibnamefont{Vinogradova}}, \bibinfo{journal}{Soft Matter}
  pp.~\bibinfo{pages}{--} (\bibinfo{year}{2017}).

\bibitem[{\citenamefont{Saleh and Sohn}(2003)}]{Saleh2003}
\bibinfo{author}{\bibfnamefont{O.~A.} \bibnamefont{Saleh}} \bibnamefont{and}
  \bibinfo{author}{\bibfnamefont{L.~L.} \bibnamefont{Sohn}},
  \bibinfo{journal}{Proc. Natl. Acad. Sci. U. S. A.}
  \textbf{\bibinfo{volume}{100}}, \bibinfo{pages}{820} (\bibinfo{year}{2003}).

\bibitem[{\citenamefont{Ito et~al.}(2004)\citenamefont{Ito, Sun, Bevan, and
  Crooks}}]{Ito2004}
\bibinfo{author}{\bibfnamefont{T.}~\bibnamefont{Ito}},
  \bibinfo{author}{\bibfnamefont{L.}~\bibnamefont{Sun}},
  \bibinfo{author}{\bibfnamefont{M.~A.} \bibnamefont{Bevan}}, \bibnamefont{and}
  \bibinfo{author}{\bibfnamefont{R.~M.} \bibnamefont{Crooks}},
  \textbf{\bibinfo{volume}{20}}, \bibinfo{pages}{6940} (\bibinfo{year}{2004}).

\bibitem[{\citenamefont{Heins et~al.}(2005)\citenamefont{Heins, Siwy, Baker,
  and Martin}}]{Heins2005}
\bibinfo{author}{\bibfnamefont{E.~A.} \bibnamefont{Heins}},
  \bibinfo{author}{\bibfnamefont{Z.~S.} \bibnamefont{Siwy}},
  \bibinfo{author}{\bibfnamefont{L.~A.} \bibnamefont{Baker}}, \bibnamefont{and}
  \bibinfo{author}{\bibfnamefont{R.~C.} \bibnamefont{Martin}},
  \bibinfo{journal}{Nano Lett.} \textbf{\bibinfo{volume}{5}},
  \bibinfo{pages}{1824} (\bibinfo{year}{2005}).

\bibitem[{\citenamefont{Arjmandi et~al.}(2012)\citenamefont{Arjmandi, Van~Roy,
  L., and Borghs}}]{Arjmandi2012}
\bibinfo{author}{\bibfnamefont{N.}~\bibnamefont{Arjmandi}},
  \bibinfo{author}{\bibfnamefont{W.}~\bibnamefont{Van~Roy}},
  \bibinfo{author}{\bibfnamefont{L.}~\bibnamefont{L.}}, \bibnamefont{and}
  \bibinfo{author}{\bibfnamefont{G.}~\bibnamefont{Borghs}},
  \bibinfo{journal}{Anal. Chem.} \textbf{\bibinfo{volume}{84}},
  \bibinfo{pages}{8490} (\bibinfo{year}{2012}).

\bibitem[{\citenamefont{\ifmmode~\check{S}\else \v{S}\fi{}amaj and
  Trizac}(2016)}]{Trizac2016}
\bibinfo{author}{\bibfnamefont{L.}~\bibnamefont{\ifmmode~\check{S}\else
  \v{S}\fi{}amaj}} \bibnamefont{and}
  \bibinfo{author}{\bibfnamefont{E.}~\bibnamefont{Trizac}},
  \bibinfo{journal}{Phys. Rev. E} \textbf{\bibinfo{volume}{93}},
  \bibinfo{pages}{012601} (\bibinfo{year}{2016}).

\bibitem[{\citenamefont{Reindl et~al.}(2017{\natexlab{a}})\citenamefont{Reindl,
  Bier, and Dietrich}}]{Reindl2017_1}
\bibinfo{author}{\bibfnamefont{A.}~\bibnamefont{Reindl}},
  \bibinfo{author}{\bibfnamefont{M.}~\bibnamefont{Bier}}, \bibnamefont{and}
  \bibinfo{author}{\bibfnamefont{S.}~\bibnamefont{Dietrich}},
  \bibinfo{journal}{The Journal of Chemical Physics}
  \textbf{\bibinfo{volume}{146}}, \bibinfo{pages}{154703}
  (\bibinfo{year}{2017}{\natexlab{a}}).

\bibitem[{\citenamefont{Reindl et~al.}(2017{\natexlab{b}})\citenamefont{Reindl,
  Bier, and Dietrich}}]{Reindl2017_2}
\bibinfo{author}{\bibfnamefont{A.}~\bibnamefont{Reindl}},
  \bibinfo{author}{\bibfnamefont{M.}~\bibnamefont{Bier}}, \bibnamefont{and}
  \bibinfo{author}{\bibfnamefont{S.}~\bibnamefont{Dietrich}},
  \bibinfo{journal}{The Journal of Chemical Physics}
  \textbf{\bibinfo{volume}{146}}, \bibinfo{pages}{154704}
  (\bibinfo{year}{2017}{\natexlab{b}}).

\bibitem[{\citenamefont{Brogioli}(2009)}]{Brogioli2009}
\bibinfo{author}{\bibfnamefont{D.}~\bibnamefont{Brogioli}},
  \bibinfo{journal}{Phys. Rev. Lett.} \textbf{\bibinfo{volume}{103}},
  \bibinfo{pages}{058501} (\bibinfo{year}{2009}).

\bibitem[{\citenamefont{Yeh et~al.}(2015)\citenamefont{Yeh, Chang, and
  Yang}}]{Yeh2015}
\bibinfo{author}{\bibfnamefont{H.-C.} \bibnamefont{Yeh}},
  \bibinfo{author}{\bibfnamefont{C.-C.} \bibnamefont{Chang}}, \bibnamefont{and}
  \bibinfo{author}{\bibfnamefont{R.-J.} \bibnamefont{Yang}},
  \bibinfo{journal}{Phys. Rev. E} \textbf{\bibinfo{volume}{91}},
  \bibinfo{pages}{062302} (\bibinfo{year}{2015}).

\bibitem[{\citenamefont{Siwy et~al.}(2005)\citenamefont{Siwy,
  Kosi\ifmmode~\acute{n}\else \'{n}\fi{}ska, Fuli\ifmmode~\acute{n}\else
  \'{n}\fi{}ski, and Martin}}]{Siwy2005}
\bibinfo{author}{\bibfnamefont{Z.}~\bibnamefont{Siwy}},
  \bibinfo{author}{\bibfnamefont{I.~D.}
  \bibnamefont{Kosi\ifmmode~\acute{n}\else \'{n}\fi{}ska}},
  \bibinfo{author}{\bibfnamefont{A.}~\bibnamefont{Fuli\ifmmode~\acute{n}\else
  \'{n}\fi{}ski}}, \bibnamefont{and} \bibinfo{author}{\bibfnamefont{C.~R.}
  \bibnamefont{Martin}}, \bibinfo{journal}{Phys. Rev. Lett.}
  \textbf{\bibinfo{volume}{94}}, \bibinfo{pages}{048102}
  (\bibinfo{year}{2005}).

\bibitem[{\citenamefont{Kosinska et~al.}(2008)\citenamefont{Kosinska, Goychuk,
  Kostur, Schmidt, and H\"{a}nggi}}]{hanggi}
\bibinfo{author}{\bibfnamefont{I.}~\bibnamefont{Kosinska}},
  \bibinfo{author}{\bibfnamefont{I.}~\bibnamefont{Goychuk}},
  \bibinfo{author}{\bibfnamefont{M.}~\bibnamefont{Kostur}},
  \bibinfo{author}{\bibfnamefont{G.}~\bibnamefont{Schmidt}}, \bibnamefont{and}
  \bibinfo{author}{\bibfnamefont{P.}~\bibnamefont{H\"{a}nggi}},
  \bibinfo{journal}{Phys. Rev. E} \textbf{\bibinfo{volume}{77}},
  \bibinfo{pages}{031131} (\bibinfo{year}{2008}).

\bibitem[{\citenamefont{Gomez et~al.}(2015)\citenamefont{Gomez, Ramirez,
  Cervera, Nasir, Ali, Ensinger, and Mafe}}]{Gomez2015}
\bibinfo{author}{\bibfnamefont{V.}~\bibnamefont{Gomez}},
  \bibinfo{author}{\bibfnamefont{P.}~\bibnamefont{Ramirez}},
  \bibinfo{author}{\bibfnamefont{J.}~\bibnamefont{Cervera}},
  \bibinfo{author}{\bibfnamefont{S.}~\bibnamefont{Nasir}},
  \bibinfo{author}{\bibfnamefont{M.}~\bibnamefont{Ali}},
  \bibinfo{author}{\bibfnamefont{W.}~\bibnamefont{Ensinger}}, \bibnamefont{and}
  \bibinfo{author}{\bibfnamefont{S.}~\bibnamefont{Mafe}},
  \bibinfo{journal}{Scientific Reports} \textbf{\bibinfo{volume}{5}},
  \bibinfo{pages}{9501 EP } (\bibinfo{year}{2015}), \bibinfo{note}{article}.

\bibitem[{\citenamefont{Laohakunakorn and Keyser}(2015)}]{Keyser2015}
\bibinfo{author}{\bibfnamefont{N.}~\bibnamefont{Laohakunakorn}}
  \bibnamefont{and} \bibinfo{author}{\bibfnamefont{U.~F.}
  \bibnamefont{Keyser}}, \bibinfo{journal}{Nanotechnology}
  \textbf{\bibinfo{volume}{26}}, \bibinfo{pages}{275202}
  (\bibinfo{year}{2015}).

\bibitem[{\citenamefont{Lairez et~al.}(2016)\citenamefont{Lairez, Clochard, and
  Wegrowe}}]{Wegrowe2016}
\bibinfo{author}{\bibfnamefont{D.}~\bibnamefont{Lairez}},
  \bibinfo{author}{\bibfnamefont{M.-C.} \bibnamefont{Clochard}},
  \bibnamefont{and} \bibinfo{author}{\bibfnamefont{J.-E.}
  \bibnamefont{Wegrowe}}, \bibinfo{journal}{Sci. Rep.}
  \textbf{\bibinfo{volume}{6}}, \bibinfo{pages}{38966} (\bibinfo{year}{2016}).

\bibitem[{\citenamefont{Park et~al.}(2006)\citenamefont{Park, Russo, Branton,
  and Stone}}]{Park2006}
\bibinfo{author}{\bibfnamefont{S.~Y.} \bibnamefont{Park}},
  \bibinfo{author}{\bibfnamefont{C.~J.} \bibnamefont{Russo}},
  \bibinfo{author}{\bibfnamefont{D.}~\bibnamefont{Branton}}, \bibnamefont{and}
  \bibinfo{author}{\bibfnamefont{H.~A.} \bibnamefont{Stone}},
  \bibinfo{journal}{Journal of Colloid and Interface Science}
  \textbf{\bibinfo{volume}{297}}, \bibinfo{pages}{832} (\bibinfo{year}{2006}).

\bibitem[{\citenamefont{Mani et~al.}(2009)\citenamefont{Mani, Zangle, and
  Santiago}}]{mani2009propagation}
\bibinfo{author}{\bibfnamefont{A.}~\bibnamefont{Mani}},
  \bibinfo{author}{\bibfnamefont{T.~A.} \bibnamefont{Zangle}},
  \bibnamefont{and} \bibinfo{author}{\bibfnamefont{J.~G.}
  \bibnamefont{Santiago}}, \bibinfo{journal}{Langmuir}
  \textbf{\bibinfo{volume}{25}}, \bibinfo{pages}{3898} (\bibinfo{year}{2009}).

\bibitem[{\citenamefont{Malgaretti et~al.}(2014)\citenamefont{Malgaretti,
  Pagonabarraga, and Rubi}}]{Malgaretti2014}
\bibinfo{author}{\bibfnamefont{P.}~\bibnamefont{Malgaretti}},
  \bibinfo{author}{\bibfnamefont{I.}~\bibnamefont{Pagonabarraga}},
  \bibnamefont{and} \bibinfo{author}{\bibfnamefont{J.~M.} \bibnamefont{Rubi}},
  \bibinfo{journal}{Phys. Rev. Lett} \textbf{\bibinfo{volume}{113}},
  \bibinfo{pages}{128301} (\bibinfo{year}{2014}).

\bibitem[{\citenamefont{Chinappi and Malgaretti}(2018)}]{Chinappi2018}
\bibinfo{author}{\bibfnamefont{M.}~\bibnamefont{Chinappi}} \bibnamefont{and}
  \bibinfo{author}{\bibfnamefont{P.}~\bibnamefont{Malgaretti}},
  \bibinfo{journal}{Soft Matter} \textbf{\bibinfo{volume}{14}},
  \bibinfo{pages}{9083} (\bibinfo{year}{2018}).

\bibitem[{\citenamefont{Malgaretti et~al.}(2015)\citenamefont{Malgaretti,
  Pagonabarraga, and Rubi}}]{Malgaretti2015}
\bibinfo{author}{\bibfnamefont{P.}~\bibnamefont{Malgaretti}},
  \bibinfo{author}{\bibfnamefont{I.}~\bibnamefont{Pagonabarraga}},
  \bibnamefont{and} \bibinfo{author}{\bibfnamefont{J.~M.} \bibnamefont{Rubi}},
  \bibinfo{journal}{Macromol. Symposia} \textbf{\bibinfo{volume}{357}},
  \bibinfo{pages}{178} (\bibinfo{year}{2015}).

\bibitem[{\citenamefont{Malgaretti et~al.}(2016)\citenamefont{Malgaretti,
  Pagonabarraga, and Miguel~Rubi}}]{Malgaretti2016}
\bibinfo{author}{\bibfnamefont{P.}~\bibnamefont{Malgaretti}},
  \bibinfo{author}{\bibfnamefont{I.}~\bibnamefont{Pagonabarraga}},
  \bibnamefont{and}
  \bibinfo{author}{\bibfnamefont{J.}~\bibnamefont{Miguel~Rubi}},
  \bibinfo{journal}{The Journal of Chemical Physics}
  \textbf{\bibinfo{volume}{144}}, \bibinfo{pages}{034901}
  (\bibinfo{year}{2016}).

\bibitem[{\citenamefont{Janssen et~al.}(2017)\citenamefont{Janssen, Griffioen,
  Biesheuvel, van Roij, and Ern\'e}}]{janssen2017coulometry}
\bibinfo{author}{\bibfnamefont{M.}~\bibnamefont{Janssen}},
  \bibinfo{author}{\bibfnamefont{E.}~\bibnamefont{Griffioen}},
  \bibinfo{author}{\bibfnamefont{P.~M.} \bibnamefont{Biesheuvel}},
  \bibinfo{author}{\bibfnamefont{R.}~\bibnamefont{van Roij}}, \bibnamefont{and}
  \bibinfo{author}{\bibfnamefont{B.}~\bibnamefont{Ern\'e}},
  \bibinfo{journal}{Phys. Rev. Lett.} \textbf{\bibinfo{volume}{119}},
  \bibinfo{pages}{166002} (\bibinfo{year}{2017}).

\bibitem[{\citenamefont{Werkhoven et~al.}(2018)\citenamefont{Werkhoven, Everts,
  Samin, and van Roij}}]{Werkhoven2018}
\bibinfo{author}{\bibfnamefont{B.~L.} \bibnamefont{Werkhoven}},
  \bibinfo{author}{\bibfnamefont{J.~C.} \bibnamefont{Everts}},
  \bibinfo{author}{\bibfnamefont{S.}~\bibnamefont{Samin}}, \bibnamefont{and}
  \bibinfo{author}{\bibfnamefont{R.}~\bibnamefont{van Roij}},
  \bibinfo{journal}{Phys. Rev. Lett.} \textbf{\bibinfo{volume}{120}},
  \bibinfo{pages}{264502} (\bibinfo{year}{2018}).

\bibitem[{\citenamefont{Russel et~al.}(1989)\citenamefont{Russel, Saville, and
  Schowalter}}]{RusselBook}
\bibinfo{author}{\bibfnamefont{W.~B.} \bibnamefont{Russel}},
  \bibinfo{author}{\bibfnamefont{W.~B.} \bibnamefont{Saville}},
  \bibnamefont{and} \bibinfo{author}{\bibfnamefont{W.~R.}
  \bibnamefont{Schowalter}}, \emph{\bibinfo{title}{Colloidal Dispersions}}
  (\bibinfo{publisher}{Cambridge University Press}, \bibinfo{year}{1989}).

\bibitem[{\citenamefont{Luo et~al.}(2015)\citenamefont{Luo, Xing, Ling,
  Kleinhammes, and Wu}}]{luo2015electroneutrality}
\bibinfo{author}{\bibfnamefont{Z.-X.} \bibnamefont{Luo}},
  \bibinfo{author}{\bibfnamefont{Y.-Z.} \bibnamefont{Xing}},
  \bibinfo{author}{\bibfnamefont{Y.-C.} \bibnamefont{Ling}},
  \bibinfo{author}{\bibfnamefont{A.}~\bibnamefont{Kleinhammes}},
  \bibnamefont{and} \bibinfo{author}{\bibfnamefont{Y.}~\bibnamefont{Wu}},
  \bibinfo{journal}{Nature communications} \textbf{\bibinfo{volume}{6}},
  \bibinfo{pages}{6358} (\bibinfo{year}{2015}).

\bibitem[{\citenamefont{Colla et~al.}(2016)\citenamefont{Colla, Girotto, dos
  Santos, and Levin}}]{colla2016charge}
\bibinfo{author}{\bibfnamefont{T.}~\bibnamefont{Colla}},
  \bibinfo{author}{\bibfnamefont{M.}~\bibnamefont{Girotto}},
  \bibinfo{author}{\bibfnamefont{A.~P.} \bibnamefont{dos Santos}},
  \bibnamefont{and} \bibinfo{author}{\bibfnamefont{Y.}~\bibnamefont{Levin}},
  \bibinfo{journal}{The Journal of chemical physics}
  \textbf{\bibinfo{volume}{145}}, \bibinfo{pages}{094704}
  (\bibinfo{year}{2016}).

\bibitem[{\citenamefont{van~der Heyden et~al.}(2005)\citenamefont{van~der
  Heyden, Stein, and Dekker}}]{Dekker2005}
\bibinfo{author}{\bibfnamefont{F.~H.~J.} \bibnamefont{van~der Heyden}},
  \bibinfo{author}{\bibfnamefont{D.}~\bibnamefont{Stein}}, \bibnamefont{and}
  \bibinfo{author}{\bibfnamefont{C.}~\bibnamefont{Dekker}},
  \bibinfo{journal}{Phys. Rev. Lett.} \textbf{\bibinfo{volume}{95}},
  \bibinfo{pages}{116104} (\bibinfo{year}{2005}).

\bibitem[{\citenamefont{Gross and Osterle}(1968)}]{gross1968membrane}
\bibinfo{author}{\bibfnamefont{R.~J.} \bibnamefont{Gross}} \bibnamefont{and}
  \bibinfo{author}{\bibfnamefont{J.}~\bibnamefont{Osterle}},
  \bibinfo{journal}{The Journal of chemical physics}
  \textbf{\bibinfo{volume}{49}}, \bibinfo{pages}{228} (\bibinfo{year}{1968}).

\bibitem[{\citenamefont{Ajdari}(2001)}]{ajdari2001transverse}
\bibinfo{author}{\bibfnamefont{A.}~\bibnamefont{Ajdari}},
  \bibinfo{journal}{Physical Review E} \textbf{\bibinfo{volume}{65}},
  \bibinfo{pages}{016301} (\bibinfo{year}{2001}).

\bibitem[{\citenamefont{Ghosal}(2002)}]{ghosal2002lubrication}
\bibinfo{author}{\bibfnamefont{S.}~\bibnamefont{Ghosal}},
  \bibinfo{journal}{Journal of Fluid Mechanics} \textbf{\bibinfo{volume}{459}},
  \bibinfo{pages}{103} (\bibinfo{year}{2002}).

\bibitem[{\citenamefont{Bruus}(2008)}]{bruus2008theoretical}
\bibinfo{author}{\bibfnamefont{H.}~\bibnamefont{Bruus}},
  \emph{\bibinfo{title}{Theoretical microfluidics}}, vol.~\bibinfo{volume}{18}
  (\bibinfo{publisher}{Oxford university press Oxford}, \bibinfo{year}{2008}).

\bibitem[{\citenamefont{Yoshida et~al.}(2016)\citenamefont{Yoshida, Kinjo, and
  Washizu}}]{yoshida2016analysis}
\bibinfo{author}{\bibfnamefont{H.}~\bibnamefont{Yoshida}},
  \bibinfo{author}{\bibfnamefont{T.}~\bibnamefont{Kinjo}}, \bibnamefont{and}
  \bibinfo{author}{\bibfnamefont{H.}~\bibnamefont{Washizu}},
  \bibinfo{journal}{Computers \& Fluids} \textbf{\bibinfo{volume}{124}},
  \bibinfo{pages}{237} (\bibinfo{year}{2016}).

\bibitem[{\citenamefont{Delgado et~al.}(2007)\citenamefont{Delgado,
  Gonz{\'a}lez-Caballero, Hunter, Koopal, and
  Lyklema}}]{delgado2007measurement}
\bibinfo{author}{\bibfnamefont{{\'A}.~V.} \bibnamefont{Delgado}},
  \bibinfo{author}{\bibfnamefont{F.}~\bibnamefont{Gonz{\'a}lez-Caballero}},
  \bibinfo{author}{\bibfnamefont{R.}~\bibnamefont{Hunter}},
  \bibinfo{author}{\bibfnamefont{L.}~\bibnamefont{Koopal}}, \bibnamefont{and}
  \bibinfo{author}{\bibfnamefont{J.}~\bibnamefont{Lyklema}},
  \bibinfo{journal}{Journal of colloid and interface science}
  \textbf{\bibinfo{volume}{309}}, \bibinfo{pages}{194} (\bibinfo{year}{2007}).

\bibitem[{\citenamefont{Peters et~al.}(2016)\citenamefont{Peters, van Roij,
  Bazant, and Biesheuvel}}]{VanRoij2016}
\bibinfo{author}{\bibfnamefont{P.~B.} \bibnamefont{Peters}},
  \bibinfo{author}{\bibfnamefont{R.}~\bibnamefont{van Roij}},
  \bibinfo{author}{\bibfnamefont{M.~Z.} \bibnamefont{Bazant}},
  \bibnamefont{and} \bibinfo{author}{\bibfnamefont{P.~M.}
  \bibnamefont{Biesheuvel}}, \bibinfo{journal}{Phys. Rev. E}
  \textbf{\bibinfo{volume}{93}}, \bibinfo{pages}{053108}
  (\bibinfo{year}{2016}).

\bibitem[{\citenamefont{de~Groot and Mazur}(1983)}]{DeGroot}
\bibinfo{author}{\bibfnamefont{S.~R.} \bibnamefont{de~Groot}} \bibnamefont{and}
  \bibinfo{author}{\bibfnamefont{P.}~\bibnamefont{Mazur}},
  \emph{\bibinfo{title}{Non-Equilibrium Thermodynamics}}
  (\bibinfo{publisher}{Dover}, \bibinfo{address}{Amsterdam},
  \bibinfo{year}{1983}).

\end{thebibliography}

\end{document}